\documentclass[aps,prb,superscriptaddress,preprintnumbers,twocolumn,showpacs]{revtex4}

\usepackage{psfrag,graphicx}
\usepackage{dcolumn}
\usepackage{amsmath,amssymb}
\usepackage{bm}
\usepackage{amsfonts,amssymb,amsmath}        
\usepackage{multirow}

\usepackage{epstopdf}

\newcommand{\be}{\begin{equation}}
\newcommand{\ee}{\end{equation}}
\newcommand{\bq}{\begin{eqnarray}}
\newcommand{\eq}{\end{eqnarray}}

\newcommand{\rf}[1]{(\ref{#1})}

\newcommand{\p}{\mathbf{p}}

\graphicspath{{figures/}}

\begin{document}

\title{Perturbed vortex lattices and the stability of nucleated topological phases}

\author{Ville Lahtinen}
\affiliation{Institute for Theoretical Physics, University of Amsterdam, Science Park 904, NL-1090 GL Amsterdam, Netherlands}
\affiliation{Institute-Lorentz for Theoretical Physics, Leiden University, PO Box 9506, NL-2300 RA Leiden, Netherlands}

\author{Andreas W. W. Ludwig} 
\affiliation{Department of Physics, University of California, Santa Barbara, CA 93106, USA}

\author{Simon Trebst}
\affiliation{Institute for Theoretical Physics, University of Cologne, 50937 Cologne, Germany}

\date{\today}

\begin{abstract}

We study the stability of nucleated topological phases that can emerge when interacting non-Abelian anyons form a regular array. The studies are carried out in the context of Kitaev's honeycomb model, where we consider three distinct types of perturbations in the presence of a lattice of Majorana mode binding vortices -- spatial anisotropy of the vortices, dimerization of the vortex lattice and local random disorder. While all the nucleated phases are stable with respect to weak perturbations of each kind, strong perturbations are found to result in very different behavior. Anisotropy of the vortices stabilizes the strong-pairing like phases, while dimerization can recover the underlying non-Abelian phase. Local random disorder, on the other hand, can drive all the nucleated phases into a gapless thermal metal state. We show that all these distinct behavior can be captured by an effective staggered tight-binding model for the Majorana modes. By studying the pairwise interactions between the vortices, i.e. the amplitudes for the Majorana modes to tunnel between vortex cores, the locations of phase transitions and the nature of the resulting states can be predicted. We also find that due to oscillations in the Majorana tunneling amplitude, lattices of Majorana modes may exhibit a Peierls-like instability, where a dimerized configuration is favored over a uniform lattice. As the nature of the nucleated phases depends only on the Majorana tunneling, our results are expected to apply also to other system supporting localized Majorana mode arrays, such as Abrikosov lattices in $p$-wave superconductors, Wigner crystals in Moore-Read fractional quantum Hall states or arrays of topological nanowires.

\end{abstract}

\pacs{05.30.Pr, 73.43.Nq, 74.25.Uv, 75.10.Jm}

\maketitle

\section{Introduction}

Topologically ordered states of matter can be characterized by topological invariants. In the bulk of a two dimensional topologically ordered system, their non-trivial values can imply that the vortex or quasiparticle excitations exhibit anyonic statistics. Of particular interest are the so called non-Abelian anyons. Their presence leads to a macroscopic ground state degeneracy, which has been proposed as a robust memory to store and manipulate quantum information \cite{Pachos12}. While the details of such topological quantum computing schemes depend on the specific platform, the basic idea is to have a large number of anyons, usually arranged in a regular array to enable systematic control, which are locally manipulated to implement robust quantum gates \cite{Alicea11, Halperin12,Hyart13}. 

As quasiparticles the non-Abelian anyons will always be interacting due to system specific microscopics. These interactions mean that the topological degeneracy will only be exact in the limit of infinite quasiparticle separation \cite{Bonderson09}. Microscopics of this degeneracy lifting have been analyzed in several systems including Moore-Read fractional quantum Hall states \cite{Baraban09}, $p$-wave superconductors \cite{Cheng09}, Kitaev's honeycomb model \cite{Lahtinen11} and topological nanowires \cite{Sau10}. As any realization of topological order will ultimately be in a finite system, the anyons are forced to be in proximity to each other and thus the interactions are rarely negligible. For instance, it has been recently appreciated that they can have direct consequences for the experiments to detect non-Abelian anyons \cite{Rosenow12,Ben-Shach13}. A more dramatic consequence is the possibility of interacting non-Abelian anyons to form a new collective topological state \cite{Gils09,Ludwig11}. This mechanism of \emph{topological liquid nucleation} has been postulated to occur when the anyons form a regular array, such as an Abrikosov vortex lattice in a topological superconductor \cite{Grosfeld06, Silaev13,Biswas13} or a quasiparticle Wigner crystal in a quantum Hall state \cite{Zhu10}. It has been recently realized that topological phase transition in quasi 1D systems can also be related to this mechanism \cite{Kells12}.

While the interactions underlie the nucleation mechanism, there is no critical interaction strength -- if the vortex lattice is uniform nucleation should always occur. However, as the interaction strength decays with increasing anyon separation, so does the interaction induced energy gap. Thus disorder effects are expected to become more and more relevant as the vortex lattices get sparser. In this paper we study microscopically three distinct types of perturbations in non-Abelian anyon arrays: anisotropic interactions due to anisotropic vortices, dimerization of the vortex lattice and random local vortex position disorder. We perform the studies in the context of Kitaev's honeycomb lattice model \cite{Kitaev06}, where the microscopics of nucleation have been previously studied \cite{Lahtinen12}. In particular, it was shown that the collective state of the vortices in the non-Abelian phase of the model could be fully understood from a model of free Majorana fermions tunneling between the vortex cores. Here we generalize the corresponding tight-binding model to include anisotropic, staggered and random couplings, which we show to be sufficient to fully capture the behavior of the perturbed vortex lattices in the honeycomb model. Our main result is that while the nucleated phases are stable to moderate perturbations of all types -- as expected for a gapped topological phase -- very different physics is obtained for strong perturbations. Anisotropy in the interactions stabilizes the strong-pairing-like phases,  while dimerization can drive the system back to the underlying non-Abelian phase. Local random disorder, on the other hand,  is found to drive the nucleated phases to a recently predicted thermal metal state \cite{Laumann12}. Furthermore, we find that the Majorana tunneling can give rise to a Peierls-like instability, where the Majorana mode binding vortices may energetically prefer to dimerize over forming an uniform lattice.

Qualitatively similar collective behavior is expected to occur in any system where localized Majorana modes can tunnel on a regular array. By studying the system specific pairwise tunneling amplitudes\cite{Lahtinen11, Cheng09, Baraban09}, the collective state of the system can be predicted from the generalized Majorana tight-binding model presented here. In addition to the honeycomb model \cite{Kitaev06} and it is generalizations \cite{Yao07,Kells11}, Majorana arrays may naturally occur as Abrikosov vortex lattices in $p$-wave superconductors \cite{Read00} or as quasihole Wigner crystals in Moore-Read fractional quantum Hall states \cite{Moore91}. They could also be engineered in two dimensional arrays of topological nanowires \cite{Alicea11,Kells13} or realized in optical lattice experiments for fractional quantum Hall physics \cite{Dalibard13}. The latter offers a particularly exciting prospect for probing the collective behavior of Majorana modes due to the precise control one can have over the vortex lattice geometry \cite{Komineas12}. The disorder effects considered in the present work are relevant to all these microscopicallly distinct systems due to the ever present impurities. 

This paper is structured as follows. In Section \ref{sec:honeycomb} we review the solution of the honeycomb model. In Section \ref{sec:PD} the phase diagram and the characteristic band structure in the presence of vortex lattices is studied. In Section \ref{sec:Majmodel} we review the vortex-vortex interactions and introduce the tight-binding Majorana model to study perturbed vortex lattices. In Section \ref{sec:disorder} we apply this model study the effects of vortex anisotropy, dimerization of the vortex lattice and random local disorder. This section contains the main results of our work. In Appendix \ref{App_stagger} we present an analytic solution to a staggered Majorana model and study its general properties. Appendix \ref{App_data} contains supporting supplementary data.

\section{Vortices in the honeycomb lattice model}
\label{sec:honeycomb}

Kitaev's honeycomb model is a lattice model of spin $1/2$-particles residing on the vertices of a honeycomb lattice \cite{Kitaev06}. The spins interact according to the Hamiltonian
\be \label{H_honey}
	H = \sum_{\alpha =x,y,z} \sum_{\langle i,j \rangle} J_\alpha \sigma_i^\alpha \sigma_j^\alpha + K \sum_{\langle i,j,k \rangle} \sigma_i^\alpha \sigma_j^\beta \sigma_k^\gamma,
\ee
where $J_\alpha$ are nearest neighbour spin exchange couplings along links of type $\alpha$ and $K$ is the magnitude of a three spin term that explicitly breaks time reversal symmetry. The latter is required for the model to support gapped topological phases characterized by non-zero Chern numbers. For every hexagonal plaquette $p$ one can associate a $Z_2$ valued six spin operator $\hat{W}_p=\sigma_1^x \sigma_2^y \sigma_3^z \sigma_4^x \sigma_5^y \sigma_6^z$ that describes a local symmetry $[H,\hat{W}_p]=0$. The Hilbert space of the spin model thus breaks into sectors labeled by the patterns $W=\{ W_p \}$ of the eigenvalues of $\hat{W}_p$. We refer to these sectors as {\it vortex sectors}, because, as we argue below, $W_p=-1$ corresponds to having a $\pi$-flux vortex on plaquette $p$.

The interacting spin system \rf{H_honey} can be mapped to a system of Majorana fermions $c_i=c_i^\dagger$ coupled to a $Z_2$ gauge field $\hat{u}_{ij}$ \cite{Kitaev06,Lahtinen08}. The corresponding Hamiltonian is then given by
\be \label{H_honey_Maj}
	H = \frac{i}{2}\sum_{\langle i,j \rangle} J_{ij} \hat{u}_{ij} c_i c_j + \frac{i}{2} K \sum_{\langle \langle i,j\rangle\rangle} \hat{u}_{ik} \hat{u}_{kj} c_i c_j,
\ee
where the first sum is over nearest neighbour sites $\langle i,j \rangle$, $J_{ij}=J_x, J_y$ or $J_z$ depending on the type of link, and the second over next nearest neighbours $\langle\langle i,j \rangle\rangle$ with $k$ denoting the connecting site. The gauge field is static, i.e. the local gauge potentials satisfy $[H,\hat{u}_{ij}]=0$. The plaquette operators become $Z_2$ valued Wilson loop operators $\hat{W}_p=\prod_{(i,j)\in p} \hat{u}_{ij}$, which justifies the interpretation of the eigenvalues $W_p=-1$ corresponding to the presence of a $\pi$-flux vortex on plaquette $p$. By choosing a gauge, i.e. replacing the operators $\hat{u}_{ij}$ with their eigenvalues $u_{ij}=\pm1$, one restricts to a particular vortex sector $W(u)$. In each sector the Hamiltonian $H_{W(u)}$ is quadratic in the $c_i$'s and hence readily diagonalizable, with the resulting spectrum of free fermions depending only on the vortex sector $W$. This spectrum encodes the physics of the underlying vortices, whose properties can be probed by studying the spectrum over various vortex sectors\cite{Lahtinen11}.

There is a subtlety in the mapping from the spins to free Majorana fermions. The gauge field $\hat{u}_{ij}$ emerges as one embeds the Hilbert space of spins into an enlargened space of fermionic modes and then subsequently projecting the states back to the physical space \cite{Kitaev06, Pedrocchi11}. This amounts to imposing a set of local constraints on the physical states, as well as a global constraint on the fermionic parity $P_W$ that depends on the vortex sector $W$. The latter can be obtained either from the eigenvectors \cite{Pedrocchi11} or through a singular value decomposition \cite{Kells09}. The fermionic parity needs to be taken into account when connecting the vortex-vortex interactions to the nucleated phases.

\subsection{Simulating vortex transport by tuning the spin exchange couplings}

The vortex sectors form a discrete set with the states in different sectors being orthogonal to each other. However, the spectra of two distinct sectors can be adiabatically connected by noticing that in \rf{H_honey_Maj} the local gauge potentials $\hat{u}_{ij}$ are always uniquely paired with the local couplings $J_{ij}$. Tuning adiabatically $J_{ij} \to 0 \to -J_{ij}$ (and the corresponding couplings $K$) will therefore interpolate between Hamiltonians $H_{W(\{u_{ij}\})}$ and $H_{W(\{-u_{ij}\})}$ that differ by the vortex occupation on the two plaquettes sharing the link $(i,j)$. This means that while the vortex sector, as characterized by the pattern of eigenvalues $W$, does not change under such adiabatic process, the spectrum will smoothly interpolate between the two orthogonal vortex sectors.

This process can be viewed as simulating adiabatic vortex transport, which has been employed to verify the non-Abelian statistics of the vortices \cite{Lahtinen09,Bolukbasi12} and to uncover the oscillating interactions between them \cite{Lahtinen11}. Here we will employ this equivalence between coupling and gauge configurations to simulate perturbations in vortex lattices. From now on, when talking about perturbing vortex lattices, we thus mean that we perturb the couplings $J_{ij}$ in the corresponding manner.

\section{The phase diagram and the vortex sectors}
\label{sec:PD}

In this section we first review the phase diagram of the honeycomb model both in the absence and presence of vortex lattices. Then we discuss the characteristic band structure that arises in the presence of vortex lattices.

\subsection{Vortex-free sector: the non-Abelian phase}

\begin{figure}[t]
\begin{tabular}{c}
\includegraphics[width=7cm]{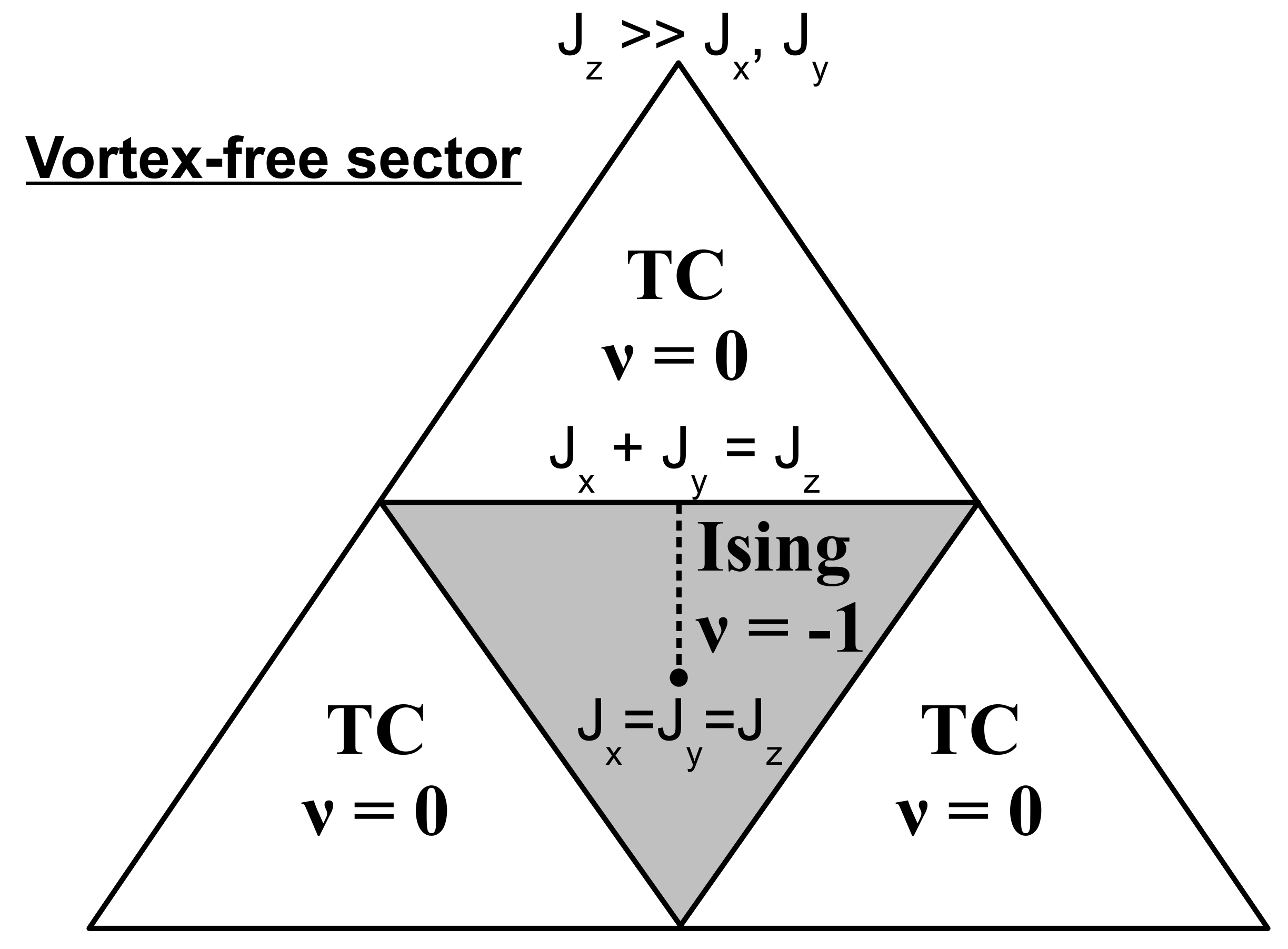} \\ (a) \\ \includegraphics[width=7cm]{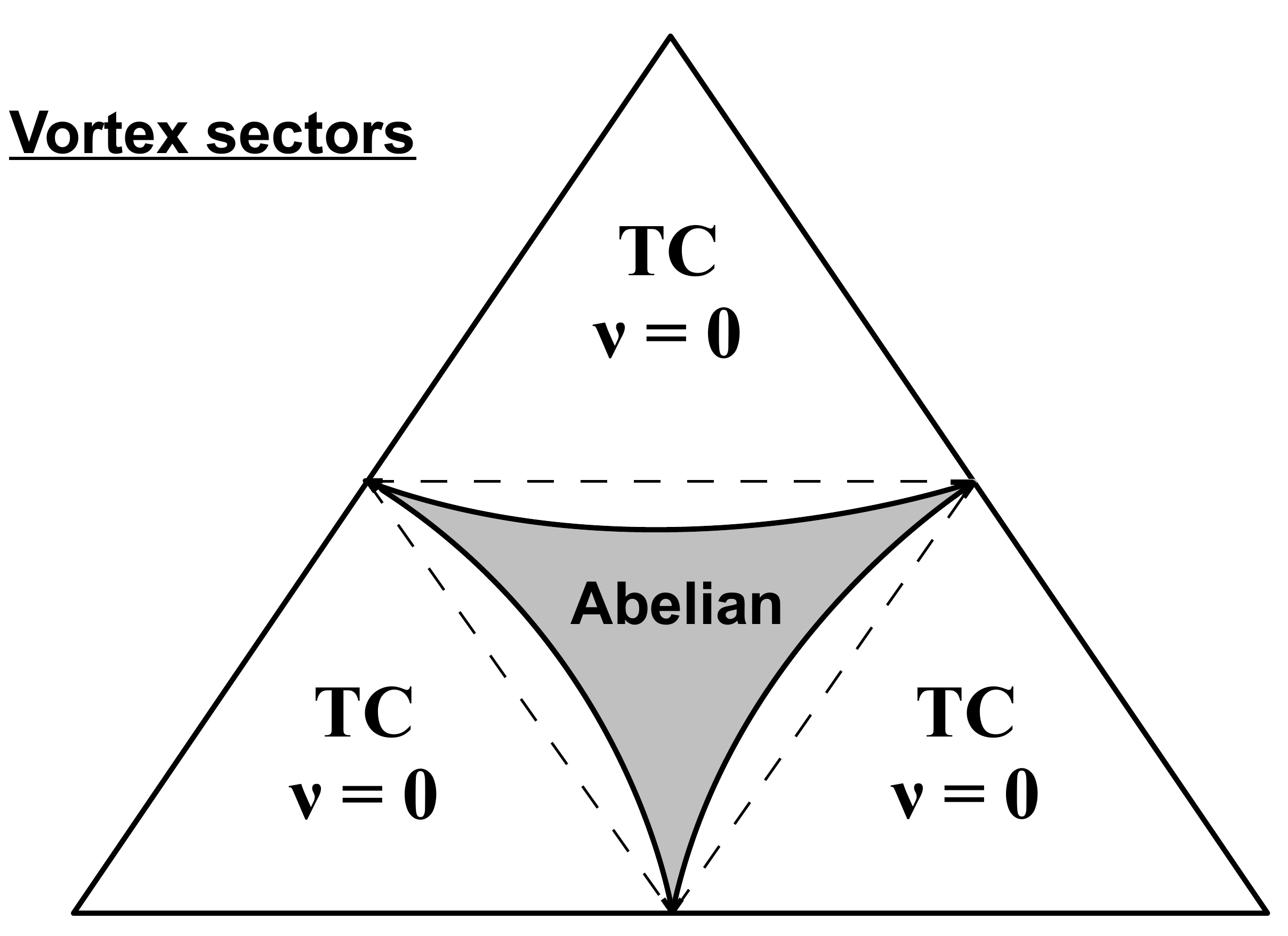} \\ (b)
\end{tabular}
\caption{\label{PD}  The phase diagram in (a) the vortex-free sector and in (b) the vortex lattice sectors when $K > 0$. In the presence of a vortex lattice, the non-Abelian phase characterized by Chern number $\nu=-1$ is always replaced by one or more Abelian phases characterized by even Chern numbers. The strong-pairing like TC phases are also always stabilized. The data how the phase diagram is modified for a particular vortex lattice of spacing $D$ is given in Appendix \ref{App_data}. }
\end{figure}

The ground state over all vortex sectors resides in the vortex-free sector ($W_p=1$ on all plaquettes) \cite{Lieb94}. The phase diagram of this sector, as illustrated in Fig.~\ref{PD}, admits an analytic solution that has been obtained by Kitaev in his original work \cite{Kitaev06}. When $J_\alpha < J_\beta + J_\gamma$ for all permutations of $\alpha,\beta,\gamma=x,y,z$ and $K \neq 0$, the system is in a gapped topological phase characterized by Chern number $\nu=-\textrm{sign}(K)=\pm1$. This means if one introduces few vortices, i.e. considers vortex sectors with few $W_p=-1$ eigenvalues, one finds isolated exponentially localized modes on the sites around these plaquettes.\cite{Kells10-2} These modes are Majorana modes and the vortices binding them exhibit non-Abelian statistics of Ising anyons\cite{Lahtinen09}. In the language of $p$-wave superconductors, this phase is adiabatically connected to the weak-pairing phase\cite{Yu08}.

In the three opposing limits $J_\alpha > J_\beta + J_\gamma$ the spins dimerize and the system will be in gapped topological phases with Chern number $\nu=0$. In these strong-pairing like phases the vortices behave as Abelian anyons with mutually semionic statistics (often referred to as toric code anyons). We denote these spin dimerized phases by TC to distinguish them from the $\nu=0$ phases that can emerge in the presence of vortex lattices.

Unless otherwise mentioned, we use the parametrization $J=J_x=J_y$ and $J_z=1$. In this notation the TC phase spans the regime $0 < J < \frac{1}{2}$, while the non-Abelian Ising phase spans the $\frac{1}{2} < J \leq 1$ regime. We often consider the isotropic $J=1$ couplings at the center of the non-Abelian phase, where the gap is maximized and the model has rotational $C_3$ symmetry.

\subsection{Vortex sectors: nucleated Abelian phases}

\begin{figure}[t]
\begin{tabular}{cc}
\includegraphics[width=4.1cm]{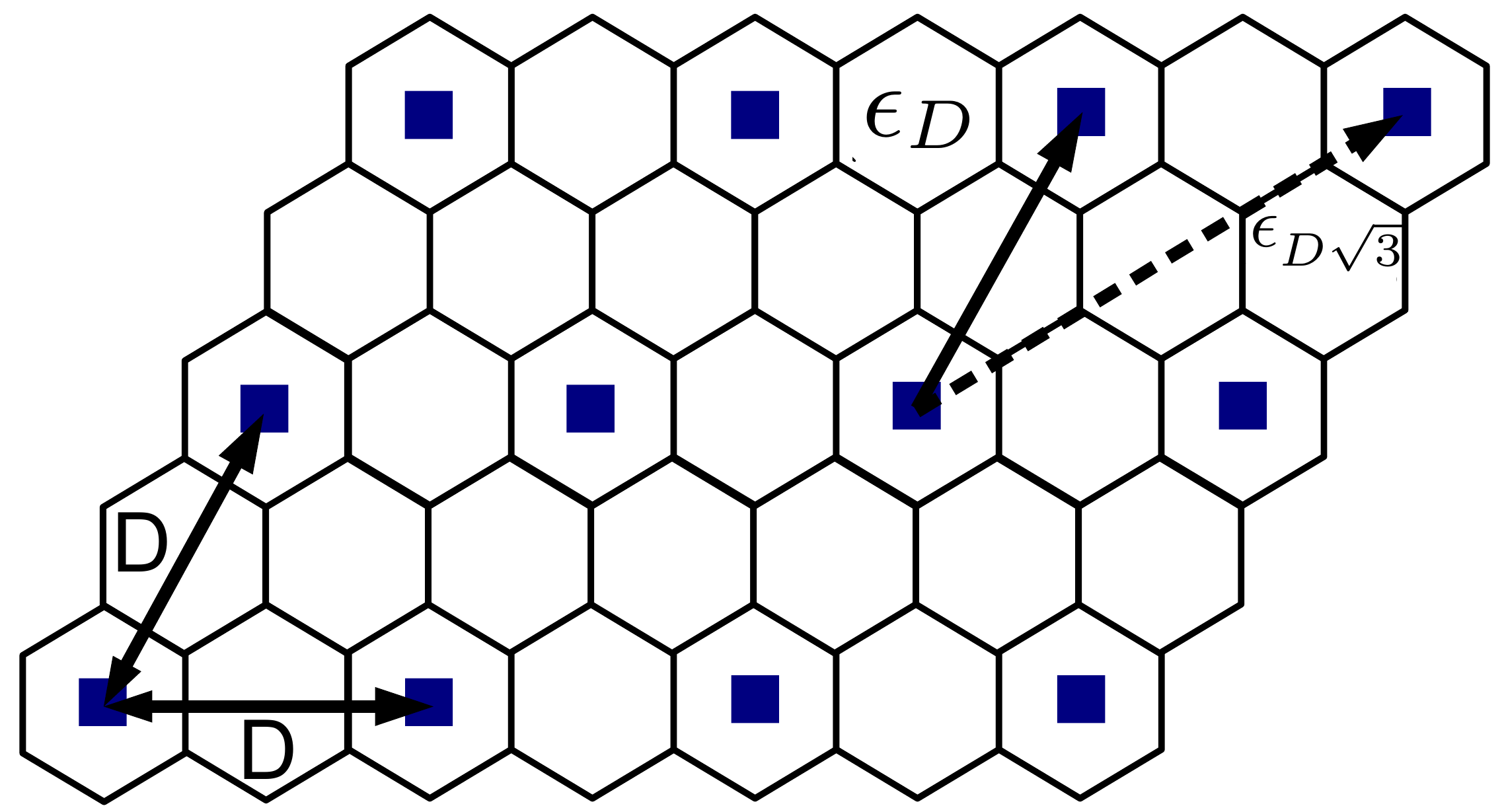} 
& \includegraphics[width=4.1cm]{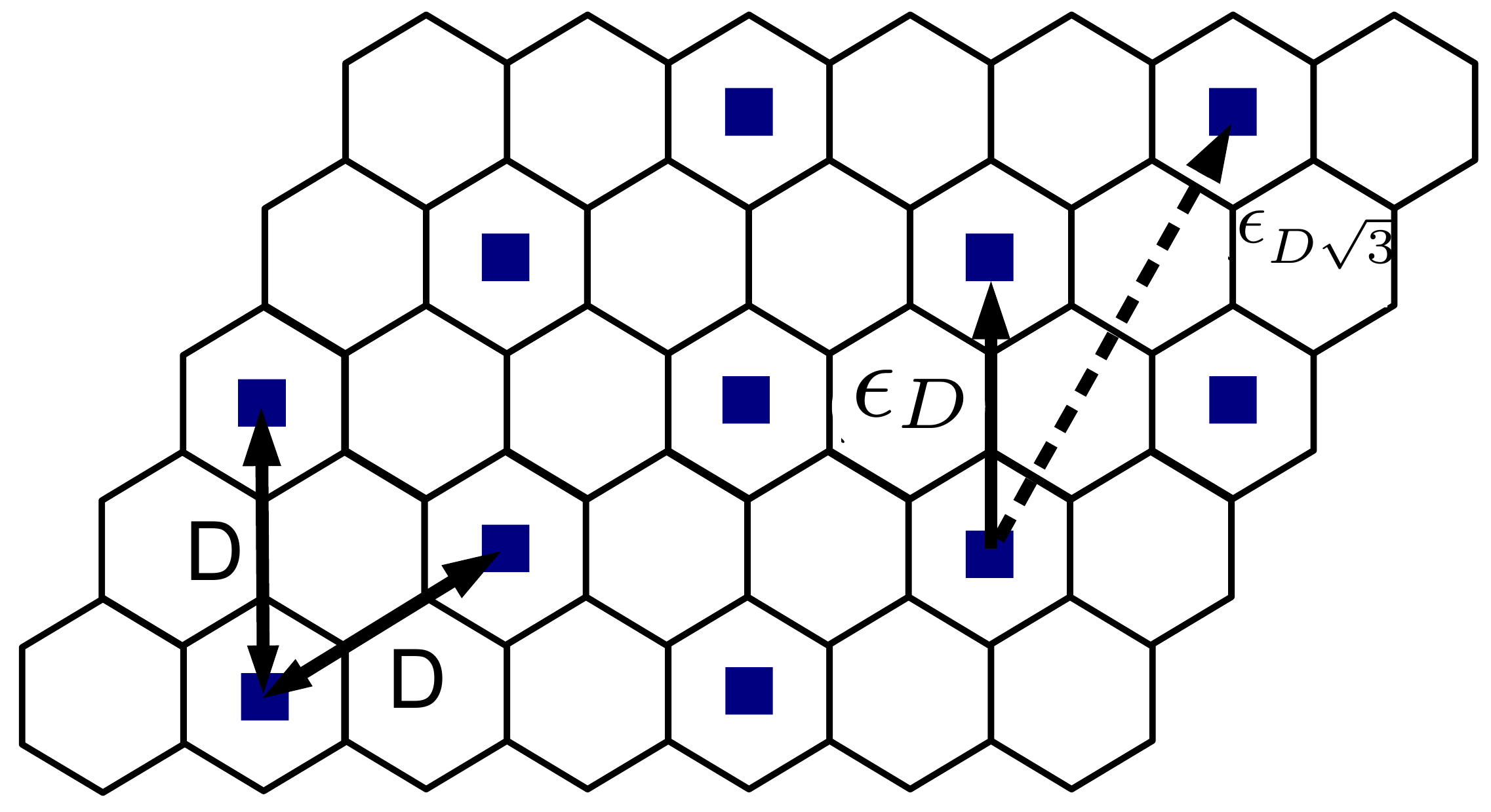} 
\end{tabular}
\caption{\label{vlattice} Two different types of equidistant triangular vortex lattices. {\it Left:} The shortest vortex separation is orthogonal to the links of the lattice, which gives integer superlattice spacings $D=1,2,3,\ldots$. {\it Right:} The shortest vortex separation is parallel to the links giving superlattice spacings $D=\sqrt{3},2\sqrt{3},\ldots$. For each type of vortex lattice, the nearest neighbour ($\epsilon_D$) and next nearest neighbour ($\epsilon_{D\sqrt{3}}$) interaction energies are shown.  }
\end{figure}

To study the phase diagram across the vortex sectors, we restrict to considering equidistant triangular vortex lattices, which we parametrize by $D$ -- the vortex superlattice spacing in the units of plaquette spacings between all nearest neighbour vortices. In particular, we consider lattices with spacings $D=1,2,\ldots$ and $D=\sqrt{3},2\sqrt{3},\ldots$, as illustrated in Fig.~\ref{vlattice}. While the presence of the vortex lattice reduces translational symmetry, the systems with homogenous vortex lattices are still translationally invariant with respect to larger (magnetic) unit cells. In general, a vortex superlattice of spacing $D$ will give rise to a Bloch Hamiltonian that is a $4D^2 \times 4D^2$ matrix. While an analytic solution is available only for the densest case of $D=1$ \cite{Lahtinen10,Pachos07}, sparser vortex lattice systems can still be readily solved numerically.\cite{Lahtinen08}  Unless explicitly mentioned that a finite-size system has been used, all the data in the manuscript has been obtained by Fourier transforming with respect to these vortex lattice unit cells.

As illustrated in Fig.~\ref{PD}, there are two general ways the presence of a vortex lattice modifies the phase diagram: (i) the non-Abelian phase is always replaced by one (or more) Abelian phases characterized by even Chern numbers $\nu=0,-2$ or $-4$ (see Fig.~\ref{bands_data}) and (ii) the strong-pairing like TC phases are always enlarged, i.e. the transition out of the TC phase occurs always for some $J_c > \frac{1}{2}$. It has been shown that the character of the Abelian phases around isotropic couplings ($J=1$) can be fully traced back to pairwise vortex-vortex interactions,\cite{Lahtinen12} i.e. they emerge through the mechanism of topological liquid nucleation.\cite{Ludwig11} Here we study the stability of these phases when the vortex lattices are perturbed away from the uniform triangular configurations.

Before proceeding, we note that it has been observed that the phase diagram of the honeycomb model can also be modified in the absence of a vortex lattice if, for instance, the spin exchange couplings $J_\alpha$ are suitably staggered \cite{Kamfor09, Nash09}. As we show in this paper, the picture of nucleation applies also to such staggered couplings configurations, which suggests that these results could also be understood in terms of a collective state of non-Abelian vortices.

\subsection{The band structure of the nucleated phases}

\begin{figure}[t]
\includegraphics[width=8cm]{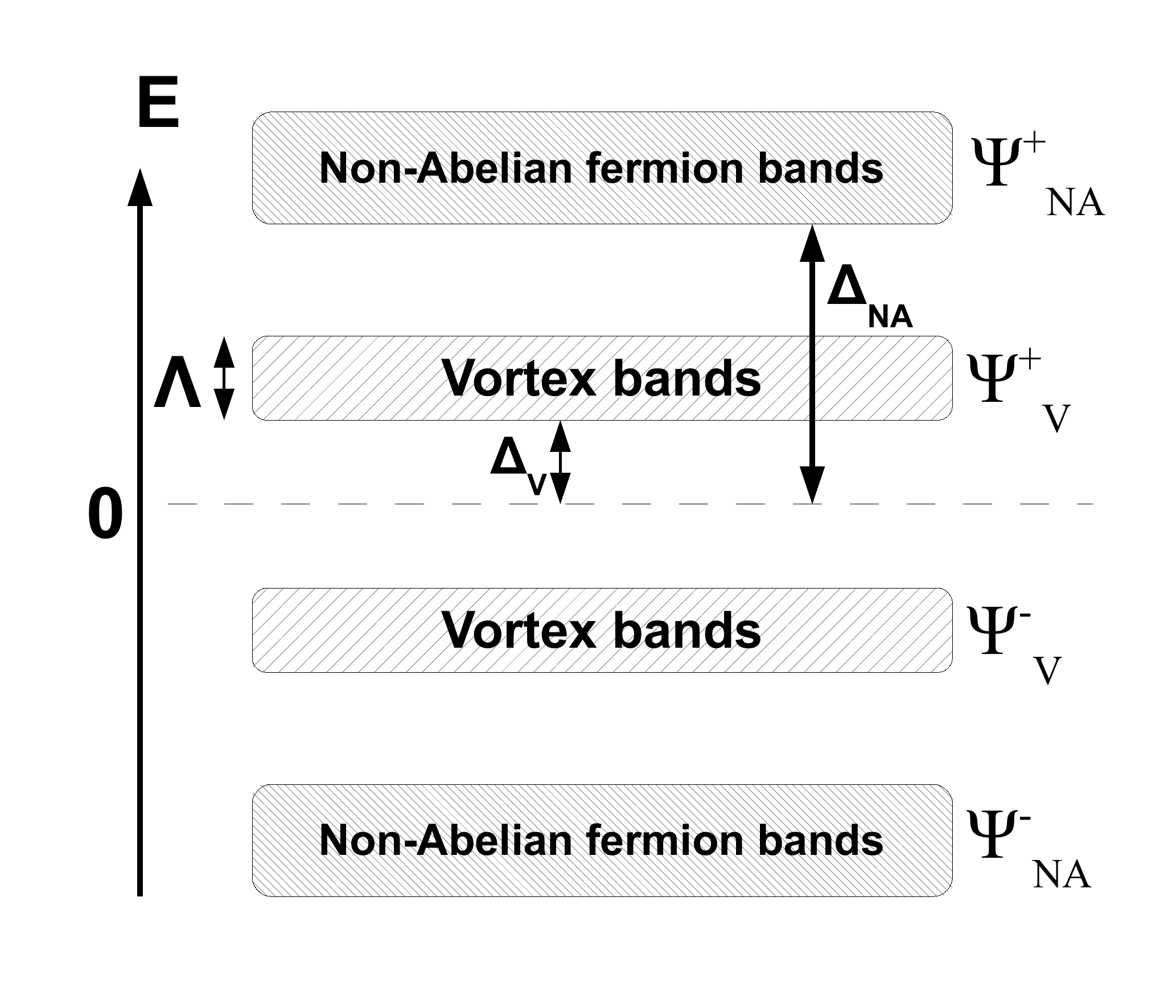}
\caption{\label{bands} The band structure of nucleated phases consists of the high-energy bands $\Psi_{NA}^{\pm}$ that originate from the underlying non-Abelian phase and the hybridized low-energy bands $\Psi_{V}^{\pm}$ that live effectively on the vortex lattice. $\Delta_V$ is the energy gap of the nucleated phase, $\Delta_{NA}$ is the gap to $\Psi_{NA}^{\pm}$ and $\Lambda$ is the bandwidth of the hybridized vortex band. }
\end{figure}

As schematically illustrated in Fig.~\ref{bands}, all the nucleated phases have a characteristic band structure: they consists of a set of high-energy bands $\Psi^\pm_{NA,i}$ that are an artefact the underlying non-Abelian Ising phase, and (a set of) low-energy bands $\Psi^\pm_{V,i}$ that emerge due to the presence of the vortex superlattice. In the presence of $2N$ vortices, the latter will contain $N$ modes that have support only on the sites around the vortices, while the modes in $\Psi^\pm_{NA,i}$ will in general have support on all lattice sites. As the bands are always separated in energy by a band gap, the Chern number $\nu$ for the ground state, that consists of the occupied negative energy bands $\Psi^-_{V,i}$ and $\Psi^-_{NA,i}$, can be written as
\bq 
	\nu & = & \nu_{NA}+\nu_V, \label{nu}
\eq	
where $\nu_{NA}$ and $\nu_V$ are the corresponding the band Chern numbers. For $K\neq0$ one always finds $\nu_{NA}=-\textrm{sign}(K)=\pm 1$, consistent with $\Psi^\pm_{NA,i}$ originating from the non-Abelian phase. On the other hand, $\nu_V$ depends on the superlattice spacing $D$. 

\begin{figure}[t]
\begin{tabular}{cc}
\includegraphics[width=4.2cm]{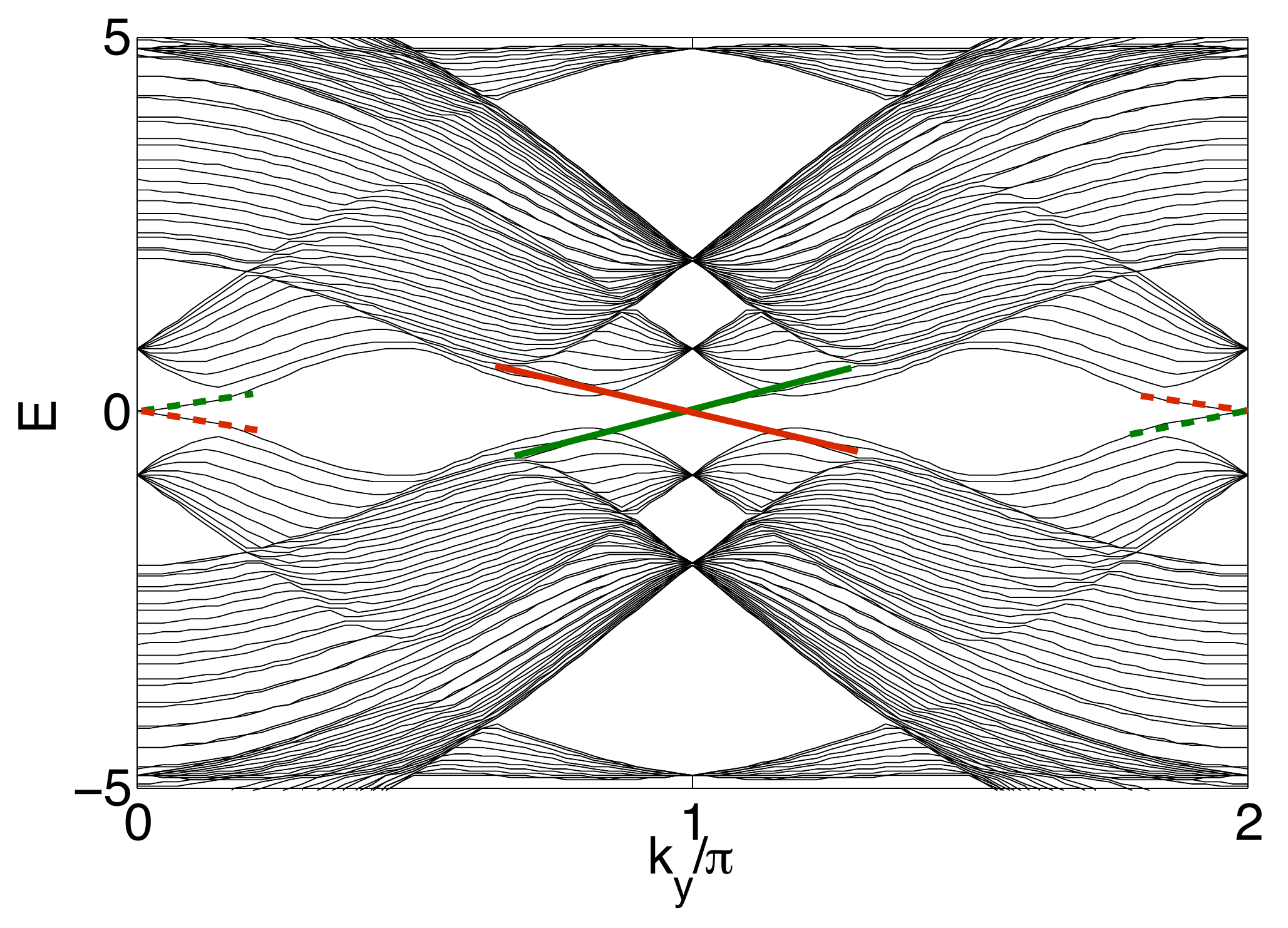} & \includegraphics[width=4.2cm]{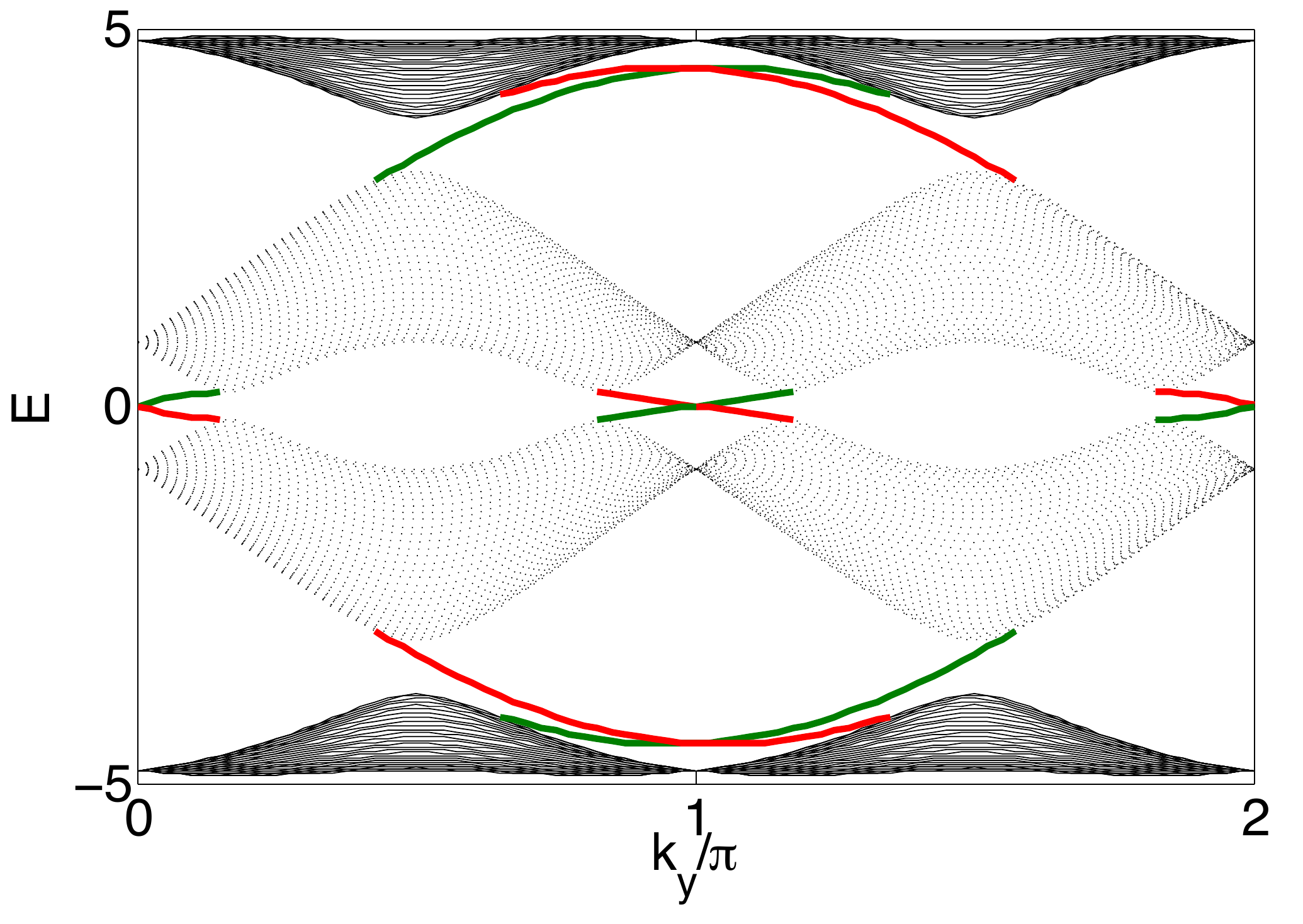} \\
\includegraphics[width=3cm]{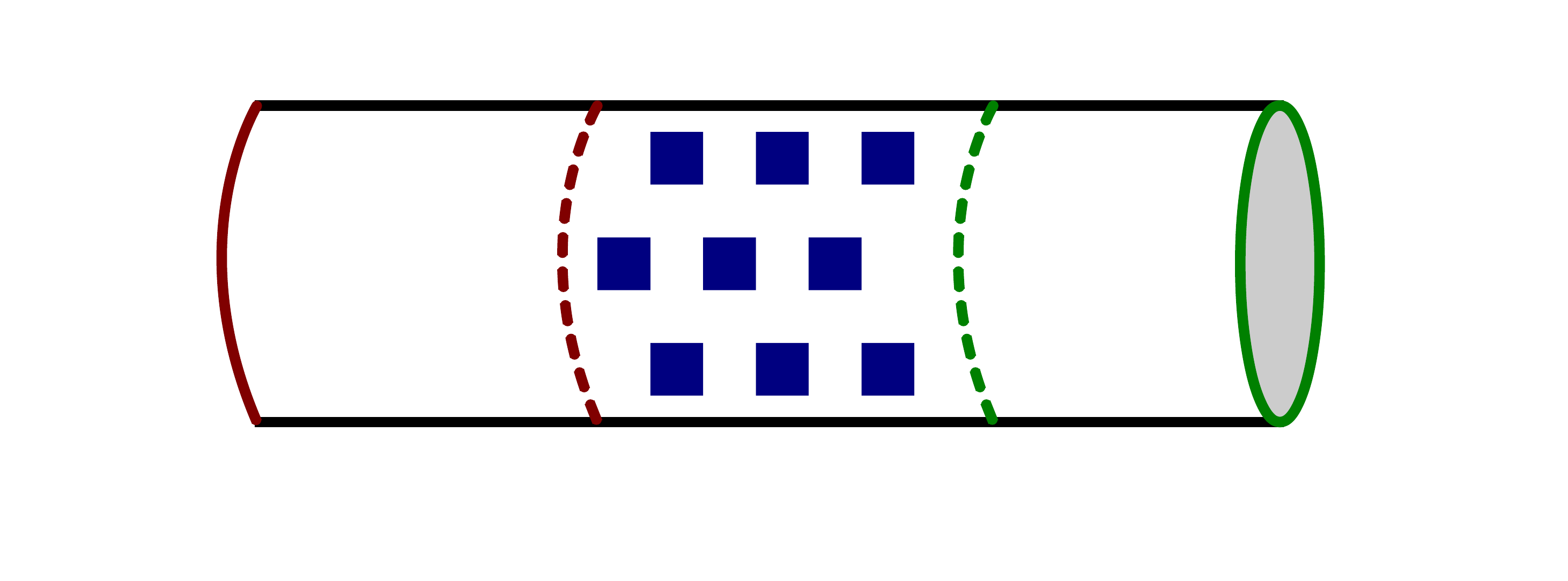} & \includegraphics[width=3cm]{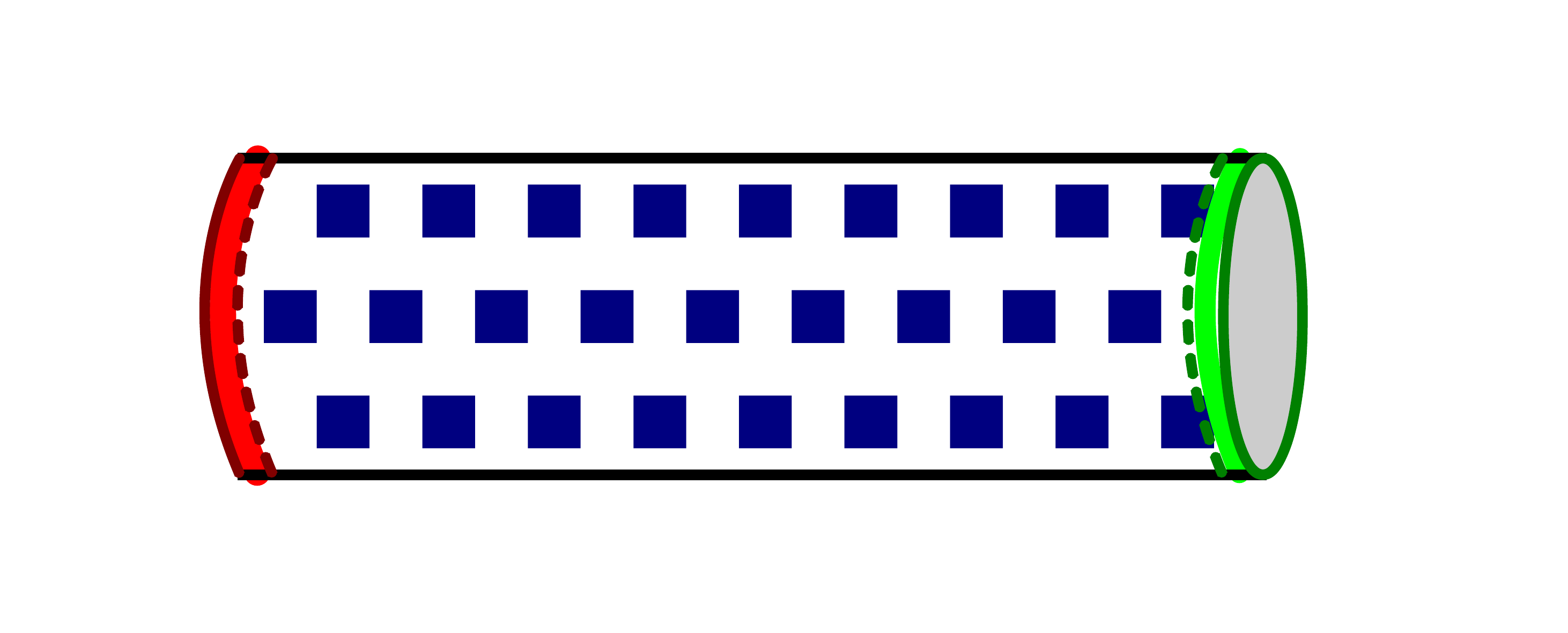} \\
(a) & (b)
\end{tabular}
\caption{\label{edge}
Composition of the nucleated edge states. (a) When the $D=1$ vortex lattice spans only part of the cylinder, the spectrum consists of the overlapping spectra of the nucleated $\nu=-2$ phase in the area covered by the vortex lattice and the non-Abelian $\nu=-1$ phase on the sides of the cylinder. Due to the Chern number difference, there are $|\nu_V|=1$ edge states on each domain wall between the nucleated and the non-Abelian phase (dashed lines crossing $E=0$ at $k_y=0$), and $|\nu_{NA}|=1$ edge states on each physical edge between the non-Abelian phase and the vacuum (solid lines crossing $E=0$ at $k_y=\pi$). (b)  When the vortex lattice covers the whole cylinder, the domain walls are brought close to the physical edge and the edge states can tunnel between them. Depending on the direction of propagation (the slope of the edge modes), they either cancel or add up giving $|\nu|=|\nu_V+\nu_{NA}|$ low-energy edge states crossing $E=0$ at the same momenta. The figure also shows the characteristic band structure with the high-energy (low-energy vortex) bands denoted by solid (dashed) lines. The plots are for $J=1$ and $K=0.1$ and calculated on a cylinder of length $L=80$ ($160$ sites in $x$-direction, $L/\xi \approx 60$). }
\end{figure}

This characteristic band structure suggests that the nucleated phases can be viewed as consisting of two ``layers'':  the underlying non-Abelian $|\nu_{NA}|=1$ phase living on the honeycomb lattice and an emergent $\nu_V$ theory living effectively on top of it on the vortex lattice. This picture is supported by Figure \ref{edge}, which shows how the edge states in a nucleated phase can indeed be understood as the composite of the edge states of the two layers. We note that when viewed as a two-layer system, the transition to an Abelian phase due to the vortex lattice is consistent with a condensate-induced transition \cite{Bais09}.

The two-layer picture enables to separate the contributions from the two distinct types of dynamics in the system: The bands $\Psi_{V,i}^-$  desribe the microscopics of the Majorana modes bound to the vortex cores, whereas the  bands $\Psi_{NA,i}^+$ describe microscopics of the vortices themselves. Since only $\nu_{V}$ depends on the vortex lattice spacing $D$, it is the first energy band that describes the topological behavior of the system for a given vortex configuration. The dynamics associated with the latter, such as an electrostatic repulsion between vortices in actual superconductors, can still affect the behavior of the system though. Some vortex configurations can be energetically favoured over others and, as we will show below, the collective state of the Majorana modes depends in general on the vortex lattice geometry.

\section{Vortex-vortex interactions and the effective Majorana model}
\label{sec:Majmodel}

In this section we first review the interactions between vortices and then define a generalization of the effective Majorana model that has been shown to fully capture the behavior of the uniform vortex lattices \cite{Lahtinen12}.  

\subsection{Pairwise vortex-vortex interactions}

\begin{figure}[t]
\begin{tabular}{c}
\includegraphics[width=8cm]{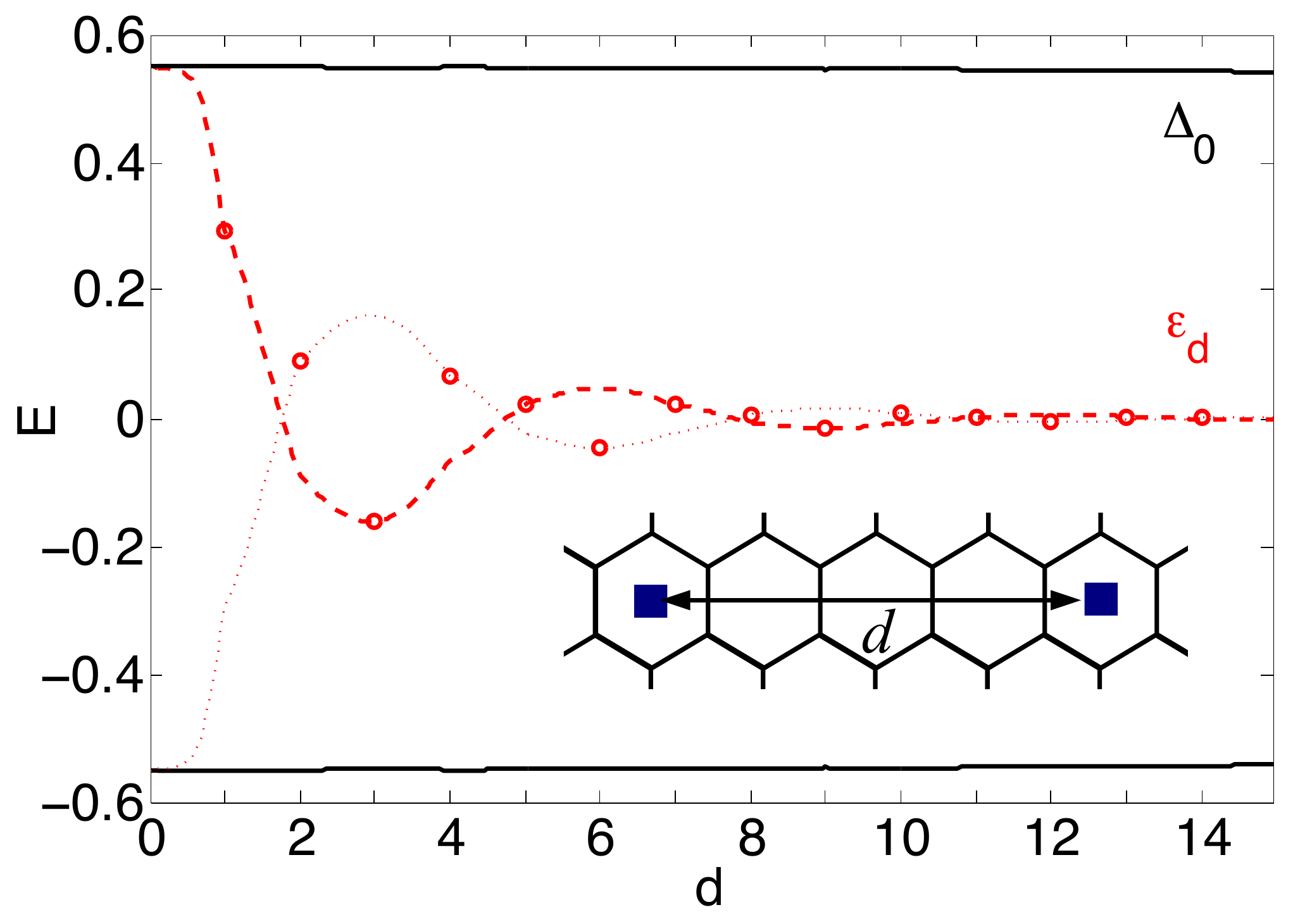}
\end{tabular}
\caption{\label{int} Microscopics of the interaction between a pair of vortices in Kitaev's honeycomb lattice model \cite{Lahtinen11}. In the non-Abelian Ising phase the vortices bind localized Majorana modes $\gamma_i=\gamma_i^\dagger$. Two vortices thus share a complex fermionic mode $d_i=(\gamma_i+i\gamma_{i+1})/\sqrt{2}$. This fusion degree of freedom manifests iself as the presence of a midgap mode in the spectrum (shown in red, dashed line for the particle and the dotted line for the hole conjugate). Its energy $\epsilon_d$ decreases exponentially with vortex separation $d$ and shows oscillations of the form \rf{splitting_sim}. The circles show the physical interaction energy \rf{splitting} at integer values of $d$ when the fermionic parity is taken into account. $\Delta_0$ is the energy gap of the extended states which is insensitive to $d$. The plot is for $J=1$ and $K=0.05$ and calculated using a finite $L \times L$ plaquette system ($L=40$, $3.2 \cdot 10^3$ sites, $L/\xi \approx 15$). }
\end{figure}

The vortices have been shown to be interacting in the non-Abelian phase \cite{Lahtinen11}, which means both the ground state energy and degeneracy depend on the relative vortex positions. As illustrated in Fig.~\ref{int}, the presence of a vortex pair gives rise to two ground states which are exponentially degenerate as the vortex separation $d$ increases. As they are brought closer (simulating continuous vortex transport by tuning the couplings), one finds the energy splitting $\epsilon_d$ between the two states oscillating. This energy splitting can be approximated by 
\be \label{splitting_sim}
 \epsilon_d^{sim} \sim \Delta_0 \cos \left( \omega d \right) e^{-\frac{d}{\xi}}.
\ee
Here $\Delta_0$ is the energy gap of the non-Abelian Ising phase, $\omega \sim k_F^+ - k_F^-$ is the difference of the two Fermi momenta (for $\Delta_0=0$ the spectrum exhibits\cite{Kitaev06} two Dirac cones at $k_F^\pm$) and $\xi$ is the coherence length. For isotropic $J=1$ couplings the energy gap scales as $\Delta_0 = 6\sqrt{3}K$ and the coherence length is well approximated\cite{Lahtinen11} by $\xi \approx \frac{1.4}{\Delta_0}$.

Oscillating degeneracy splitting of the form \rf{splitting_sim} has been shown to originate from Majorana modes bound to the vortex cores.\cite{Cheng09} Their wavefunctions, which, while being exponentially localized at the vortex cores, have oscillating tails \cite{Gurarie07,Kells10-2}. Depending on the vortex separation, the interference between two such wavefunctions can thus be either constructive or destructive. This leads to oscillations in the amplitude for the Majorana mode to tunnel between the vortex cores and thus also to oscillations in the degeneracy splitting. 

When we simulate continuous vortex transport, we restrict first to the vortex-free sector and then tune the couplings in a way that corresponds to creation and transport of the vortices. The spectrum evolves as if the vortices were continuously moved between plaquettes, but the state of the system remains in the vortex-free sector. This constrasts with the situation when one restricts to a two-vortex vortex sectors that corresponds vortex spacing $d$. In this case one would find that the degeneracy of the two exponentially degenerate states is split by $|\epsilon_d^{sim}|$, but that the sign of the splitting now depends on the fermionic parity $P_d$ of the respective vortex sector\cite{Pedrocchi11}. For even (odd) parity the mid-gap mode has positive (negative) energy and is thus unoccupied (occupied) in the ground state. Thus the physical interaction energy between a pair of vortices at separation $d=1,2,\ldots$ is given by
\be \label{splitting}
	\epsilon_d = (-1)^{P_{d}} |\epsilon_{d}^{sim}|,
\ee
as illustrated in Fig. \ref{int}. We find that the parity is odd for linear separations $d=3,6,9,\ldots$ as well as for all diagonal separations $d=\sqrt{3},2\sqrt{3},3\sqrt{3},\ldots$. 

We should point out that this extra condition imposed by the fermionic parity on the interaction energy originates from the mapping from the spins to fermions and is thus specific only to the honeycomb model. In a $p$-wave superconductor or the $5/2$ fractional quantum Hall state, where similar oscillatory interactions have been discovered\cite{Cheng09,Baraban09}, continuous vortex transport is well defined and both the magnitude and the sign of the interaction energy can be directly obtained from the oscillating energy splitting.  


\subsection{Effective Majorana model for the vortex band}

We now connect the pairwise vortex-vortex interactions to the character of the nucleated many-vortex phases.  To do this we view the interactions as tunneling processes of the Majorana modes $\gamma_i=\gamma_i^\dagger$ bound to the vortex cores with the interaction energy $\epsilon_d$ giving the tunneling amplitude at separation $d$. The collective state of a vortex lattice can thus be modelled by a tight-binding model of Majorana modes tunneling on the lattice whose sites coincide with vortex cores.\cite{Lahtinen12} We show that by using the interaction energies $\epsilon_d$ as the only inputs, such an effective model captures the behavior of the vortex bands $\Psi_V^\pm$ with the spectral gap $\Delta_V$, band energy $E_V$ and the observed Chern number $\nu$ predicted correctly.

\begin{figure}[t]
\includegraphics[width=8cm]{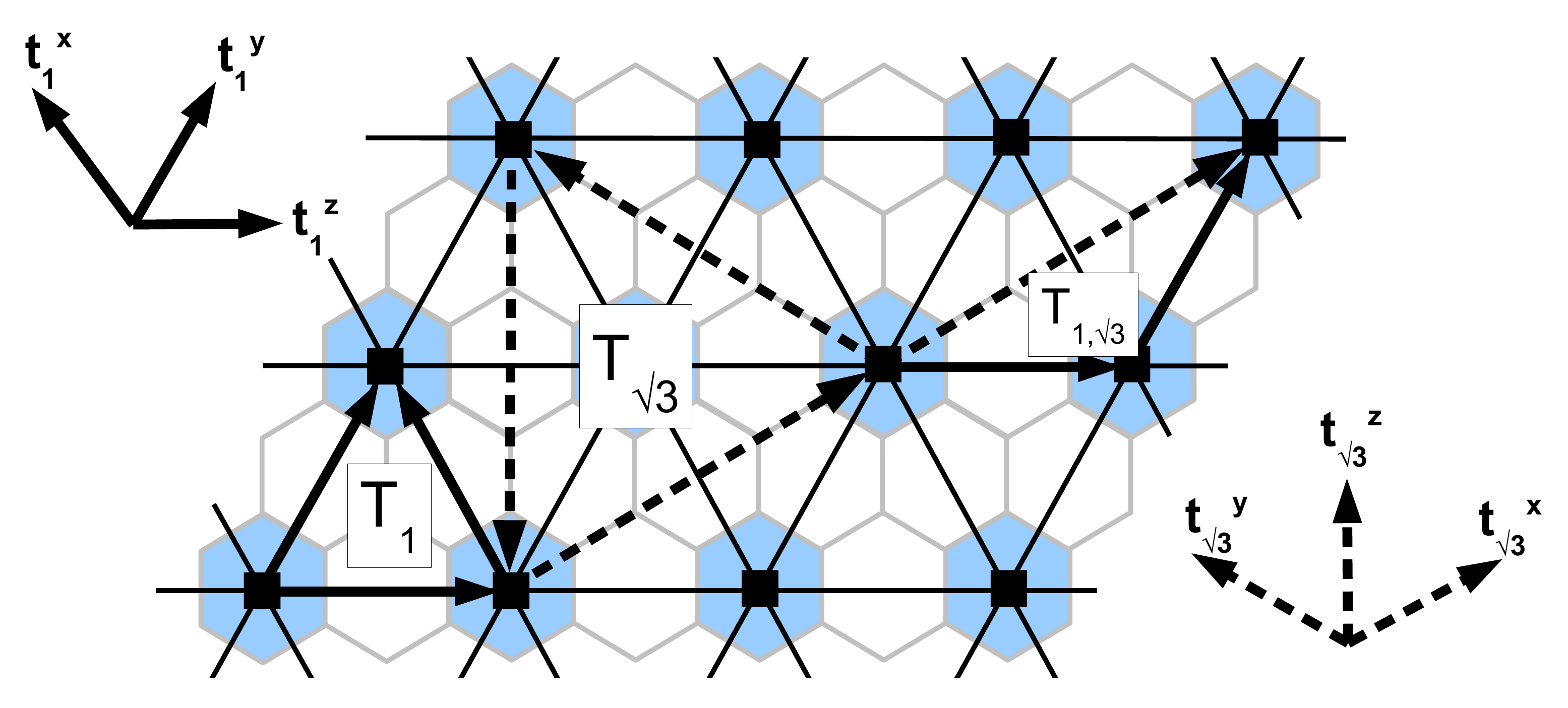}
\caption{\label{efflattice} The effective Majorana model living on the vortex lattice with the sites coinciding with the vortex cores, as shown here for the $D=2$ vortex lattice. Including both nearest $t_1$ (solid links) and next nearest $t_{\sqrt{3}}$ tunneling (dashed links), there are three distinct types of plaquettes: $T_1$ and $T_{\sqrt{3}}$ plaquettes that consists of only $t_1$ or $t_{\sqrt{3}}$, respectively, and the intermediate $T_{1,\sqrt{3}}$ plaquettes that consist of both.}
\end{figure}

The Hamiltonian for our effective Majorana model is given by 
\be \label{Heff}
	H_M = i \sum_{\langle i,j\rangle} t^1_{ij} s_{ij}^1 \gamma_i \gamma_j + i \sum_{\langle \langle i,j\rangle\rangle} t^{\sqrt{3}}_{ij} s_{ij}^{\sqrt{3}} \gamma_i \gamma_j .
\ee 
The Majorana operators satisfying $\{ \gamma_j, \gamma_j\}=2\delta_{ij}$ live on the vortex cores that coincide with the sites of the vortex lattice, as illustrated in Fig.~\ref{efflattice}. The nearest and next-nearest neighbour tunneling amplitudes $t^1_{ij}$ and $t^{\sqrt{3}}_{ij}$, respectively, are vortex lattice geometry dependent and possibly locally varying. The $Z_2$ valued gauge variables $s_{ij}^1$ and $s_{ij}^{\sqrt{3}}$, on the other hand depend only on the topology of the vortex lattice\cite{Grosfeld06}. The latter give rise to flux $\Phi_{ijk}=\ln(-is_{ij}s_{jk}s_{ki})=\pm \frac{\pi}{2}$ on every plaquette with corners $i$, $j$ and $k$.  As illustrated in Fig.~\ref{efflattice}, there are three distinct types of plaquettes: $T_1$ and $T_{\sqrt{3}}$ plaquettes that consist only of $t_1$- and $t_{\sqrt{3}}$-links, respectively, and the intermediate $T_{1,\sqrt{3}}$ plaquettes that consist of both. It has been shown\cite{Lahtinen12} that the flux on them should be fixed to be $\Phi_{1}=\frac{\pi}{2}$, $\Phi_{\sqrt{3}}=-\frac{\pi}{2}$ and $\Phi_{1,\sqrt{3}}=\frac{\pi}{2}$.
 
\begin{figure}[t]
\includegraphics[width=8.4cm]{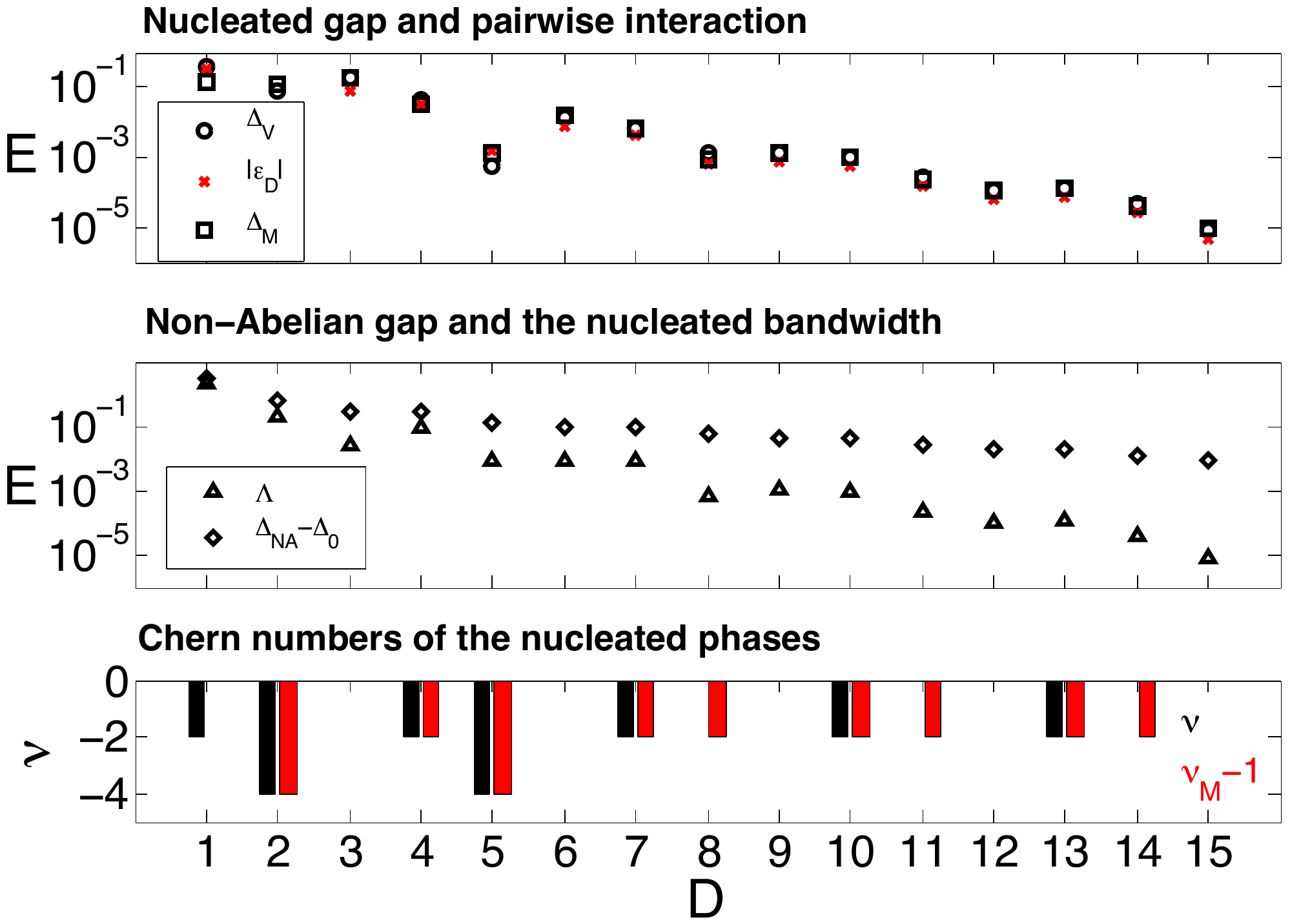}
\caption{\label{bands_data} Nucleated phases in the presence of uniform vortex lattice of spacing $D$. {\it Top:} The exponentially decaying energy gap $\Delta_V$ of the full model and the energy gap $\Delta_M$ predicted from the exponentially decaying pairwise interactions $\epsilon_D$ are in quantitative agreement. {\it Middle:} Emergence of the nucleated band structure of Fig. \ref{bands}. As the vortex lattice spacing $D$ increases, the bandwidth $\Lambda$ of the vortex band decays exponentially giving a macroscopically degenerate zero energy band at large vortex lattice spacings. In this limit the band gap $\Delta_{NA}$ converges to the spectral gap $\Delta_0$ in the absence of a vortex lattice. {\it Bottom:}  The Chern number $\nu_M-1$ (red) predicted by the Majorana model \rf{Heff}, with $\epsilon_D$ obtained from Fig. \ref{int} as the only input, show in general agreement with the observed Chern numbers $\nu$ (black). The observation of $\nu=0$ for all $D=3,6,\ldots$ is in agreement with previous studies.\cite{Kamfor11} The data is for $J=1$ and $K=0.1$. }
\end{figure}

To employ \rf{Heff} to model behavior of the vortex band of particular vortex lattice, the tunneling amplitudes $t^1_{ij}$ and $t_{ij}^{\sqrt{3}}$ are identified with the interaction energies $\epsilon_d$ corresponding to the vortex separations. It has been shown \cite{Lahtinen12} that a model with uniform couplings, i.e.
\be \label{tl_uni}
t^1_{ij} \to t_1=\epsilon_{D}, \qquad t^{\sqrt{3}}_{ij} \to t_{\sqrt{3}}=\epsilon_{D\sqrt{3}},
\ee
accurately describes vortex lattices with uniform spacing $D$ for isotropic exchange couplings $J_x=J_y=J_z$. Indeed, Fig. \ref{bands_data} shows that for a wide range of superlattice spacings $1 \leq D \leq 15$, the gap $\Delta_M$ of the effective Majorana model provides an excellent approximation of the observed gap $\Delta_V$ of the full honeycomb model. There is also systematic agreement between the vortex band Chern number $\nu_V$ and the Chern number $\nu_M$ characterizing the ground state of our effective model. Depending on the vortex lattice spacing $D$, we find phases characterized by Chern numbers $\nu=0,-2$ or $-4$. In terms of the effective model, these phases arise when $|t_1| \gg |t_{\sqrt{3}}|$ and $t_1 < 0$ or $t_1 > 0$, or $t_{\sqrt{3}} \gg t_1$ and, respectively.\cite{Lahtinen12} 

The agreement between the observed and predicted Chern numbers is exact in the range $2 \leq D \leq 7$. We attribute the disagreement for the tightly packed case of $D=1$ for the vortex lattice spacing being smaller than the coherence length $\xi$ of the underlying non-Abelian phase. The Majorana wavefunctions are thus strongly overlapping and individual vortices are no longer well defined. This means that our assumption that the energy splitting equals the tunneling amplitude breaks down and the effective model, as we defined it, no longer captures accurately the behavior of the full system. Indeed, the general form \rf{splitting_sim} for the interaction energy holds\cite{Cheng09} strictly speaking only for $d \gg \xi$. For the same reason the approximation becomes worse for parameter regimes where the gap $\Delta_0$ becomes small, i.e. near $J=1/2$ or when $K$ becomes small. Thus we take $D > \xi$ as a physical requirement for our Majorana model to accurately describe the behavior of the vortex lattices. 

On the other hand, we attribute the periodically occuring disagreeing cases $D =8,11,14,\ldots$ to finite-size errors in calculating the Majorana tunneling amplitude $t_{Dl}=\epsilon_D$. Such corrections become more significant as the interaction energy $\epsilon_D$ becomes exponentially small with increasing vortex separation $D$. They are assumed to be particularly pronounced for those vortex lattices whose spacings are in the vicinity of the nodes in the interaction oscillations. As shown in Fig.~\ref{int}, in our case these occur for $D=2,5,8,11,14$, which strongly suggests that the disagreeing cases for larger spacings are indeed due to such finite-size effects in the calculation of $\epsilon_D$.

The appearance of only even Chern numbers across the whole $1 \leq D \leq 15$ range confirms that nucleation is not a process that is driven by the interactions -- there is no critical interaction strength. Even if the energy gap $\Delta_V$ and the bandwidth $\Lambda$ become exponentially small, nucleation occurs for uniform vortex lattices of arbitrary spacing. However, this is a highly idealized case as real physical systems always come with impurities. These pin the vortices, which results in perturbations in the vortex lattice. We now turn to a quantitative study of the robustness and the fate of nucleated topological phases under anisotropy, dimerization and random local disorder.

\section{Perturbed nucleated phases}
\label{sec:disorder}

In this section we consider three distinct types of perturbations in the vortex lattices: Anisotropy in the interactions arising from anisotropic vortices, dimerization of the vortex lattice and random local disorder of the vortex positions. For each case we show how to choose the effective tunneling amplitudes such that the effect Majorana model \rf{Heff} captures the behavior of the full honeycomb model.

\subsection{Spatially anisotropic vortices}

\begin{figure}[t]
\begin{tabular}{cc}
\includegraphics[width=4cm]{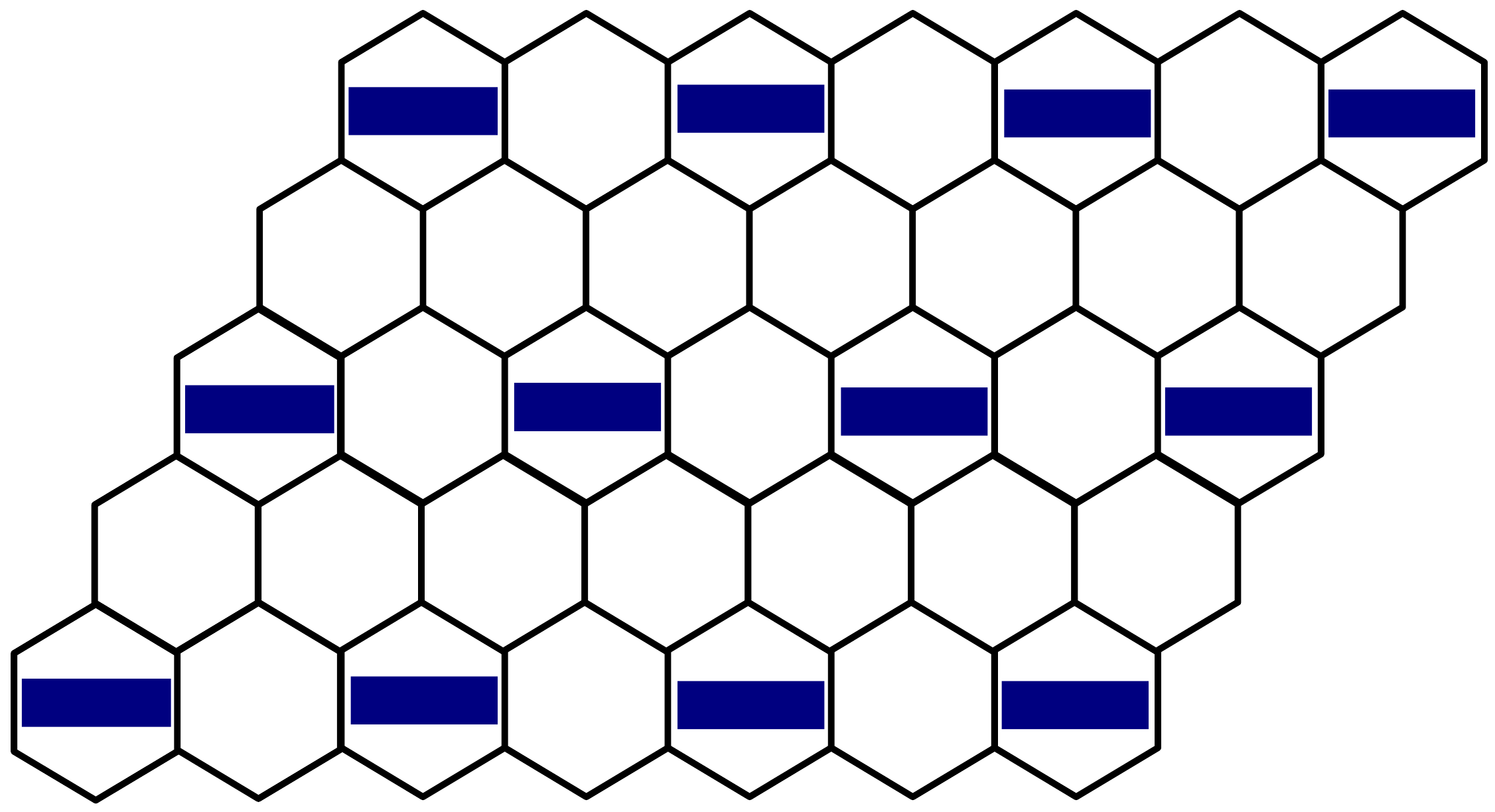} & \includegraphics[width=4cm]{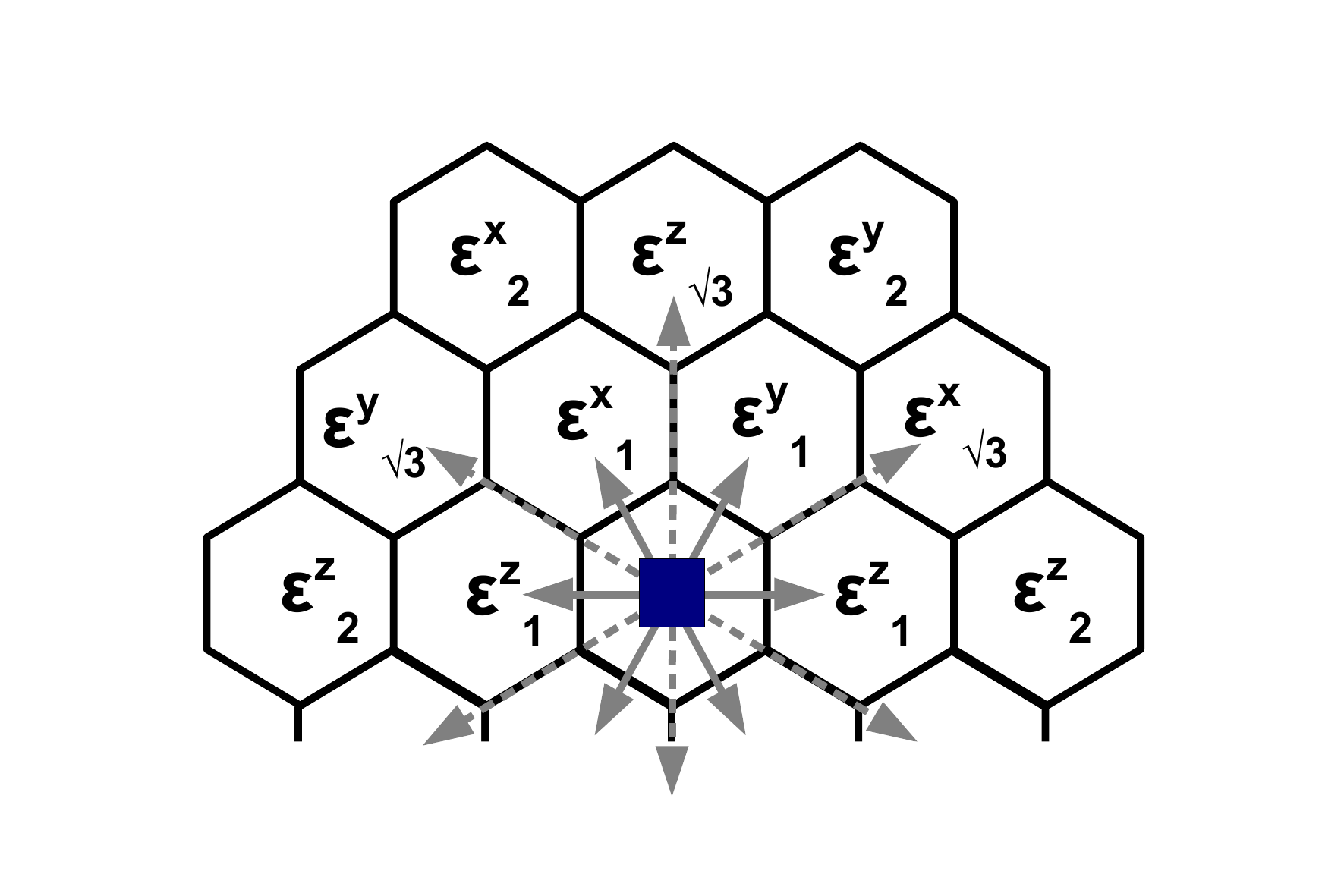} \\
(a) & (b)
\end{tabular}
\caption{\label{int_anisotropy}Spatial vortex anisotropy and the effective Majorana model. (a) When $J_z<J_x,J_y$, the vortices and the Majorana wave functions bound to them are effectively stretched in the direction of the weaker vertical $J_z$ links. (b) This leads to different anisotropic overlaps, and thus anisotropic tunneling amplitudes for the Majoranas. When $J_x \neq J_y \neq J_z$ the $C_3$ rotational symmetry is broken and shown the vortex-vortex interaction energies are anisotropic such that $\epsilon_d^x \neq \epsilon^y_d \neq \epsilon^z_d$. The solid (dashed) arrows show the directions identified with the (next) nearest neighbour interactions of the effective Majorana model.  }
\end{figure}

As the first type of perturbation we consider anisotropic interactions which occur when the vortices are not spatially isotropic.  In the honeycomb lattice model this happens when the spin exchange couplings are tuned away from the $C_3$ rotationally symmetric $J=1$ couplings. In $p$-wave superconductors it could arise when impurities deform the magnetic field through a vortex or, more generally, when the Fermi surface is anisotropic.

\begin{figure}[t]
\includegraphics[width=8.2cm]{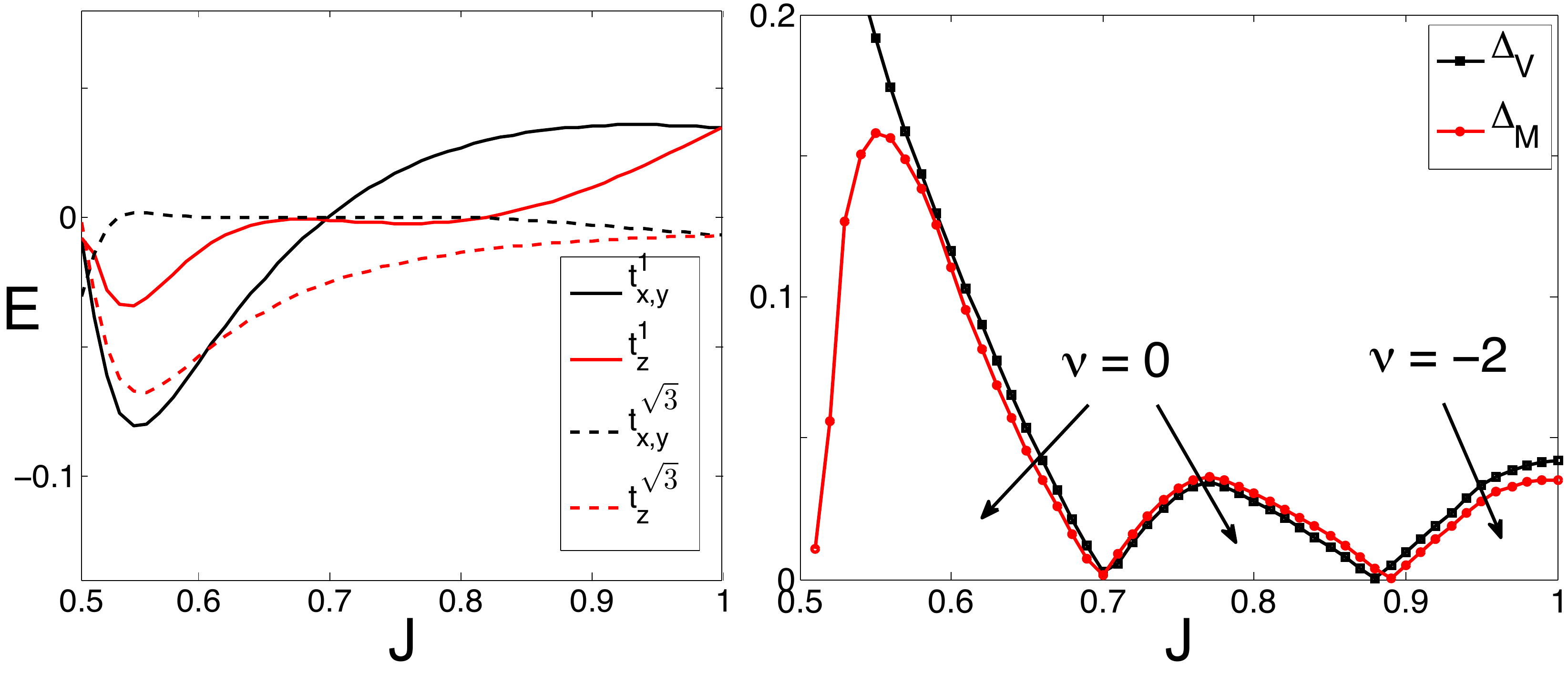}
\caption{\label{Jdeform} The prediction by the Majorana model \rf{Heff} with the anisotropic tunneling couplings \rf{tl_anis} for the $D=4$ vortex lattice. {\it Left:} The behavior of the anisotropic tunneling couplings $t^1_{x,y}=\epsilon_4^{x,y}$, $t^1_{z}=\epsilon_4^{z}$, $t^{\sqrt{3}}_{x,y}=\epsilon_{4\sqrt{3}}^{x,y}$ and $t^{\sqrt{3}}_{z}=\epsilon_{4\sqrt{3}}^{z}$  as we tune $J$ away from the isotropic $J=1$ spin exchange couplings towards the non-Abelian phase boundary at $J=1/2$. {\it Right:} The observed ($\Delta_V$) and the predicted ($\Delta_M$) energy gaps when the anisotropic couplings on the left are input to \rf{Heff}. We find the effective model predicting correctly both the locations of the phase transitions as well as the Chern numbers $\nu=\nu_M-1$ of the various nucleated phases, including the transition to the extended TC phase  $J \lesssim 0.7$. The data is for $K=0.1$. }
\end{figure}

As schematically illustrated in Fig.~\ref{int_anisotropy}, anisotropy means that the bound Majorana wavefunctions become spatially deformed, which in turn gives unequal overlaps and thus unequal tunneling amplitudes in different spatial directions. To account for anisotropic interactions in the effective model \rf{Heff}, we define the tunneling amplitudes by 
\be \label{tl_anis}
	t^l_{ij} \to t^l_{\alpha} = \epsilon^\alpha_{lD}, \qquad l = 1, \sqrt{3},
\ee
when the tunneling between sites $i$ and $j$ is in the direction $\alpha$ (see Fig.~\ref{efflattice}). For $J=1$ all the couplings of the same range will be equal, but for $J<1$ we find $t^l_{z}$ couplings acquiring different behavior from $t^l_{x}$ and $t^l_{y}$ that in turn will behave identically. This different $J$ dependance is shown in Fig.~\ref{Jdeform}, which also shows for the case $D=4$ that when these couplings are inserted into \rf{Heff}, we find excellent agreement with the observed and predicted gaps and Chern numbers. Further data presented in Appendix \ref{App_data} shows that this holds for also other vortex lattice spacings. This confirms that the description by our Majorana model is valid in the presence of spatial anisotropy.

The interactions between the non-Abelian vortices are only defined in the coupling regime $\frac{1}{2} < J \leq 1$, which supports the non-Abelian Ising phase. One might thus expect that this regime would be fully covered by the nucleated phases with any $\nu=0$ phase in this regime being distinct from the TC phase in the $J < \frac{1}{2}$ regime. Instead, as illustrated in Fig.~\ref{PD} and shown by the actual data in Appendix \ref{App_data}, we find that the TC phase is always enlarged into the $J < \frac{1}{2}$ region. As this occurs for all vortex lattice spacings $D$, we conclude that the strong-pairing like TC phases are always stabilized in the presence of a vortex lattice. 

\subsection{Dimerization of the vortex lattice}

Above we kept the vortex lattice uniform while tuning the couplings to induce anisotropic interactions. A somewhat more complicated situation arises when the vortex lattice is periodically deformed such that the vortices dimerize. As some vortices are now closer to each other while being further away from others, the effective Majorana tunneling couplings become not only anisotropic, but also staggered. This can in general occur in the presence of a periodic background potential, but it could also arise due to multi-scale interactions in superconductors \cite{Varney12} or be induced in optical lattice setting through long-range dipolar interactions \cite{Komineas12}. As we will show below, it may also occur spontaneously due to a possible dimerizing instability arising from the oscillations in the pairwise vortex interactions.

\begin{figure}[t]
\begin{tabular}{cc}
\includegraphics[width=4.2cm]{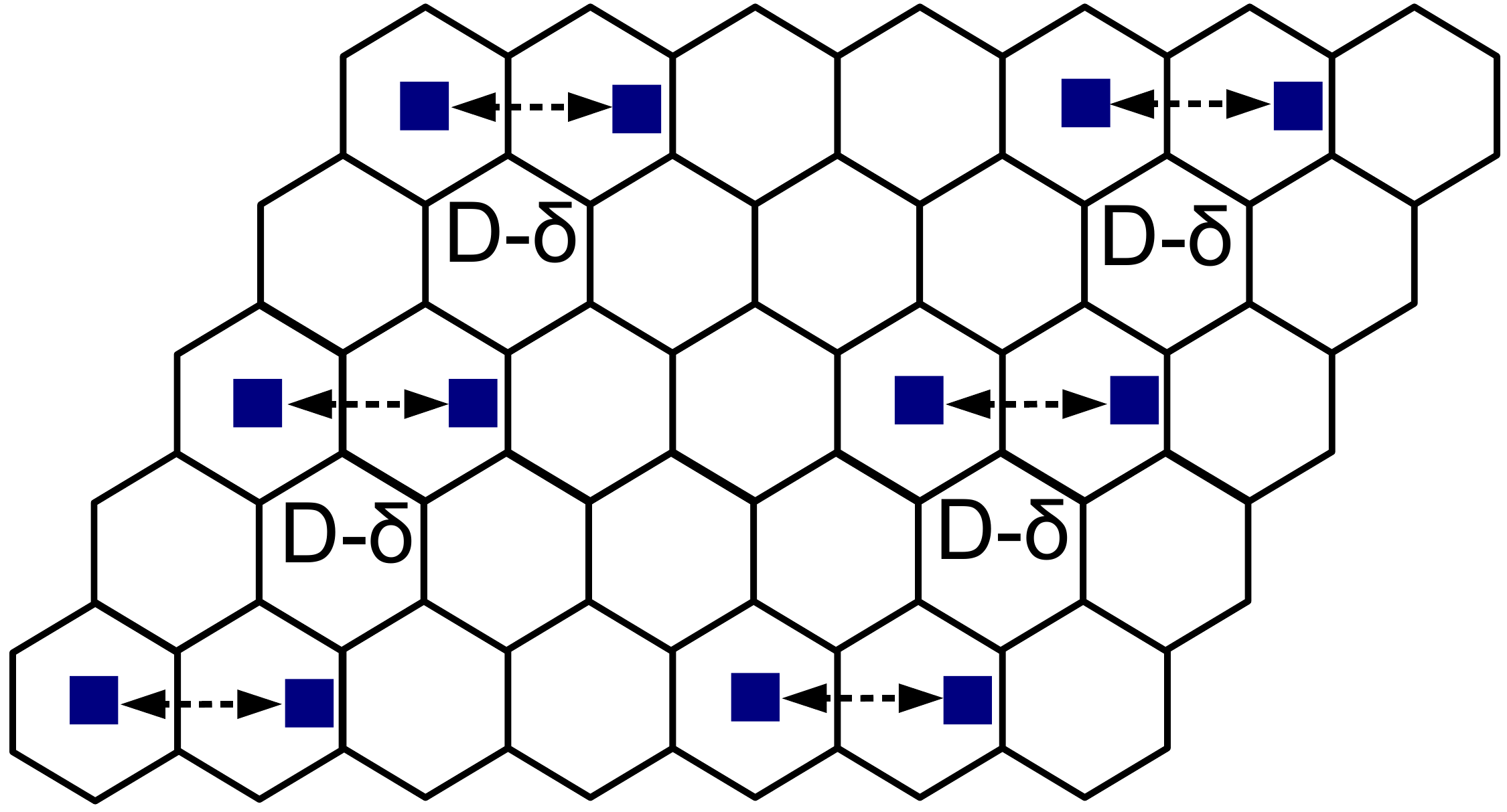} & \includegraphics[width=4.2cm]{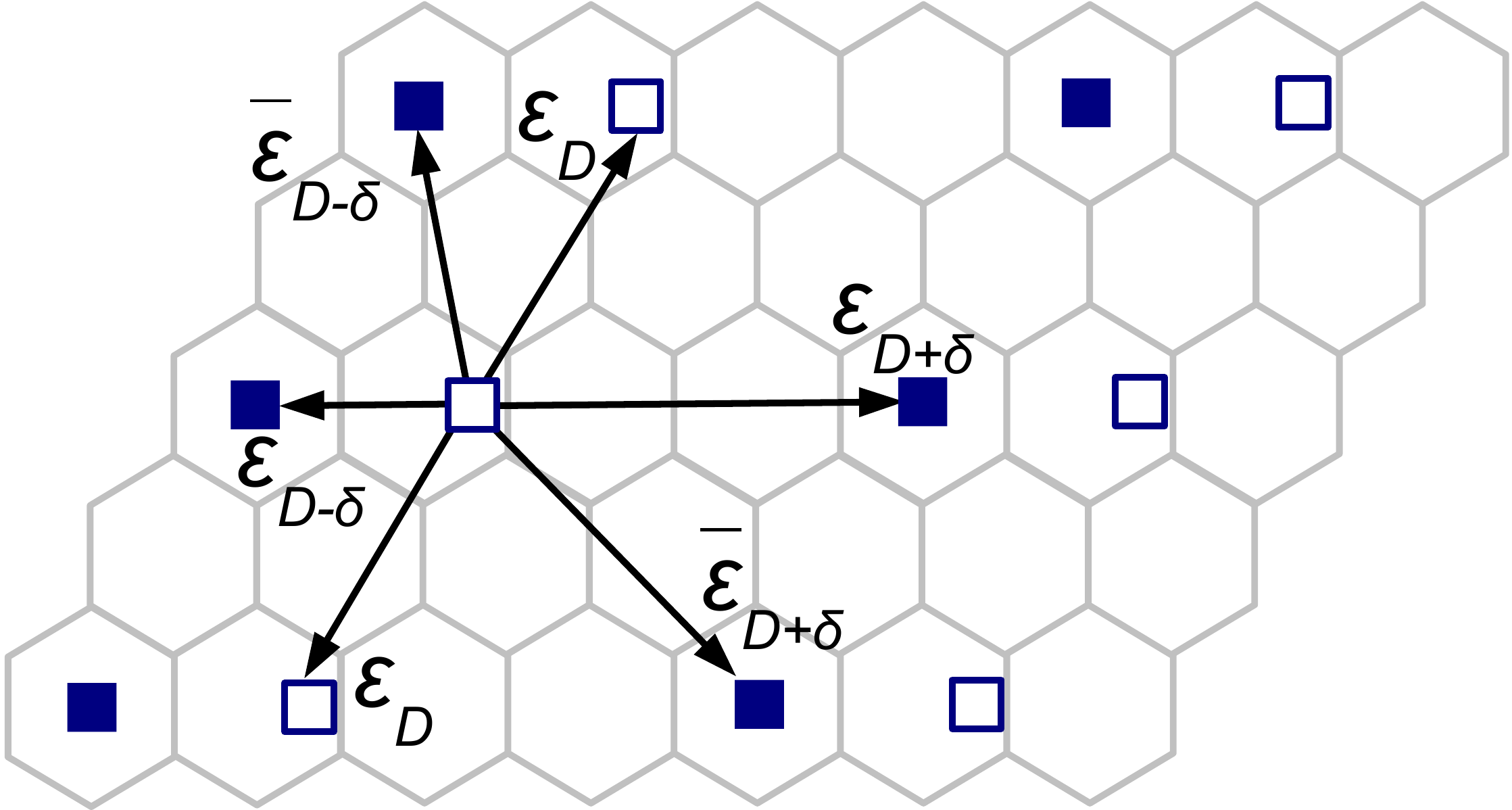} \\
(a) & (b)
\end{tabular}
\caption{\label{deform} Dimerization of the vortex lattice. (a) Deforming the uniform vortex lattice of spacing $D$ periodically by $\delta$ leads to a dimerization of the vortices into pairs of separation $D-\delta$. In our parametrization $\delta=0$ corresponds to no deformation (uniform vortex lattice), while $\delta=D$ corresponds to a fusion of all vortices. As described in Sec. II.A, this can be achieved by staggering the local couplings $J_{ij}$ suitably. (b) The corresponding staggered nearest neighbour interactions energies that are identified with the tunneling amplitudes according to \rf{tl_stagger}.}
\end{figure}

To study the effect of such staggering on the Majorana model \rf{Heff}, we simplify the model in two ways. First, we set $t^{\sqrt{3}}_{ij}=0$ and consider staggered nearest neighbour $t^1_{ij}$ tunneling only. This amounts to the model neglecting the $\nu=-4$ phases\cite{Lahtinen12}. These, however, are rare special cases, which we do not consider now. Second, we consider a strongly modulated deformation where, as illustrated in Fig. \ref{deform}(a), the vortex separation varies periodically such that in $z$-direction we have an alternating pattern of separations $D+\delta$ and $D-\delta$. In the case of $\delta=0$ there is no deformation, while for $\delta=D$ all the vortices are fused and one recovers the vortex-free sector. The assumption is that when vortices become strongly paired ($\delta \to D$), the system locks into a localized configuration of complex fermions bound to the dimerized vortex pairs. This suppresses any collective state of the Majorana modes and is thus expected to drive a transition back to the underlying non-Abelian Ising phase. 

\begin{figure}[t]
\begin{tabular}{cc}
\includegraphics[width=4.2cm]{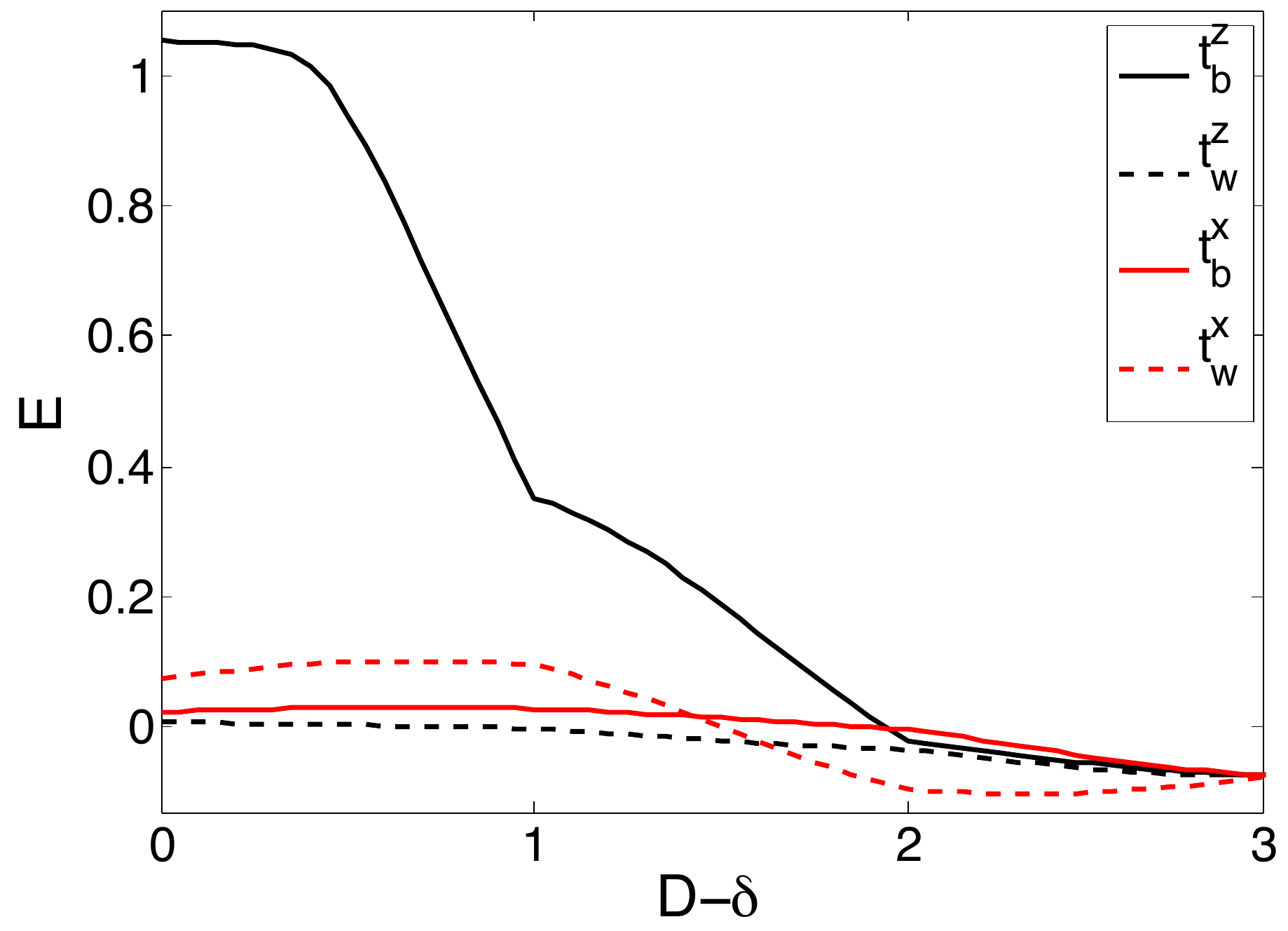} & \includegraphics[width=4.2cm]{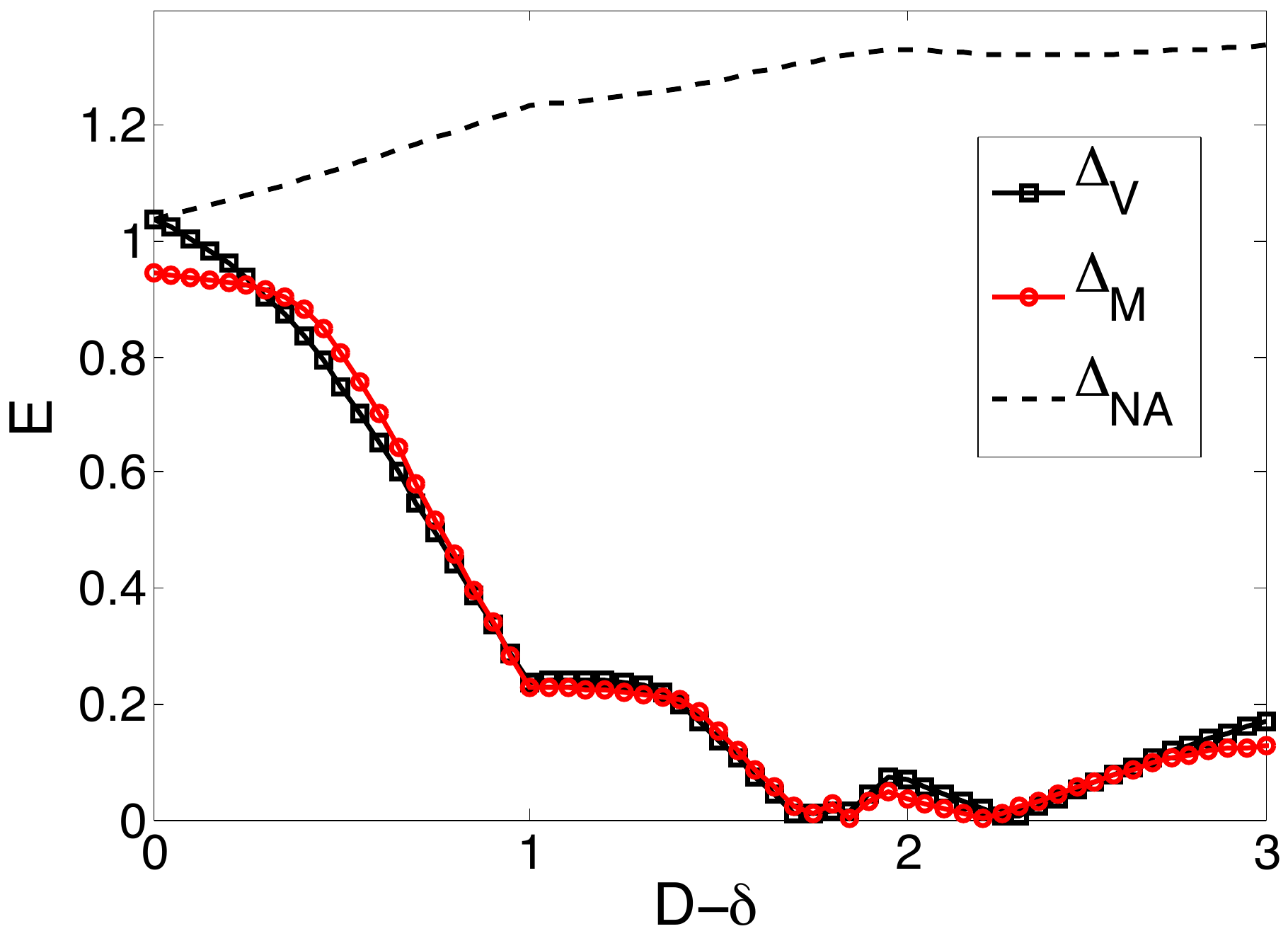} \\
(a) & (b) \\
\includegraphics[width=4.2cm]{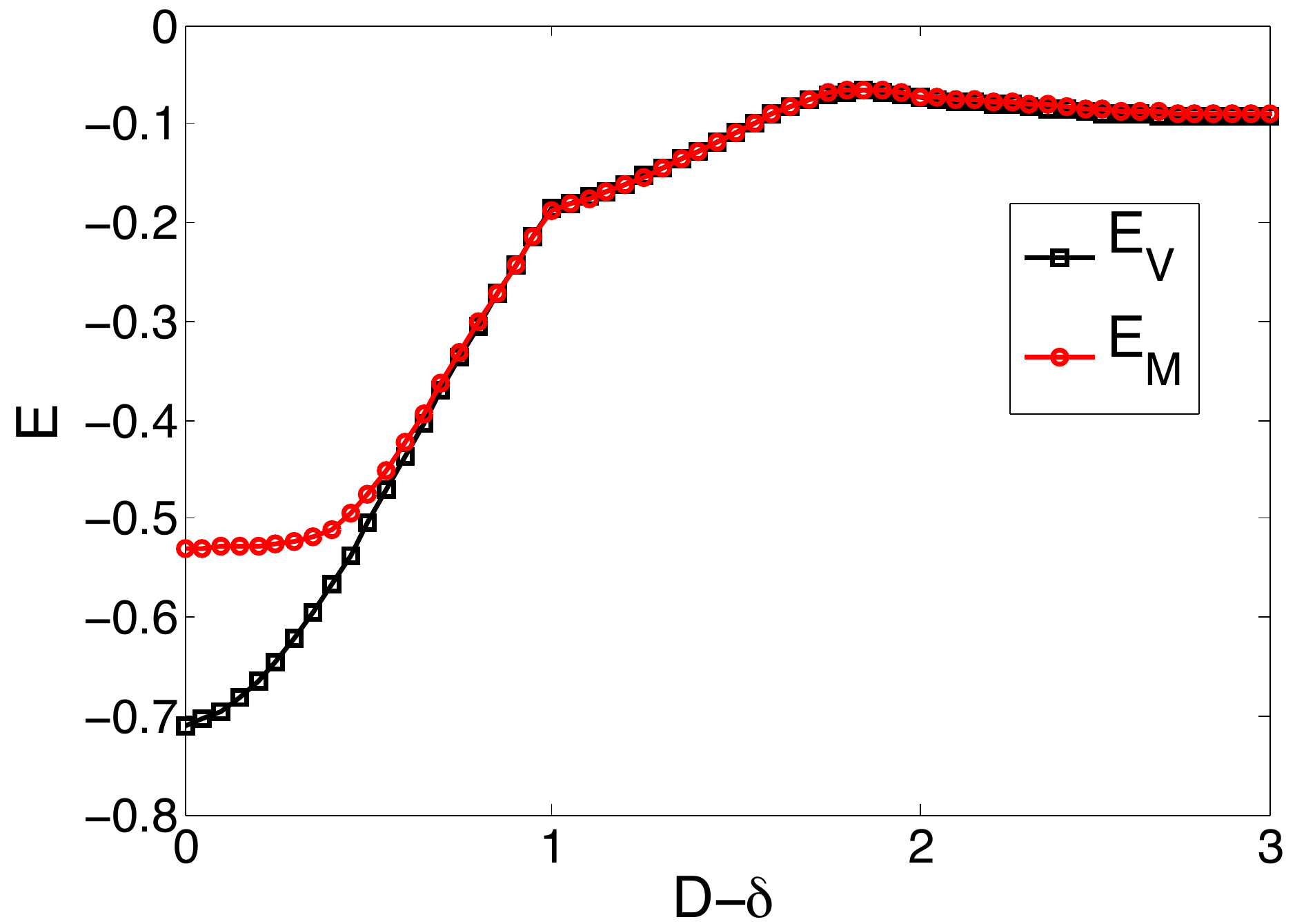} & \includegraphics[width=4.2cm]{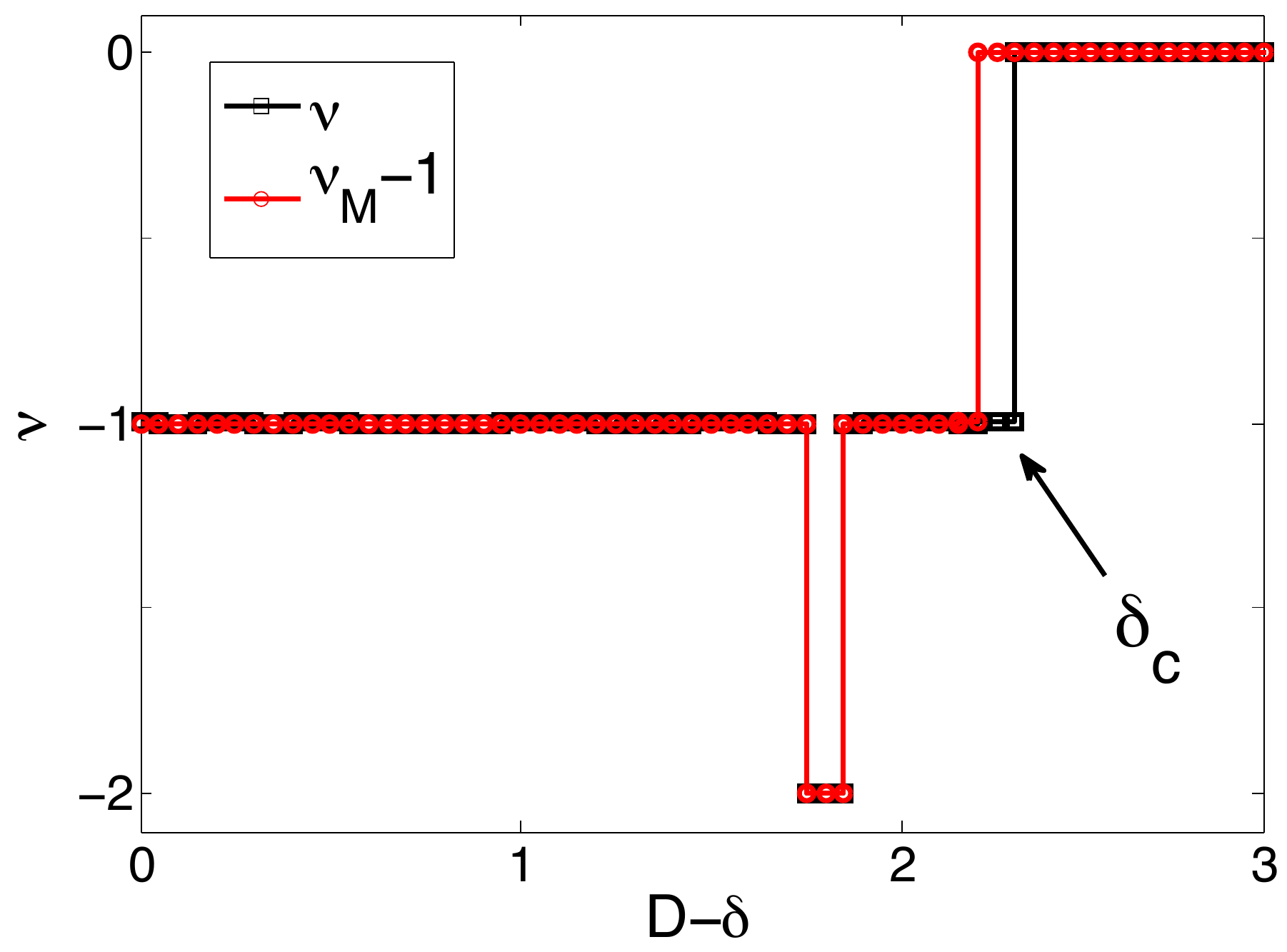} \\
(c) & (d)
\end{tabular}
\caption{\label{predstagger} Dimerized $D=3$ vortex lattice and the effective description by the staggered Majorana model. (a) The staggered tunneling amplitudes \rf{tl_stagger}. $t^y_b=t^y_w$ does not change with $\delta$ and takes always the $\delta=0$ value. When these are input to \rf{Heff}, we find agreement in the observed and predicted (b) energy gaps $\Delta_V$ and $\Delta_M$, (c) the band energies $E_V$ and $E_M$ and (d) Chern numbers $\nu$ and $\nu_M-1$, respectively. The data is for $K=0.1$, while the couplings $|J_{ij}| \leq 1$ are suitably locally staggered for each $\delta$ to simulate the dimerization. }
\end{figure}

In the absence of $t^{\sqrt{3}}$ tunneling the effective Majorana model \rf{Heff} has a two site unit cell. Labeling these sites as black ($b$) and white ($w$), the unit cell accommodates up to six independent nearest neighbour tunneling amplitudes. To account for the staggering, we will identify these couplings with the interaction energies as follows
\be \label{tl_stagger}
\begin{array}{rclcrcl}
	t_b^x & = & \bar{\epsilon}_{D-\delta}, & \qquad & t_w^x & = & \bar{\epsilon}_{D+\delta}, \\
	t_b^y & = & \epsilon_{D}, & \qquad & t_w^y & = & \epsilon_{D}, \\
	t_b^z & = & \epsilon_{D-\delta}, & \qquad & t_w^z & = & \epsilon_{D+\delta},
\end{array}
\ee  
where the $\bar{\epsilon}_{\delta}$ is the staggered interaction energy in $x$-direction, as illustrated in Fig.~\ref{deform}(b). The comparison between the honeycomb model data and the prediction by our effective model with amplitudes \rf{tl_stagger} is shown in Fig.~\ref{predstagger}. We find that as the vortex lattice dimerizes, there will indeed be a transition at some $\delta_c$ away from the nucleated phase in the uniform ($\delta \to 0$) vortex lattice limit. However, this transition will in general be to another nucleated phase and only when the dimerization is sufficiently strong is the underlying non-Abelian phase recovered. As Fig.~\ref{predstagger} shows, this series of transitions is fully captured by the effective Majorana model. 

To systematically study how much vortex lattice dimerization the nucleated phases tolerate, we have studied a wide range vortex lattices with spacings $1 \leq D \leq 10$ (some of the data can be found in Appendix \ref{App_data}). We find that the critical dimerization $\delta_c$ oscillates with $D$ such that those lattices whose spacing coincides with oscillations minima/maxima (nodes) are in general more stable (unstable). The critical deformations we find are bounded from above, such that a dimerization by $\delta \gtrsim \lambda/8$, where $\lambda$ is the wavelength of the interaction oscillations \rf{splitting_sim}, will always destroy the nucleated phase. The proportionality of the critical deformation $\delta_c$ to the interaction wavelength can be understood in terms of the effective Majorana model. As we show in Appendix \ref{App_stagger} by analytically solving the staggered nearest neighbour Majorana model, dimerization larger than $\delta_c = \lambda/4$ will always result in some of the Majorana tunneling amplitudes changing signs. This corresponds to a change of the flux sector of the effective model, which we find to be associated with a phase transition. The smaller upper bound observed in the honeycomb model suggests that microscopics not accounted for by the Majorana model, such as the neglected longer range interactions, make the nucleated phases more unstable with respect to vortex lattice dimerization.

The dimerization required to drive the system back to the non-Abelian phase does not exhibit such systematic behavior, but the data presented in Appendix \ref{App_data} suggests that this occurs in general for dimerizations $\delta \gtrsim D/2$. The effective model can again be used to understand why such strong dimerization is required. In Appendix \ref{App_stagger} we find that the staggered Majorana model can be driven into a $\nu_M=0$ phase in two distinct ways. Either the effective couplings are sign staggered such that one obtains a stripey flux sector with $\pm \pi/2$ flux alternating on adjacent triangular plaquettes, or one remains in the same flux sector, but one of the tunneling amplitudes becomes at least twice as large compared to the other amplitudes. These correspond to different phases of the effective model, with the spectrum of the latter exhibiting localization and degeneracy as the dominant coupling increases. This motivates us to interpret the intermediate $\nu=-1$ phase in Fig.~\ref{predstagger} to arise due to the first mechanism (the magnitude of the tunneling amplitudes is roughly the same). On the other hand, the underlying non-Abelian phase in the $\delta \to D$ limit is recovered due to the second mechanism as $t_b^z$ amplitude becomes much larger than the others. The Majoranas are paired into localized complex fermion modes bound to dimerized vortex pairs and the collective state of the vortex lattice is suppressed.

\subsubsection*{Possible Peierls instability due to Majorana tunneling}

We found above that nucleated phases are in general stable with respect to a dimerization that is small compared to the wavelength of the interaction oscillations. In general, vortex lattice dimerization could be externally induced by subjecting the system to a periodic (impurity) potential, but it could also occur spontaneously in a clean system if the system energetically favours the pairing of vortices over forming a uniform triangular lattice.

\begin{figure}[t]
\begin{tabular}{c}
\includegraphics[width=8cm]{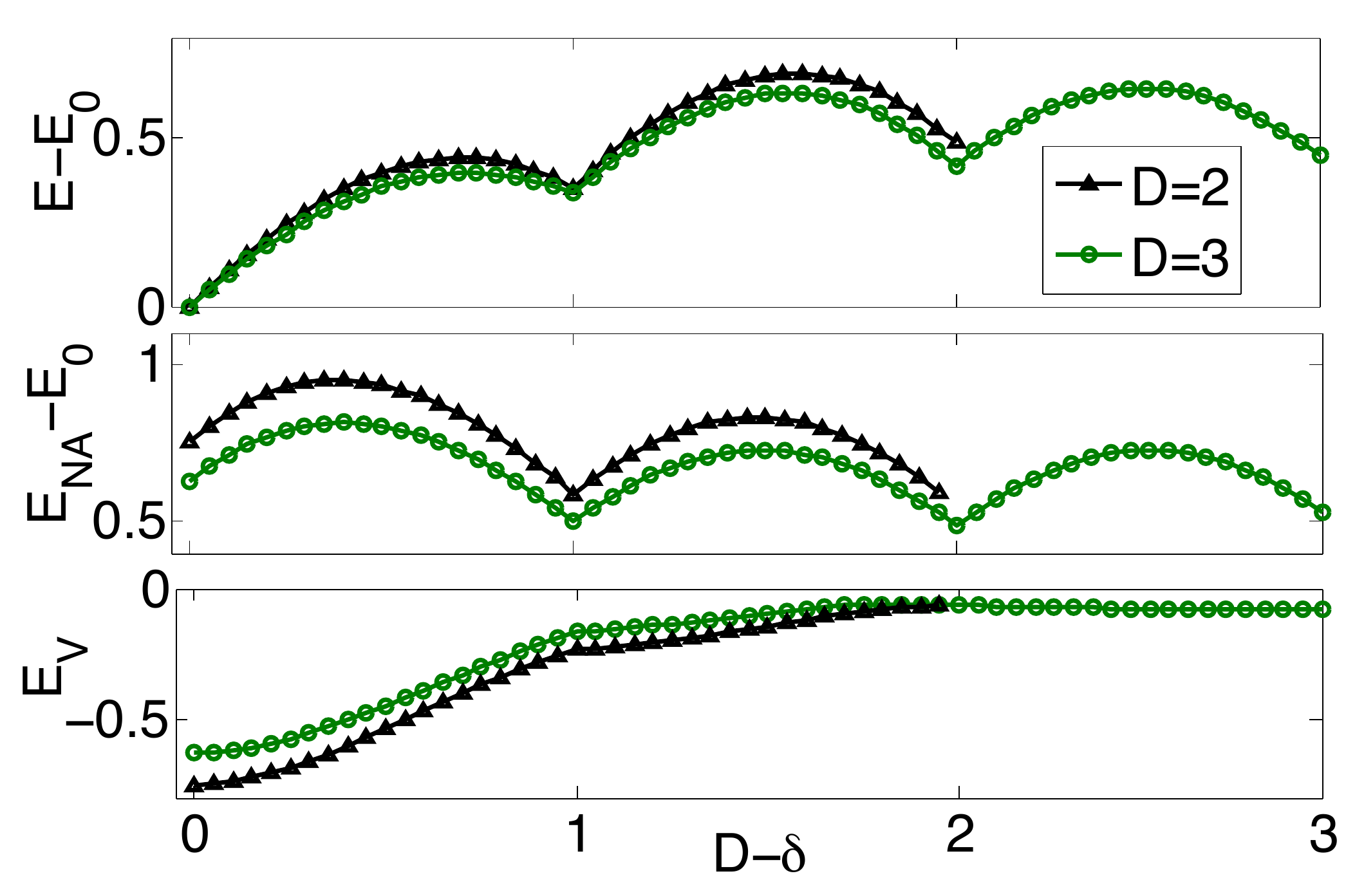}
\end{tabular}
\caption{\label{gs_deform} The band energies $E_V$ and $E_{NA}$ per plaquette as a function of the deformation $\delta$ for $D=2$ and $3$ vortex lattices. {\it Top:} The total ground state energy $E=E_{NA}+E_V$ with respect to the vortex-free sector ground state energy $E_0$. While the energy minimum is reached always when the vortices are fused ($\delta \to D$), moving vortices between plaquettes is penalized by a potential with a period of a plaquette spacing. {\it Middle:} This potential is encoded in $E_{NA}$ and it is independent of the interplay between the vortices. {\it Bottom:} $E_V$ encodes the instability due to Majorana tunneling. For the $D=3$ lattice the energy is at a local minimum for $\delta=0$ implying that the uniform lattice is stable. On the other hand, for the $D=2$ lattice the energy is at a local maximum, which means that from the point of view of pure Majorana tunneling, the system energetically prefers to dimerize.  The data is for $K=0.1$.}
\end{figure}

We investigate the existence of such {\it Peierls-like instability} by studying the ground state energy when the vortex lattice is dimerized. The band structure of the nucleated phases, as illustrated in Fig.~\ref{bands}, allows the ground state energy to be decomposed as $E=E_V+E_{NA}$, where $E_{NA}$ and $E_V$ are the energies corresponding to the high-energy non-Abelian band $\Psi_{NA}^-$ and the low-energy vortex band $\Psi_V^-$, respectively. The first describes the model specific microscopic contribution to the ground state energy, whereas the latter is the universal contribution due to the collective state of the Majorana modes (Fig.~\ref{predstagger}(c) shows that the ground state energy $E_M$ of our effective Majorana model provides an accurate approximation of $E_V$). We plot the band energies in Fig.~\ref{gs_deform} and observe that both exhibit periodic behavior, although of very different type. The minima of $E_{NA}$ have the periodicity of the plaquette spacing and occur always when the vortices are pinned on the plaquettes. It is also essentially independent of the dimerization, which suggests it should be interpreted as a periodic microscopic dependent vortex potential.\cite{Lahtinen11} The minima of $E_V$, on the other hand, have the periodicity of $\lambda/2$ and coincide with minima/maxima of the interaction oscillations. Moreover, $E_V$ decreases as $\delta \to D$, which is consistent with the vortex-free state being the lowest energy state over all vortex sectors.

The distinct behavior of the band energies suggest a competition of two distinct types of dynamics trying to minimize the energy of the many-vortex system. A vortex pinning background potential, as described by $E_{NA}$, favors the vortices to be pinned on plaquettes. On the other hand, the energetics of Majorana tunneling, as described by $E_V=E_M$, favor the vortices to be within distance $n\lambda/2$ $ (n=1,2,\ldots)$ of each other. Fig.~\ref{gs_deform} shows that unless the vortices are next to each other, the vortex pinning potential is in general stronger than the instability due to Majorana tunneling. This means that even if the vortices were allowed to move freely, the vortex lattice would be pinned to the uniform configuration by the background potential. 

The vortex pinning potential encoded in $E_{NA}$ is specific only to the honeycomb model, while the instability due to Majorana tunneling is universal to any system supporting localized Majorana modes. In Moore-Read fractional quantum Hall liquids or $p$-wave superconductors, such an instability would compete with the electrostatic repulsion that favours the formation of Wigner crystals or Abrikosov lattices, respectively. Since the repulsion decays polynomially, whereas the attractive tunneling decays exponentially with vortex separation, one expects that unless the system is at a very high vortex density ($D \approx \xi$), uniform vortex lattices in general minimize the ground state energy. One should note though that even if the instability did occur, it would not necessarily prevent nucleation necessarily from occurring. Assuming that the system would relax into a configuration minimizing $E_V$, such process would result in a maximum dimerization of $\delta \leq \lambda/4$. As this is the upper bound for critical dimerization, some of the stabler nucleated phases would be expected to survive.

\subsection{Random local disorder}

Finally, we turn to the effect of random spatial disorder on the vortex lattices. In real materials there always exists impurities that pin some of the anyonic quasiparticles. In the presence of an anyon lattice such impurities will lead to local random deformations away from a spatially uniform lattice. In the effective Majorana model \rf{Heff} this translates to the tunneling amplitudes $t^l_{ij}$ becoming local random variables. This problem has been studied in \cite{Laumann12,Kraus11}, where it is predicted that when the disorder is sufficiently strong to cause sign flips in the effective tunneling amplitudes, the system is driven to a thermal metal state. In this section we study the microscopics of this transition in the context of the honeycomb model.

We model the random vortex lattice disorder as local random disorder in the couplings $J_\alpha$, that we parametrize by $\delta J$. More precisely, by magnitude $\delta J$ disorder we mean that the couplings can vary locally as
\be \label{Jdis}
 J_{ij} \to (1 + \delta), \qquad -\delta J \leq \delta \leq \delta J,
\ee
where $\delta$ is the deviation from the mean $\langle J_{ij} \rangle=1$ selected randomly from a uniform distribution $-\delta J \leq \delta \leq \delta J$. We identify two distinct regimes of the disorder. For $\delta J < 1$ the disorder causes deformations of the vortex lattice that do not change the vortex number and thus preserve the triangular lattice, whereas for $\delta J > 1$ the disorder is strong enough to start effectively moving vortices between plaquettes and/or creating/annihilating them in pairwise fashion.

\subsubsection{Disorder in the vortex-free sector}

\begin{figure}[t]
\includegraphics[width=8.4cm]{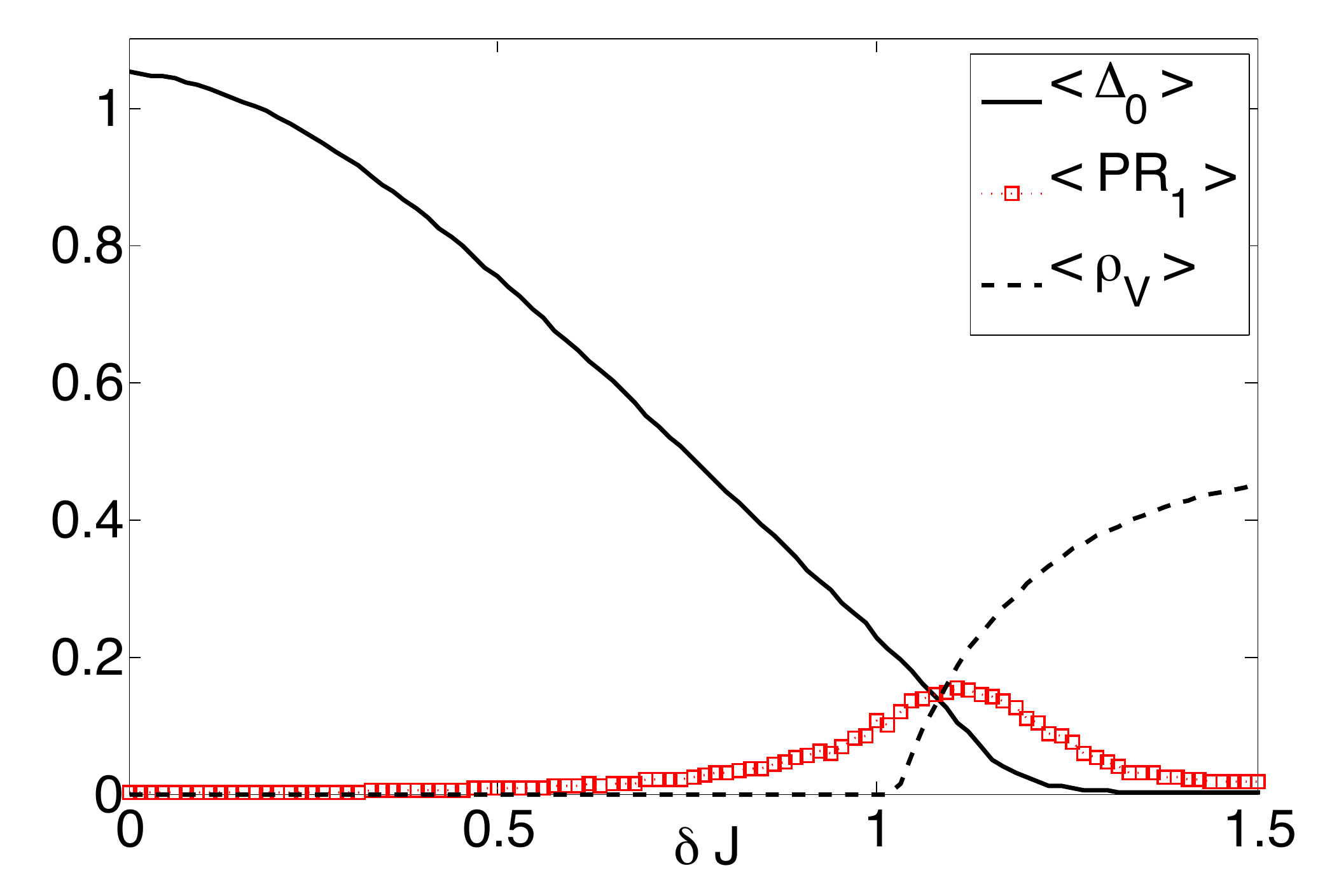} 
\caption{\label{gapNA} The disorder averaged energy gap $\langle \Delta_0 \rangle$ of the non-Abelian phase, the participation ratio of the lowest lying state $\langle PR_1 \rangle$ and the vortex density $\langle \rho_V \rangle$ as functions of the disorder strength $\delta J$. The maximum of $\langle PR_1 \rangle$ around $\delta J \approx 1.1$ at low vortex density implies the emergence of localized states bound to vortex pairs.  As disorder is further increased, vortex density increases implying an emergence of a random vortex lattice. The low-energy states re-delocalize as the bound Majorana modes can now tunnel across the whole system. The data is for $K=0.1$, averaged over 200 disorder realizations and calculated using a finite system of $L \times L$ plaquettes ($L=40$, $3.2 \cdot 10^3$ sites, $L/\xi \approx 30$ in the clean limit).}
\end{figure}

Before proceeding to study the vortex lattices, we first consider local random disorder in the non-Abelian phase in the absence of a vortex lattice. Agreeing with previous studies\cite{Willans10,Chua11} on weak disorder ($\delta J < 1$), Fig. \ref{gapNA} shows that the disorder averaged energy gap $\langle \Delta_0 \rangle$ decreases monotonously with increasing disorder strength $\delta J$. When disorder becomes sufficiently strong, the system is driven gapless around $\delta J \approx 1.2$. By looking at the participation ratio for the $n$th mode $\psi_n$,
\be
	PR_n = \int d^2r |\psi_n|^4,
\ee
we find that the lowest lying states localize in the vicinity of the transition to the gapless regime, but re-delocalize as disorder is further increased. We can understand this transition in terms of vortices which are created for $\delta J > 1$, as shown in Fig. \ref{gapNA}. Around $\delta J \approx 1.1$ only few isolated vortex pairs are present, and the Majoranas hybridize pairwise resulting in localized states bound to the pairs. When disorder is further increased, the vortices are forced into proximity of each other and a random vortex lattice is created. The Majoranas can now tunnel all across the system and the localized states delocalize again. 


\begin{figure}[t]
\includegraphics[width=8cm]{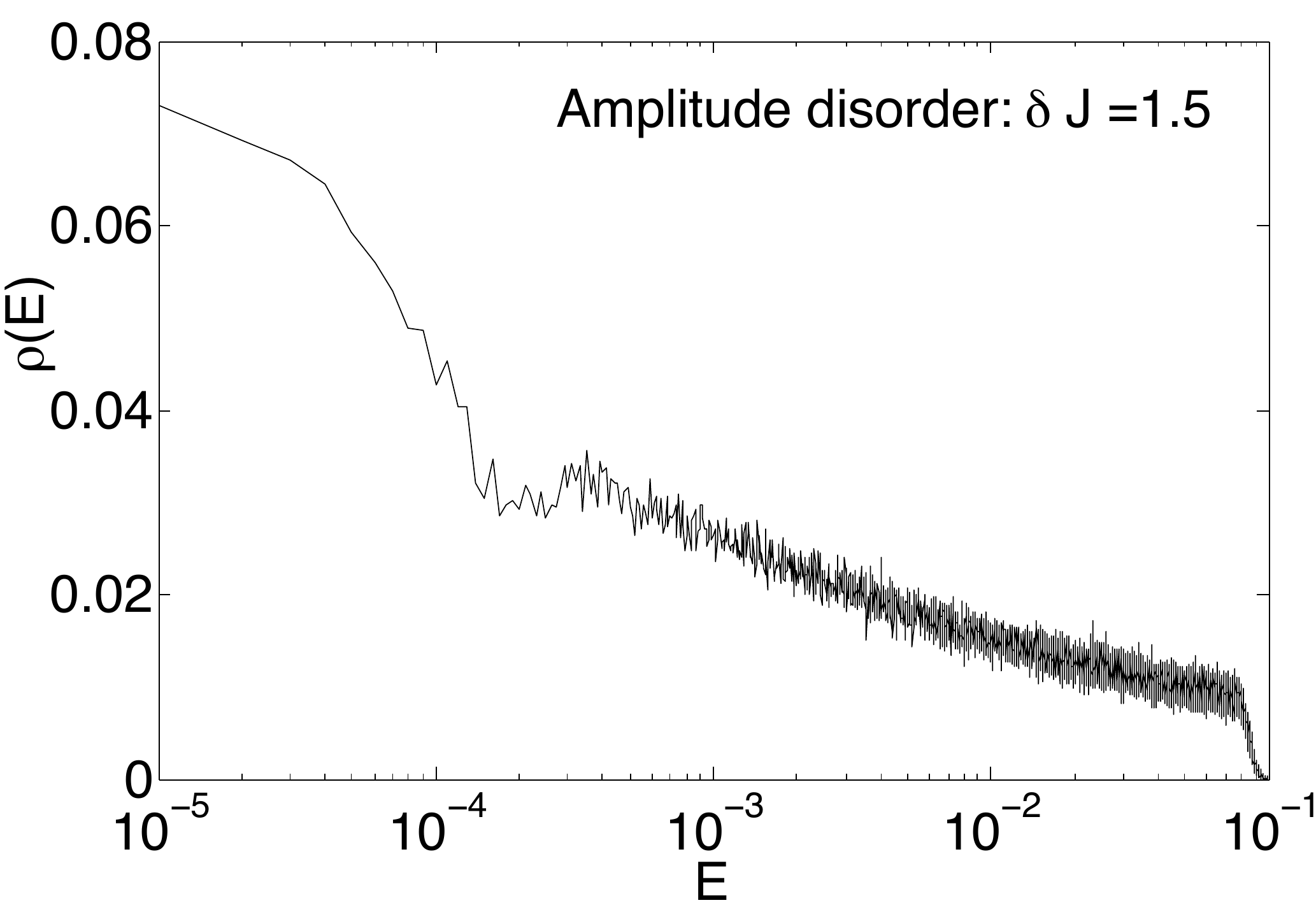} \\ (a) \\ \includegraphics[width=8cm]{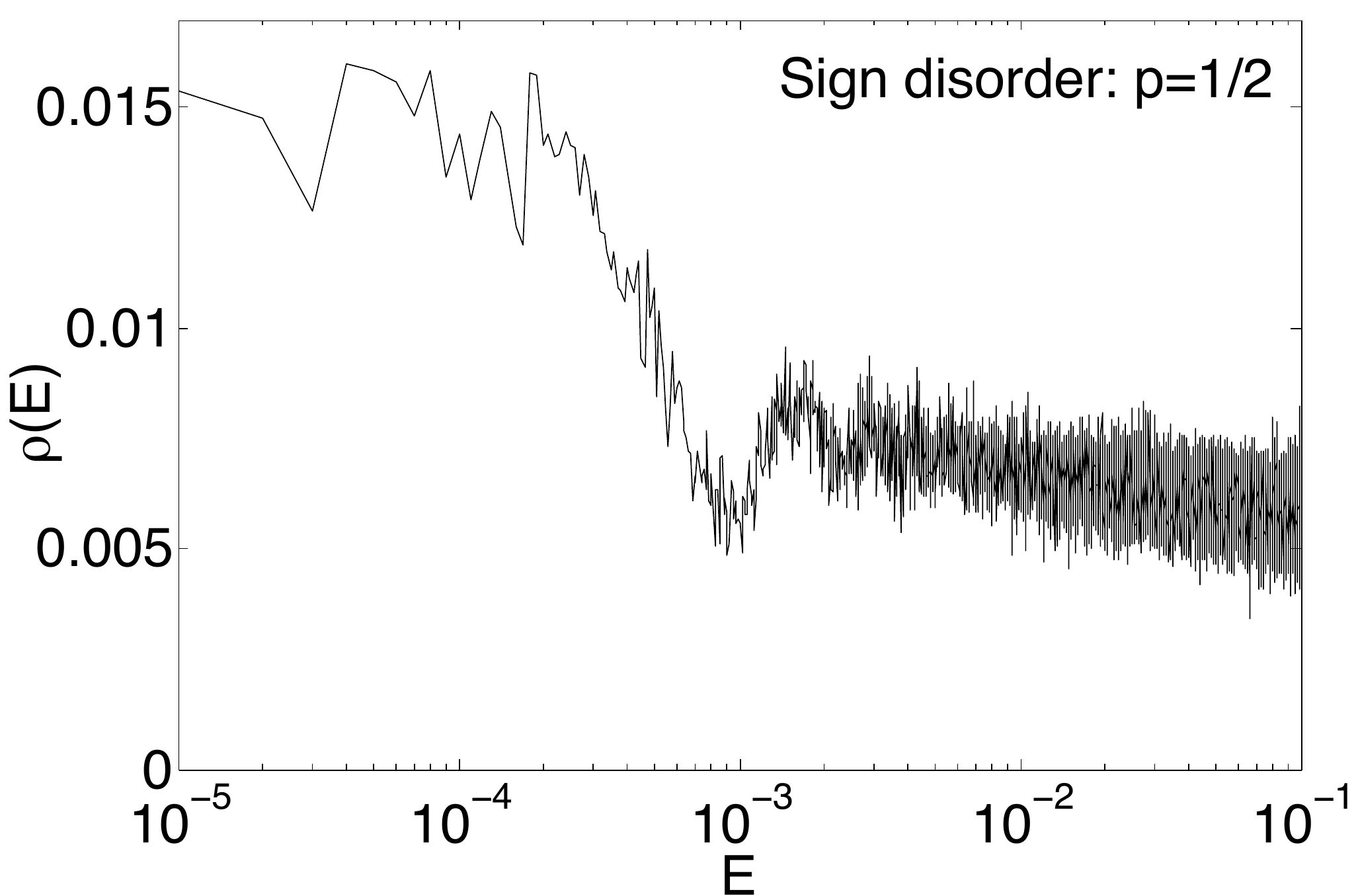} \\ (b)
\caption{\label{DOS_NA} The low-energy density of states $\rho(E)$ in the presence of (a) $\delta J = 1.5$ amplitude disorder and (b) pure sign disorder ($|J_{ij}|=1$ for all links, but the signs are completely random). Both show the random matrix theory predicted oscillation at the mean level spacings $\langle E_{i+1}-E_{i} \rangle=3 \cdot 10^{-4}$ and $1.3 \cdot 10^{-3}$, respectively. The amplitude disorder dampens the oscillations in (a), whereas for pure sign disorder they are clearly visible for several periods. The data is for $K=0.05$, averaged over $10^4$ disorder samples and calculated using a finite system of $L \times L$ plaquettes ($L=60$, $7.2 \cdot 10^3$ sites, $L/\xi \approx 22$ in the clean limit). }
\end{figure}

In terms of the effective model \rf{Heff} the random vortex lattice translates into random Majorana tunneling couplings. When the disorder is strong enough, the signs of the tunneling amplitudes become sufficiently random and the resulting gapless state is predicted to be a thermal metal \cite{Laumann12}.  This state is characterized by a logarithmically diverging density of states, which at low energies (at the order of the mean level spacing) exhibits characteristic oscillations. Indeed, Fig.~\ref{DOS_NA}(a) shows that for disorder of strength $\delta J =1.5$ the density of states diverges and displays oscillatory behavior at the energy scale of the mean level spacing. In the presence of pure sign disorder, the precise form of these oscillations is known\cite{Altland97} to be
\be \label{RMT}
	\rho(E) = \alpha + \sin(2\pi\alpha E L^2)/(2\pi\alpha E L^2),
\ee
where $\alpha$ is a non-universal constant and $L^2$ is the system size. Fig.~\ref{DOS_NA}(b) shows that when the signs of $J_{ij}$ are completely randomized (while keeping their amplitudes fixed to $|J_{ij}|=1$), the oscillations become clearly visible over several periods. The logarithmic divergence is also confirmed by studying the scaling $E_1 \sim \frac{1}{L^\gamma}$ of the lowest lying states in the gapless region. Considering systems of linear size $10 < L < 100$, we find $\gamma = 2.6$ and $2.2$ for the $\delta J = 1.5$ amplitude and pure sign disorders, respectively. Scaling faster than with the system size ($\gamma > 2$) implies at least logarithmic divergence of the density of states \cite{Kraus11}. 

We note that our method of averaging over increasing disorder is qualitatively similar to averaging over thermal fluctuations, which has been used to study the $p$-wave superconductor in a finite temperature \cite{Bauer12}. There increasing temperature also leads first to confined vortex pairs, which after some critical temperature deconfine to create a random vortex lattice that gives rise to the same thermal metal state. Thus had we sampled the couplings $J_{ij}$ from a thermal distribution instead of a uniform distribution, we expect to have discovered a different critical temperature, but otherwise similar results. 

\subsubsection{Disordered vortex lattices}

When a vortex lattice is already present, we expect the thermal metal to emerge for some $\delta J_c < 1$ that coincides with the interactions between the already present vortices becoming sufficiently disordered. We will show below that this is indeed the case by explicitly studying how the local random disorder modifies the vortex-vortex interactions.

\begin{figure}[t]
\begin{tabular}{c}
\includegraphics[width=8.4cm]{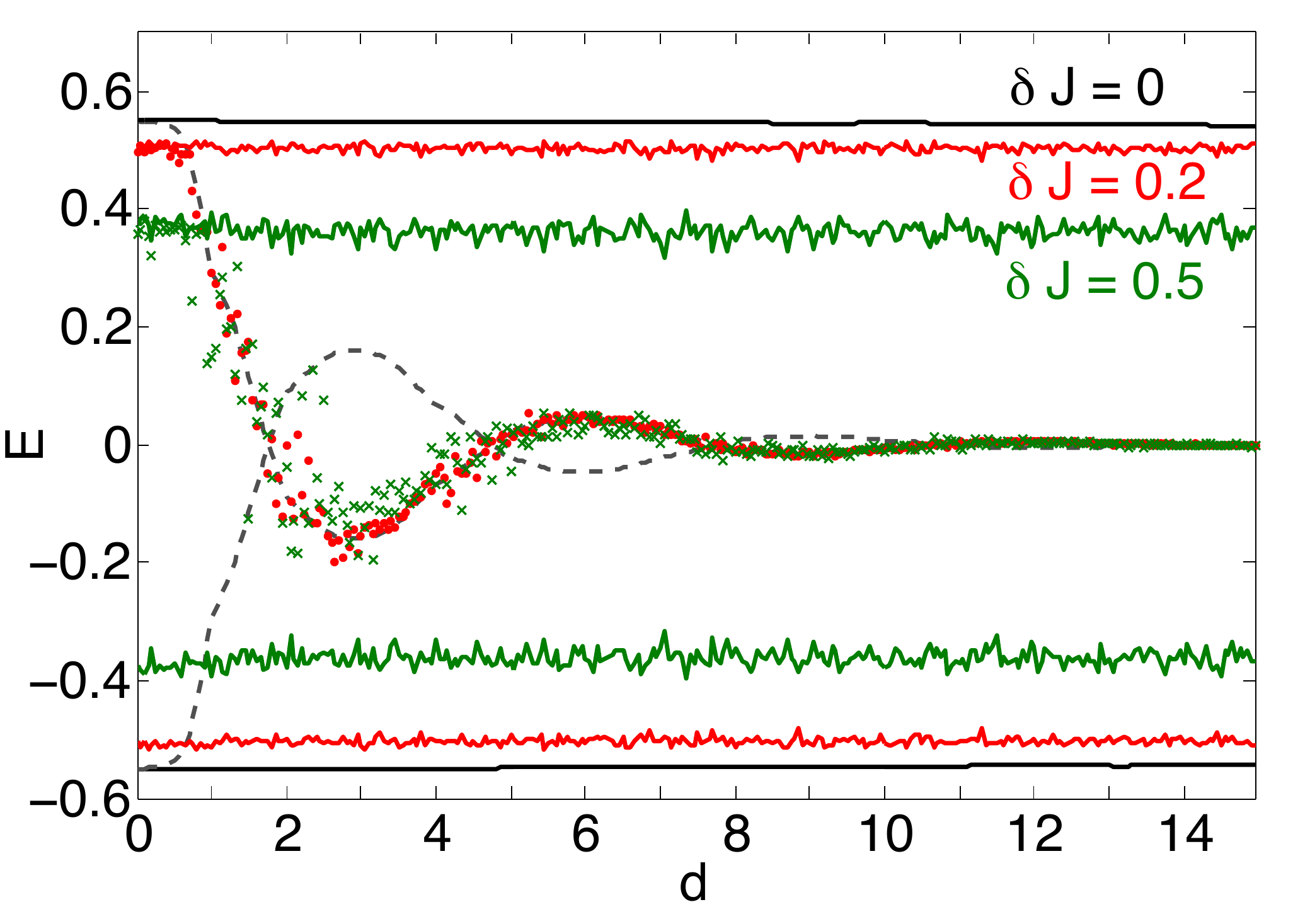} \\ (a) \\
\includegraphics[width=8.4cm]{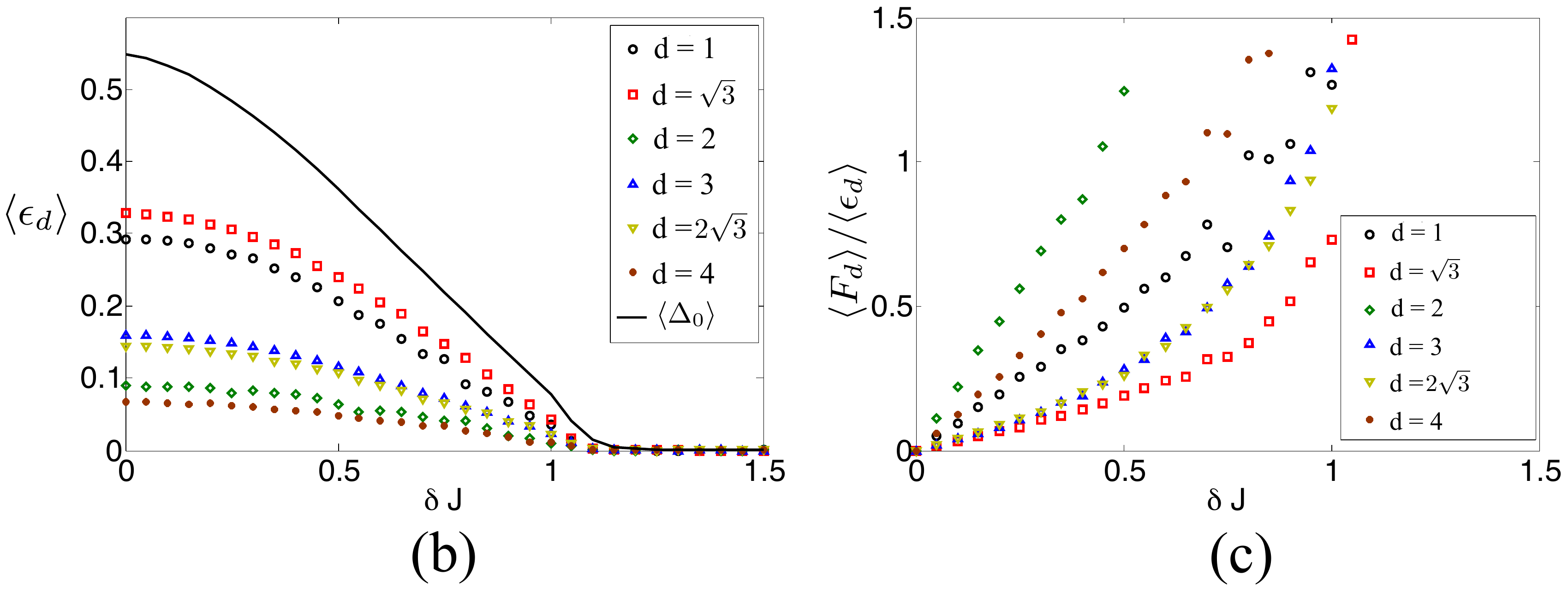}
\end{tabular}
\caption{\label{dis_int} The effect of coupling disorder on the vortex-vortex interactions. (a) The spectrum as the function of vortex separation $d$ in the presence of disorder of magnitude $\delta J = 0$ (black), $0.2$ (red) and $0.5$ (green). (b) The mean energy splitting $\langle \epsilon_d \rangle$ decreases monotonously with increasing disorder, but remains finite until the closure of the disorder averaged gap $\langle \Delta_0 \rangle$. (c) This contrasts with the mean relative fluctuation, $\langle F_d \rangle / \langle \epsilon_d \rangle = \langle \langle \epsilon_d \rangle - \epsilon_d \rangle / \langle \epsilon_d \rangle$, that increases with increasing disorder. The rate depends on the proximity to the oscillation nodes with those nearby to them increasing faster. The data is for $K=0.05$, averaged over $10^3$ disorder samples and calculated using a finite system of $L \times L$ plaquettes ($L=40$, $3.2 \cdot 10^3$ sites, $L/\xi \approx 15$ in the clean limit). }
\end{figure}

Fig. \ref{dis_int} shows that in the presence of random local disorder the energy splitting $\epsilon_d$ acquires fluctuations. Averaging over many disorder realizations, we find two general ways the interactions are modified. The mean value $\langle \epsilon_d \rangle$ decreases monotonously with with increasing disorder, while the fluctuations around the mean, $F_d=\langle \langle \epsilon_d \rangle - \epsilon_d \rangle$, increase with it. The mean value remains finite all the way up to $\delta J \approx 1.2$, where we found disorder averaged gap $\langle \Delta_0 \rangle$ of the non-Abelian phase to close. Thus while the interactions are strongly influenced by disorder, they remain well-defined throughout the non-Abelian phase. Moreover, the wavelength of the oscillating interaction energy is relatively unaffected by the disorder. This insensitivity derives from the disordered $J_{ij}$ couplings randomly shifting the two Fermi points \cite{Lahtinen10}. As the interaction oscillation wavelength in \rf{int} depends only on their difference, this effect cancels out. 

\begin{figure}[t]
\includegraphics[width=8cm]{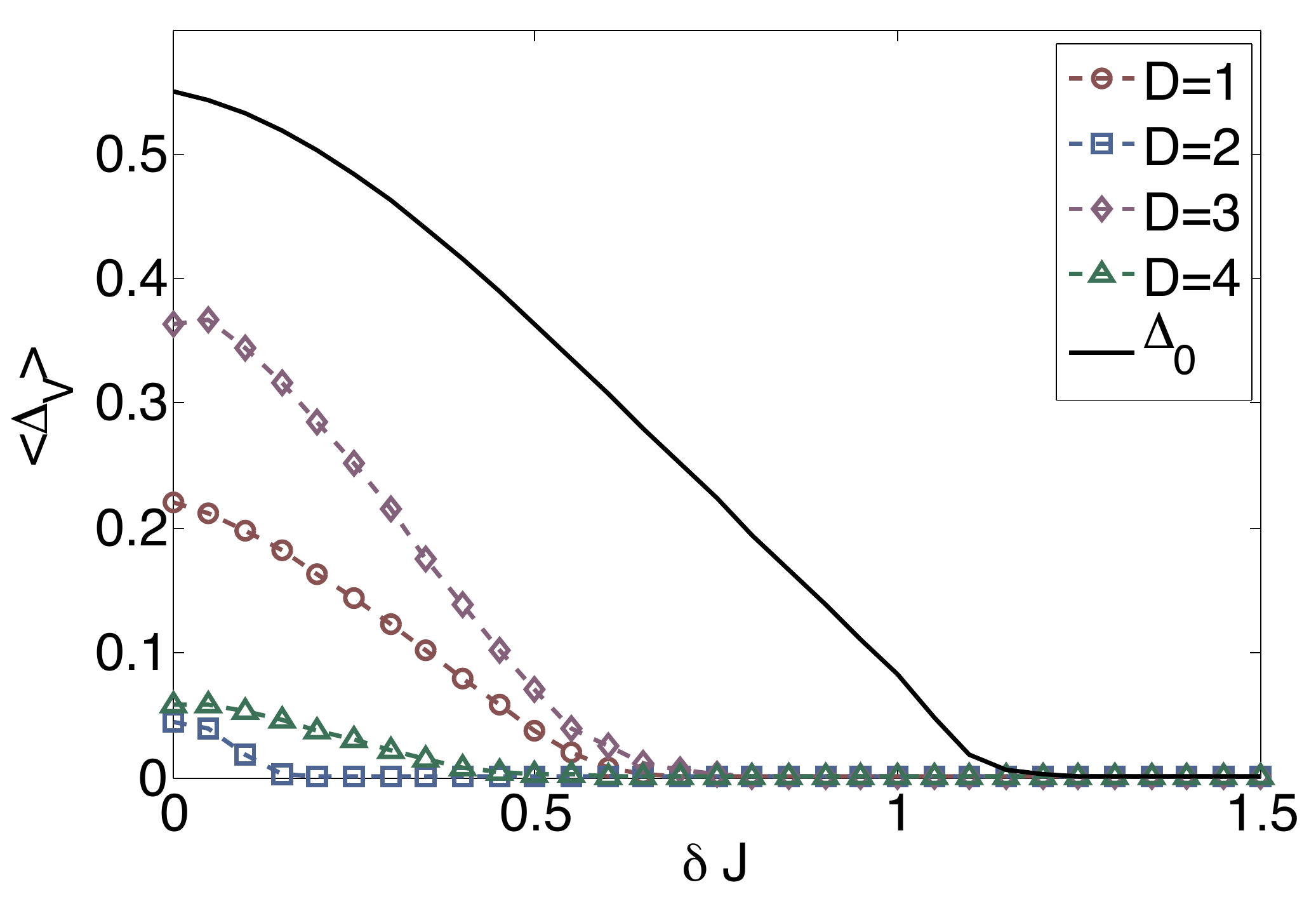}
\caption{\label{dis_D} The disorder averaged energy gaps $\langle \Delta_V \rangle$ for some of the nucleated phases as functions of $\delta J$. For all cases the nucleated gaps decrease monotonously and close at some critical value $\delta J^c < 1$, i.e. for smaller disorder than what is required to drive the non-Abelian phase gapless. The data is for $K=0.05$, averaged over $200$ disorder samples and calculated using a finite system of $L \times L$ plaquettes ($L=40$, $3.2 \cdot 10^3$ sites, $L/\xi \approx 15$ in the clean limit).}
\end{figure}

\begin{figure}[t]
\begin{tabular}{cc}
\includegraphics[width=4.2cm]{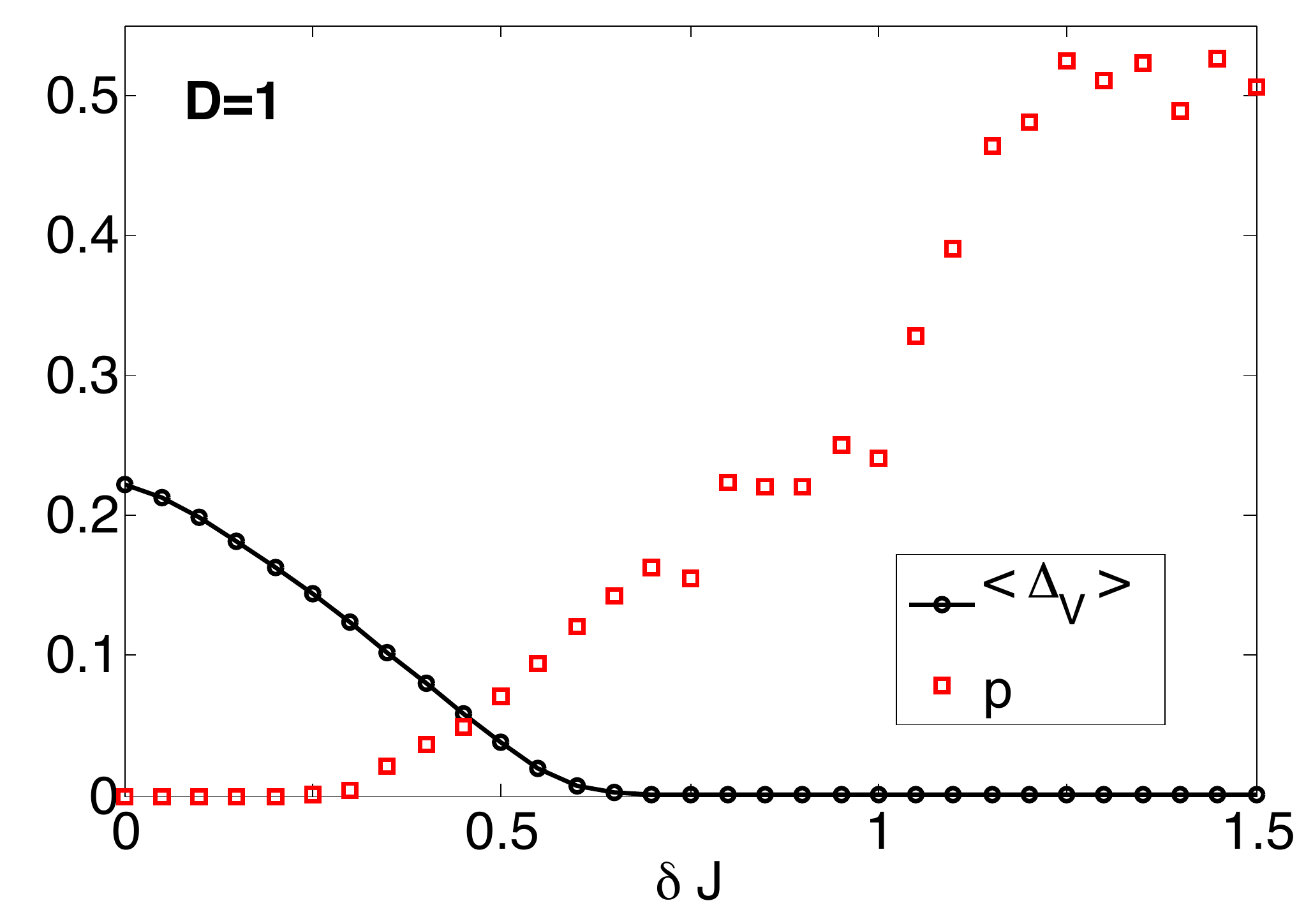} & \includegraphics[width=4.2cm]{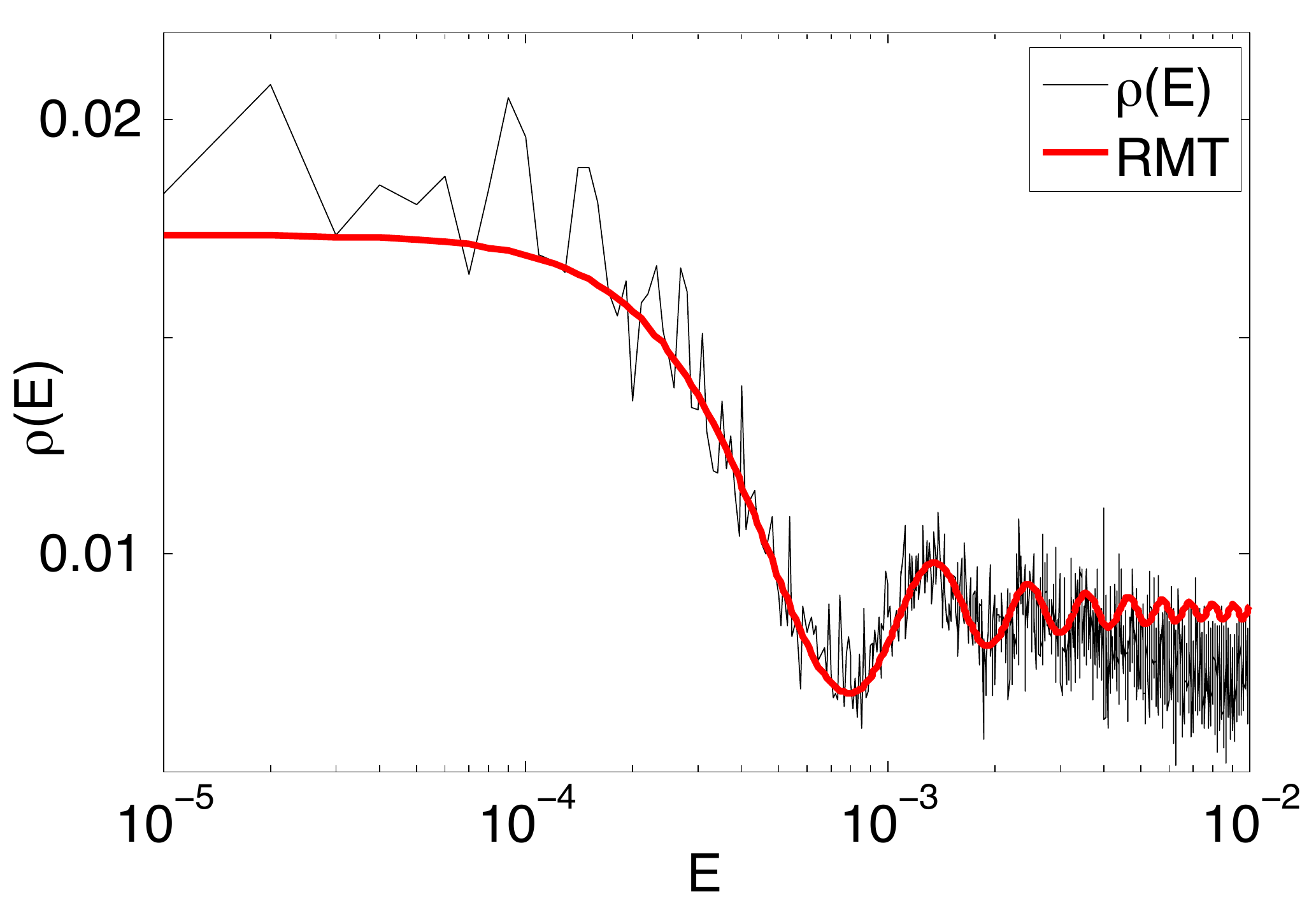} \\
(a) & (b)
\end{tabular}
\caption{\label{DOS_D} (a) The gap closures of the nucleated phases always correlate with the disorder being strong enough to give rise to a finite probability $p \gtrsim 0.1$ for the interactions to flip signs. (b) The low energy density of states in the presence of strength $\delta J=0.8$ disorder. The oscillations and the logarithmic divergence agree with the prediction of the random matrix theory\cite{Altland97} (RMT) with the first bump occurring at the mean level spacing of $\langle E_{i+1}-E_i \rangle \approx 8 \cdot 10^{-4}$. The data is for $D=1$ vortex lattice with $K=0.05$, averaged over $2 \cdot 10^4$ disorder samples and calculated using a finite system of $L \times L$ plaquettes ($L=60$, $7.2 \cdot 10^3$ sites, $L/\xi \approx 22$ in the clean limit).}
\end{figure}

The effect of local random disorder on the nucleated phases themselves is shown in Fig.~\ref{dis_D}. As expected we find all of them being driven gapless (the disorder averaged gap $\langle \Delta_V \rangle$ closes) for some critical disorder $\delta J_c < 1$ that depends on the vortex lattice spacing $D$. In general those nucleated phases whose spacing coincides with the oscillation nodes have smaller gaps and are driven gapless for weaker disorder, while the phase whose spacing coincide with the oscillations minima/maxima are more stable. When the effective Majorana tunneling amplitudes $t^l_{ij}$ are picked from the distribution $\epsilon_{lD}(\delta J)$, the transition to the thermal metal phase is predicted to correlate with the onset of finite probability $p$ for the tunneling amplitudes to have random signs.\cite{Laumann12} This is verified in Fig.~\ref{DOS_D}(a), which shows how the closure of $\langle \Delta_V \rangle$ coincides with a finite sign flip probability of $p \approx 0.1$. Data for other vortex lattices confirming this correlation can be found in Appendix \ref{App_data}.

Finally, to verify that the gapless state in the presence of a vortex lattice is indeed the thermal metal state, we plot in Fig.~\ref{DOS_D}(b) the disorder averaged low energy density of states $\langle \rho(E) \rangle$ for the $D=1$ vortex lattice in the presence of disorder of magnitude $\delta J = 0.8$. This disorder strength is sufficient to drive the nucleated phase gapless, but not strong enough to destroy the underlying non-Abelian phase. Like in the case of the sufficiently disordered non-Abelian phase, we find the characteristic logarithmic divergence and the characteristic oscillations \rf{RMT}, which again confirm that the gapless state in the presence of a disordered vortex lattice is indeed the thermal metal.

\section{Conclusions}

We have studied the stability of nucleated topological phases\cite{Ludwig11} in the context of Kitaev's honeycomb model\cite{Kitaev06} under three different kinds of perturbations: anisotropic interactions due spatially anisotropic vortices, dimerization of the underlying vortex lattice and local random disorder. While the system remains stable with respect to moderate perturbations of every type, something one expects for a gapped topological phase, for strong perturbations very different physics is obtained. Spatial anisotropy is found to stabilize the strong pairing phases and to explain how the the phase diagram of the honeycomb model is modified in the presence of vortex lattices. Dimerization of the vortex lattice, on the other hand, was found to be able to recover the underlying non-Abelian phase. The maximal dimerization tolerated by the nucleated phases was found to be directly proportional to the wavelength of the interaction oscillations. Finally, we showed that local random disorder drives all the nucleated phases into a thermal metal state. The transition to this phase could be traced back to the predicted microscopic onset of sign disorder in Majorana tunneling amplitudes \cite{Laumann12}. Our main result is to show that all these distinct behavior could be accurately described by an effective Majorana tight-binding model.

The upshot of our results is that nucleated phases are predicted to be stable with respect to various perturbations and that the simple picture provided by the effective Majorana tight-binding model applies also in the presence of disorder. However, one should keep in mind that the degree of stability is given only with respect to the interaction induced gap which decays exponentially with the vortex lattice spacing. Thus in the light of potential experiments, high vortex densities (of the order where the vortex lattice spacing is within few coherence lengths) are likely to be required for the nucleated phases to survive disorder. As the energy gaps of nucleated phases are always smaller than those protecting the parent non-Abelian phases, something that are already challenging for current technology, observing nucleated phases will require delicate control over the experiments. Thus while putative $p$-wave superconductors or Moore-Read fractional quantum Hall states seem natural places to look for them (an experimental observation of an Abelian quasiparticle Wigner crystal has been recently reported\cite{Zhu10}), a more promising route might be optical lattice realizations.\cite{Duan03,Micheli05,Alba13} Particularly promising could be optical lattice experiments on fractional quantum Hall states,\cite{Dalibard13} that due to their inherently clean nature and controllability\cite{Komineas12} are attractive for overcoming the challenges faced by the realization nucleated topological phases. Another potential route could be topological nanowires, where the experimental evidence for Majorana fermions is augmenting \cite{Mourik12,Das12,Churchill13}. A regular two dimensional array of such wires would behave much like a vortex lattice and could support collective nucleated states of Majoranas.\cite{Kells13} This possibility is also relevant to using them as topological quantum computers\cite{Alicea11} -- nucleation would constitute the ultimate failure of the computer since undergoing a phase transition would wipe out any encoded information. Our results on staggering protecting the non-Abelian phase provides a way to avoid such a scenario.

Taking the positive view that Majorana lattices could be realized in experiments, one can speculate whether the potential dimerizing instability due to the oscillations\cite{Lahtinen11, Cheng09, Baraban09, Sau10} in the Majorana tunneling amplitudes is observable. To observe it one needs a system where the quasiparticles can freely relax into a minimum energy configuration and where the energy scale of the system specific microscopics is weaker than the Majorana tunneling. Potential candidates could be Abrikosov lattices or Wigner crystals near the phase boundaries where the coherence length diverges. In this regime the vortex lattice spacing can become of the order of coherence length, which enables in principle the instability to become comparable or stronger to the Coulomb repulsion that favors uniform lattices. Our results on the emergence of the thermal metal state in the honeycomb model might also be relevant to certain class of iridates that may realize the Kitaev-Heisenberg model.\cite{Chaloupka10,Jiang11}

\section*{Acknowledgments}

VL would like to thank S. H. Simon and J. K. Pachos for inspiring discussions. VL has been supported by the Dutch Science Foundation NWO/FOM. ST acknowledges partial support from SFB TR 12 of the DFG.

\appendix

\section{Analytic solution for the staggered Majorana model}
\label{App_stagger}

In this Appendix we first analytically solve the staggered Majorana model with only nearest neighbour interactions. Then we employ the solution to study the phase diagram due to different types of staggered tunneling amplitudes. 

In the presence of only nearest neighbour tunneling, i.e. setting $t_{\sqrt{3}}^\alpha=0$ in \rf{Heff}, the unit cell consists of two sites, which we label black ($b$) and white ($w$). Allowing for arbitrary staggering, the corresponding effective Majorana model has six independent tunneling couplings. The Hamiltonian for this model in the $\Phi_1=\pi/2$ flux sector\cite{Lahtinen12} can be written as
\bq \label{Heff_staggered}
	H & = & \frac{i}{2} \sum_i \left[ \left( t_b^z b_i w_{i+y-x} - t_b^y b_i b_{i+y} - t_b^x b_i w_i \right) \right. \\
	\ & \ & + \left. \left( t_w^z w_i b_{i+y} + t_w^y w_i w_{i+y} + t_w^x w_i b_{i+x} \right) + \textrm{h.c.}  \right], \nonumber
\eq
where $b_i$ and $w_i$ denote the Majorana fermion operators on the black and white sites, respectively, and $t_{b/w}^{\alpha}$ are the nearest neighbour tunneling amplitudes in the directions illustrated in Fig.~\ref{efflattice}. 

Fourier transforming with respect to the two site magnetic unit cell, we obtain $H=\int_{BZ}\textrm{d}^2\p \psi_\p^\dagger H_\p \psi_\p$ with the Bloch Hamiltonian being given by
\be \label{Heff_Bloch}
	H_\p = \left( \begin{array}{cc} g^b_\p & if_\p \\ -if_\p^* & -g^w_\p \end{array} \right),
\ee
where
\bq
	g^\alpha_\p & = & 2 t_\alpha^y \sin(p_y), \nonumber \\	
	f_\p & = & t_b^z e^{i(p_y-2p_x)}-t_b^x-t_w^z e^{-ip_y}-t_w^x e^{-2ip_x}. \nonumber
\eq   
The Hamiltonian is written in the Fourier transformed Majorana basis $\psi^\dagger_p=(b_\p^\dagger,w_\p^\dagger)$ and the Brillouin zone (BZ) is halved to $p_x \in [0,\pi]$ and $p_y \in [-\pi,\pi]$ to avoid double counting due to $\psi_{-\p}=\psi_\p^\dagger$. The corresponding eigenvalues are given by
\be \label{Heff_E}
	E_\pm(\p) = \frac{1}{4}\left( (g^b_\p-g^w_\p) \pm \sqrt{(g^b_\p+g^w_\p)^2+4|f_\p|^2} \right),
\ee
where
\bq
	|f_\p|^2 & = & (t_b^x)^2 + (t_b^z)^2 + (t_w^x)^2 + (t_w^z)^2 \nonumber\\
	\ & \ & -2t_b^z t_w^z \cos(2p_y+2p_x) + 2 t_b^x t_w^x \cos(2p_x) \nonumber \\
	\ & \ & + 2(t_w^x t_w^z - t_b^x t_b^z) \cos(2p_x-p_y) \nonumber \\
	\ & \ & + 2(t_w^z t_b^x - t_w^x t_b^z) \cos(p_y). \nonumber
\eq
The energy gap and the ground state energy are given by $\Delta_M =\min_\p E_+(\p)$ and $E_M=\frac{1}{\pi^2}\int_{BZ}\textrm{d}^2\p E_-(\p)$, respectively. When all the couplings are equal, the solution reduces to the analytic solution for the uniform tunneling problem\cite{Grosfeld06}.

\subsection{Three distinct types of staggered tunneling }

We study first separately the three distinct types of staggering that can occur due to the oscillations in the interactions. To this end we set $t_b^x=t_w^x=t_b^y=t_w^y=1$ for the time being and study staggering only in $z$-direction. The oscillations in the interactions, as shown in Fig.~\ref{int}, can lead to three distinct types of staggering: (i) $|t_b^z|$ can increase while $|t_w^z|$ decreases, or vice versa (the general case), (ii) Both $|t_b^z|$ and $|t_w^z|$ can decrease (special case when $D$ coincides with the oscillation minima/maxima), or (iii) $|t_{b}^z|$ becomes much larger than the other couplings (strong dimerization in the $\delta \to D$ limit). The tunneling couplings corresponding to these limiting cases are given by
\bq \label{t_stagger}
	(\textrm{i}): \quad t_b^z = 1-\delta t, & \qquad & t_w^z = 1+\delta t, \nonumber \\
	(\textrm{ii}): \quad t_b^z = 1-\delta t, & \qquad & t_w^z = 1-\delta t, \\
	(\textrm{iii}): \quad  t_b^z = 1+\delta t, & \qquad & t_w^z = 1, \nonumber
\eq
where $\delta t > 0$ is the magnitude of the staggering. 

\begin{figure}[t]
\includegraphics[width=8cm]{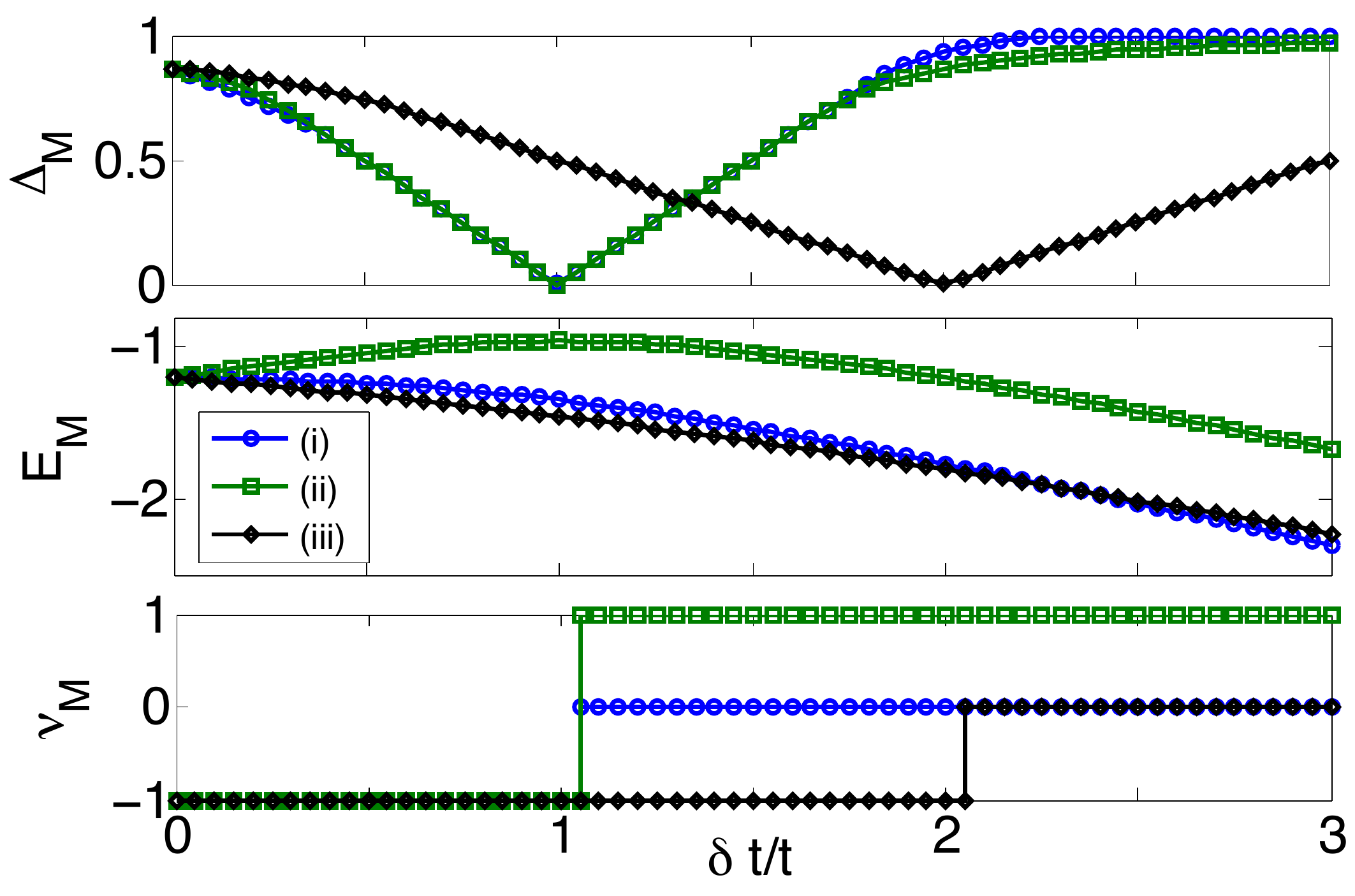}
\caption{\label{Eff_PD} The energy gap $\Delta_M$, the ground state energy $E_M$ and the Chern number $\nu_M$ for the three distinct staggerings \rf{t_stagger}. Both (i) and (iii) can drive the system into gapped phase characterized by $\nu_M=0$, and thereby recover the undelrying non-Abelian phase, whereas (ii) dimerization can only drive the system to a time-reversed $\nu_M=1$ phase. The ground state energies show that the latter is energetically penalized for small deformations. }
\end{figure}

The effect of these three distinct types of staggering is shown in Fig.~\ref{Eff_PD}. We find that both the generic staggering (i) and the strong dimerization (iii) can drive the effective model to a gapped phase characterized by $\nu_M=0$. This explains why sufficiently strong periodic deformation always recovers the underlying non-Abelian phase. Moreover, we find the ground state energies decreasing monotonously in both cases. This implies that consistent with our findings in the honeyomb model, the pure Majorana tunneling energetically favours vortex pair dimerization. However, we should note that while both types of staggering lead to a $\nu_M=0$ phase, these phases are distinct. The transition in the generic case (i) occurs when $t_w^z$ changes sign, which means that the flux sector effectively changes from the uniform to stripey one. In this phase there is no net magnetic flux as the flux alternates between time-reversed $\pi/2$ and $-\pi/2$ stripes in a columnwise fashion. On the other hand, in the latter case the $\nu_M=0$ phase emerges for uniform flux when the amplitude $t_b^z$ becomes much larger than the other couplings. In this case the Majoranas become strongly dimerized, which leads to an increasingly localized and degenerate spectrum with increasing $\delta t$. The intermediate non-Abelian $\nu=-1$ phases observed in the honeycomb model (see supplementary data in Appendix \ref{App_data}) thus correspond to stripey flux pattern in the effective Majorana model. On the other hand, the non-Abelian phases connected to the vortex-free limit correspond to the strongly dimerized scenario, where the pairing of the Majoranas suppresses the collective vortex lattice state.

The behavior of the Majorana model for the fine-tuned staggering (ii) is different. We find that it can drive the system to a time-reversed $\nu_M=1$ phase, which results from inverting all the fluxes in the effective model. However, more important is to notice that this staggering, which should only occur when the vortex lattice spacing $D$ coincides roughly with the oscillation minima/maxima, is energetically penalized for moderate values of $\delta t$. As we show in next section, this is in agreement with the our result that such vortex lattices should stable with respect to any dimerizing instability arising due to the Majorana tunneling.

\subsection{Dimerization of spherically symmetric vortices in continuum}

Having independently studied the limiting types of staggering and the transitions they can drive, we return to the more realistic case where several types staggerings are simultaneously present. To study the general behavior of the Majorana model, we consider a simplified continuum model of with full rotational symmetry (as opposed to only $C_3$ symmetry of the honeycomb lattice). This means that we assume that the interaction energy for all separations $l$ is given by 
\be \label{int_ideal}
	\hat{\epsilon}(l) = \cos(\frac{2\pi l}{\lambda}) e^{-\frac{l}{\xi}},
\ee 
where the wavelength $\lambda$ and the coherence length $\xi$ are now free parameters.  The corresponding staggered Majorana tunneling amplitudes can be identified as
\be \label{tl_stagger_ideal}
\begin{array}{rclcrcl}
	t_b^x & =& \hat{\epsilon}(\delta'), & \qquad & t_w^x & = & \hat{\epsilon}(\delta''), \\
	t_b^y & =& \hat{\epsilon}(D), & \qquad & t_w^y & = & \hat{\epsilon}(D), \\
	t_b^z & =& \hat{\epsilon}(D-\delta), & \qquad & t_w^z & = & \hat{\epsilon}(D+\delta),
\end{array}
\ee  
where
\bq
	\delta' & = & \sqrt{\left(\sqrt{3}D/2\right)^2+\left(D/2 + \delta \right)^2}, \nonumber \\
        \delta'' & = & \sqrt{D^2 + (D-\delta)^2 - D(D- \delta)}. \nonumber
\eq 
This contrasts with the staggered couplings \rf{tl_stagger} related to the full honeycomb model for which one had to study separately the interaction energy $\epsilon_d$ for each relative pairwise vortex configuration.

\begin{figure}[t]
\begin{tabular}{c}
\includegraphics[width=8cm]{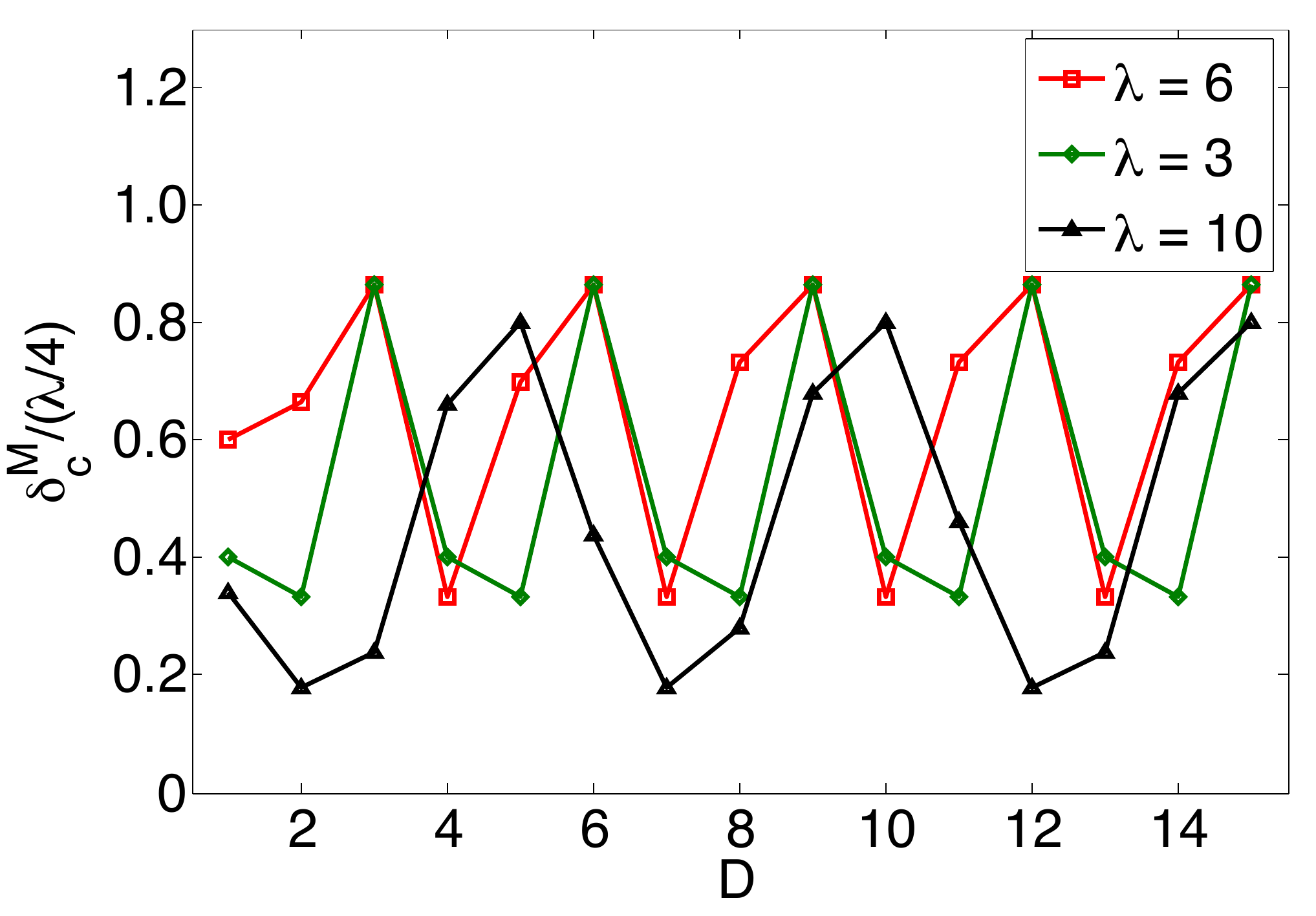}
\end{tabular}
\caption{\label{App_crit_deform} The critical deformations $\delta_c^M$ that drive the idealized Majorana model with couplings \rf{tl_stagger_ideal} out of the topological phase in the uniform coupling limit, as the functions of the vortex lattice spacing $D$.  The system is always driven out of the nucleated phase when the upper bound of $\delta\approx \lambda/4$ is breached. The stability varies with $D$ with the most stable phases occurring when $D \approx n\lambda/2$ ($n=1,2,\ldots$). The plots are all for $\xi=2$. }
\end{figure}

Like the full honeycomb model, also the idealized dimerized Majorana model exhibits numerous intermediate phases between the nucleated phase in the uniform lattice limit and the non-Abelian phase in the vortex-free limit. We plot in Fig.~\ref{App_crit_deform} the critical deformations $\delta_c^M$ required to drive the idealized effective model out of the uniform lattice limit nucleated phases for several interaction wavelengths. For each $\lambda$ we find that the most stable vortex lattices can tolerate deformations of up to $\delta \leq \lambda/4$, with the most stable phases occurring periodically when $D \approx n\lambda/2$ ($n=1,2,\ldots$). This is in agreement with our findings in the honeycomb model. The only difference is that in honeycomb model the upper bound for stability was lower ($\delta_c \leq \lambda/8$), which suggests that longer range tunneling and/or reduction of the rotational symmetry due to the lattice destabilizes the nucleated phases with respect to dimerization.

\begin{figure}[t]
\begin{tabular}{c}
\includegraphics[width=8cm]{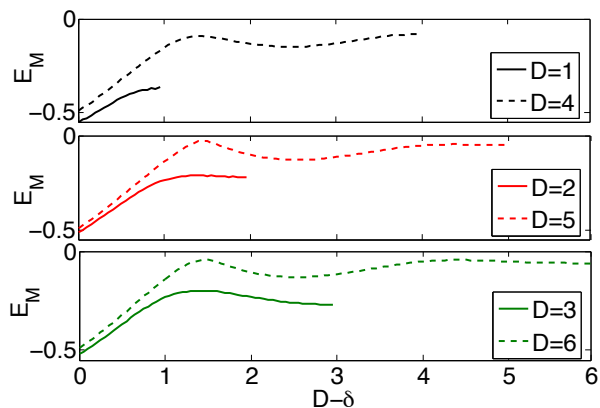} 
\end{tabular}
\caption{\label{App_EM} The ground state energies $E_M$ of the Majorana model with couplings \rf{tl_stagger_ideal} for various vortex lattice spacings $D$. The spacings $D=3,6,\ldots$ coincide with the oscillation minima/maxima and thus their corresponding ground state energies $E_M$ are in a local minimum. For the other vortex lattice spacings the ground state energies are near a local maximum, which implies that these lattices could exhibit a dimerizing instability. The plots are for $\lambda = 6$ and $\xi = 2$}
\end{figure}

We also verify the existence of the possible dimerizing Peierls-like instability using the spherically symmetric vortices. Like the full honeycomb model, Fig.~\ref{App_EM} shows that also the ground state energies $E_M$ for various vortex lattice spacings $D$ exhibit periodicity with the minima occurring for $D \approx n\lambda/2$. For other spacings the system could continuously lower its energy through dimerization. A maximum dimerization of $\delta = \lambda/4$ could occur for the most unstable cases of $D=(2n-1)\lambda/4$. 

However, dimerization is not the only instability of uniform configurations given that the vortices can freely relax into a minimal energy configuration. Instead of dimerizing, the whole lattice may uniformly shrink/expand such that the every vortex position coincides with the interaction oscillations minima/maxima. While simulating this is hard in the honeycomb model, this can be easily studied using the idealized model with couplings \rf{tl_stagger_ideal}. Studying the ground state energy under such process, we again find periodic oscillations that are qualitatively similar to dimerization. However, the global expansion/contraction will in general give a lower energy configuration than the dimerized one. This can be understood as the system minimizing the energy in all three directions, in contrast to dimerization minimizing it only in one direction. Thus given that the vortex lattice could freely relax into a minimal energy configuration and that no other dynamics were relevant, a global expansion/contraction would be favoured over dimerization. Due to the exponential decay of the Majorana tunneling amplitude, the system specific microscopics, such as a Coulomb repulsion in superconductors or fractional quantum Hall liquids, are rarely neglible though. Our study of the effective Majorana model shows that dimerizing or global expansion/contraction instabilities {\it may} occur in systems with Majorana lattices. Whether they {\it do} occur in a given system is subject to the system specific microscopics.

\section{Supplementary data for the perturbed nucleated phases} 
\label{App_data}

In this Appendix we present supplementary data for the perturbed vortex lattices.

\subsection{Anisotropic interactions}

\begin{figure}[h]
\begin{tabular}{c}
\begin{tabular}{cc}
\includegraphics[width=4.2cm]{phasediagram_0v.pdf} & \includegraphics[width=4.2cm]{phasediagram_vs.pdf} 
\end{tabular} \\
\includegraphics[width=8.4cm]{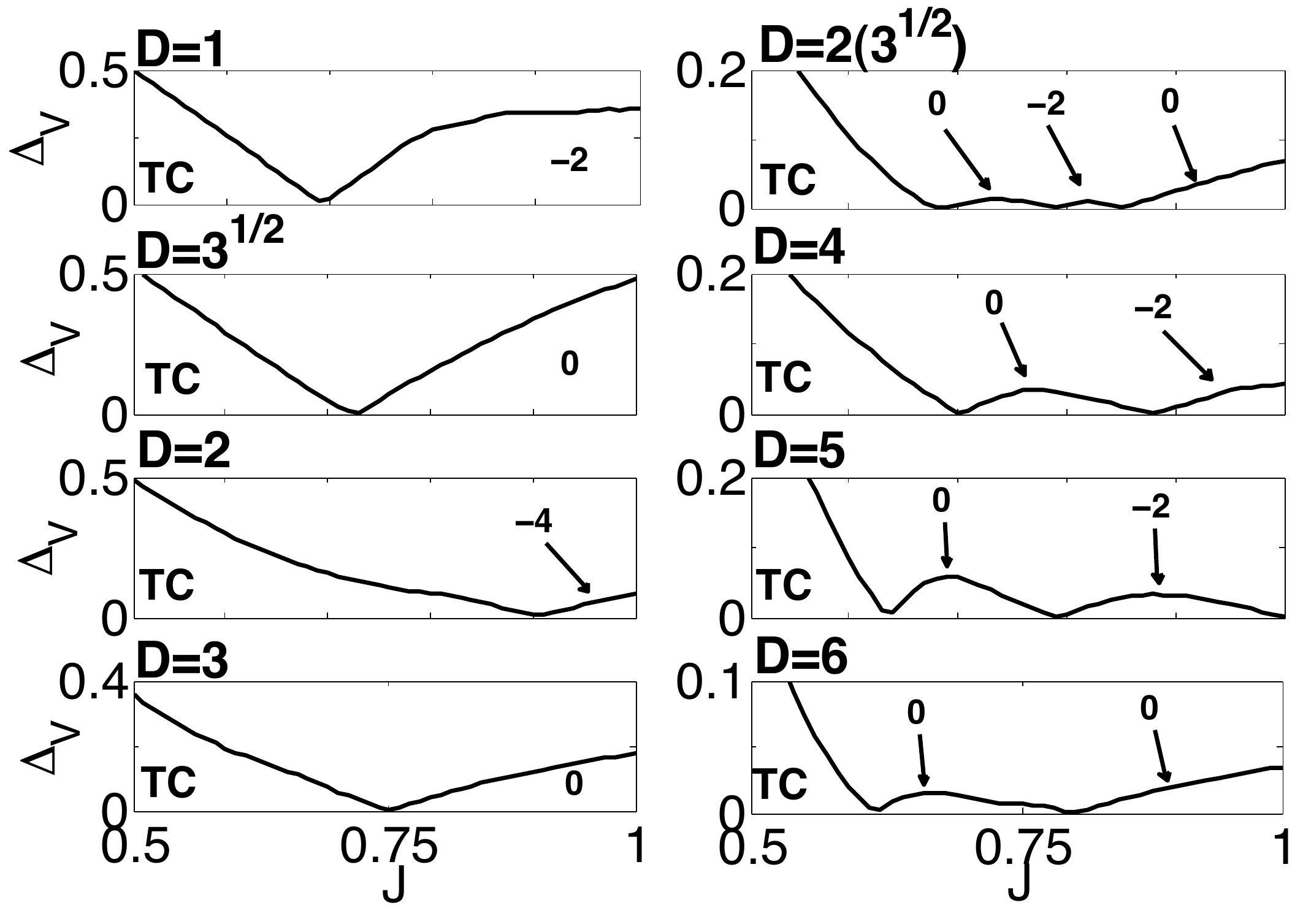}
\end{tabular}
\caption{\label{App_PD} {\it Top:} In the presence of a vortex superlattice, the non-Abelian phase ($\nu=-1$) is always replaced by one or more Abelian phases (even Chern numbers $\nu$) and the strong pairing TC phases are enlargened. {\it Bottom:} The energy gaps $\Delta_V$ and the Chern numbers $\nu$ along the cut $\frac{1}{2} \leq J \leq 1$ (we parametrize here $J=J_x=J_y$ and set $J_z=1$) shown above in the presence of a vortex lattice of spacing $D$. The data is for $K=0.1$. }
\end{figure}

Fig.~\ref{App_PD} shows data for vortex lattices with spacings $1 \leq D \leq 6$, which all show how the non-Abelian phase characterized by $\nu=-1$ is always replaced in the presence of vortex lattices by one or more nucleated Abelian phases characterized by even Chern numbers. In the absence of a vortex lattice the transition between the weak-pairing like non-Abelian phase and the strong-pairing like Abelian phase (TC) occurs for $J=1/2$, while in the presence of one the transition between the nucleated phase and the TC phase occurs always for some s$D$ dependent $J_c > 1/2$.

\begin{figure}[h]
\includegraphics[width=8.4cm]{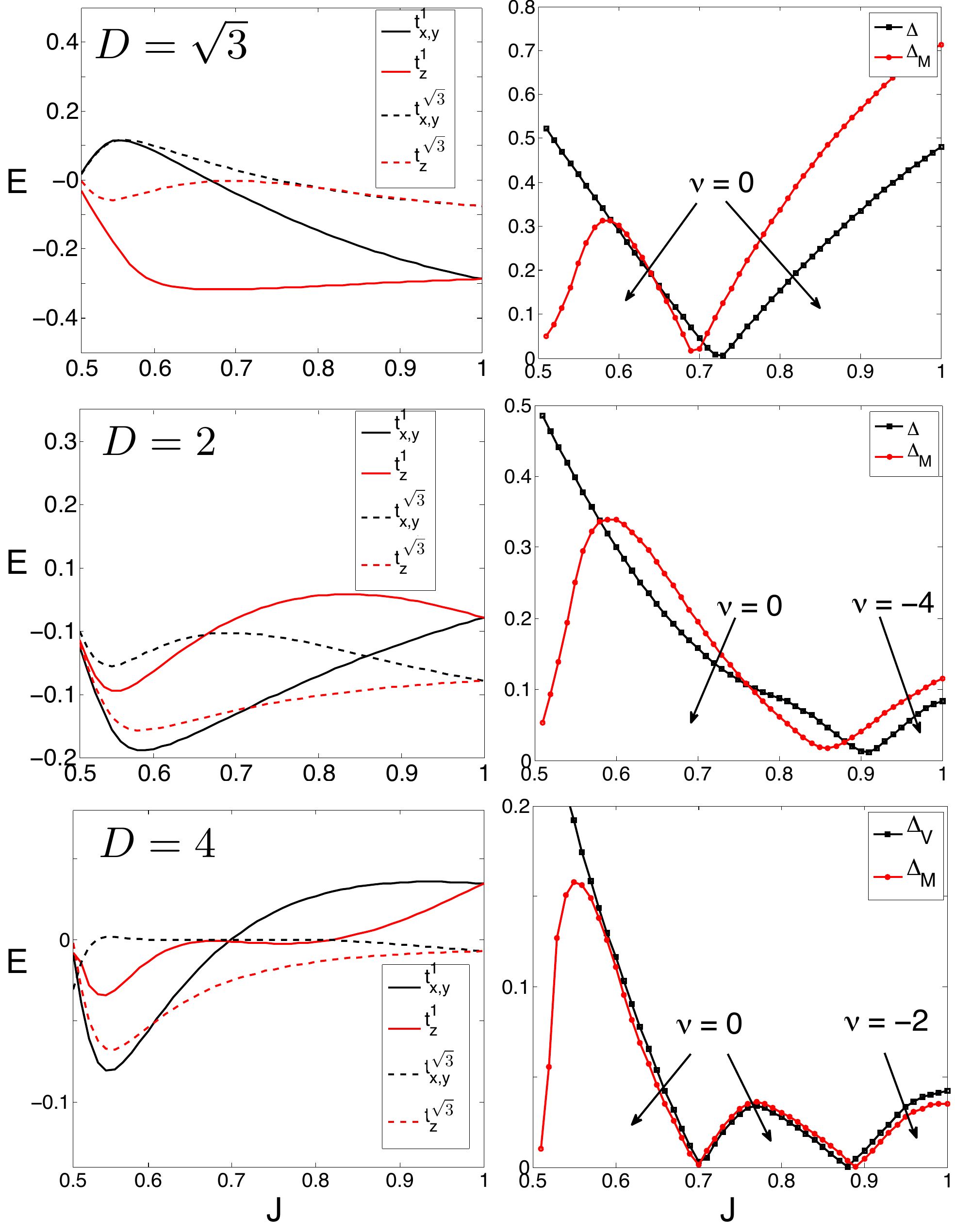}
\caption{\label{App_Jdeform} The anisotropic tunneling couplings \rf{tl_anis} (left) and the energy gaps $\Delta_M$ predicted by them (right) for $D=\sqrt{3},2$ and $4$ vortex lattices. The data is for $K=0.1$.  }
\end{figure}

Fig.~\ref{App_Jdeform} shows the prediction by the effective Majorana model for the $D=\sqrt{3},2$ and $4$ vortex lattices. While the Chern numbers are correctly predicted in all cases, the prediction for the nucleated gap and the precise location of the phase transition becomes more accurate for sparser vortex lattices and for the  $J \to 1$ regime. The reason is the same for both cases. The coherence length of the underlying non-Abelian phase increases, as the energy gap decreases, as one approaches the phase transition ($J \to 1/2$). For both tightly packed vortex lattices, as well as for regimes near the phase transition, the vortex lattice spacing becomes comparable or smaller than the coherence length $\xi$ of the underlying non-Abelian phase. In this regime individual vortices become ill-defined and the simple description by our Majorana model breaks down.

\subsection{Dimerized vortex lattices}

\begin{figure}[h]
\begin{tabular}{c}
\includegraphics[width=8cm]{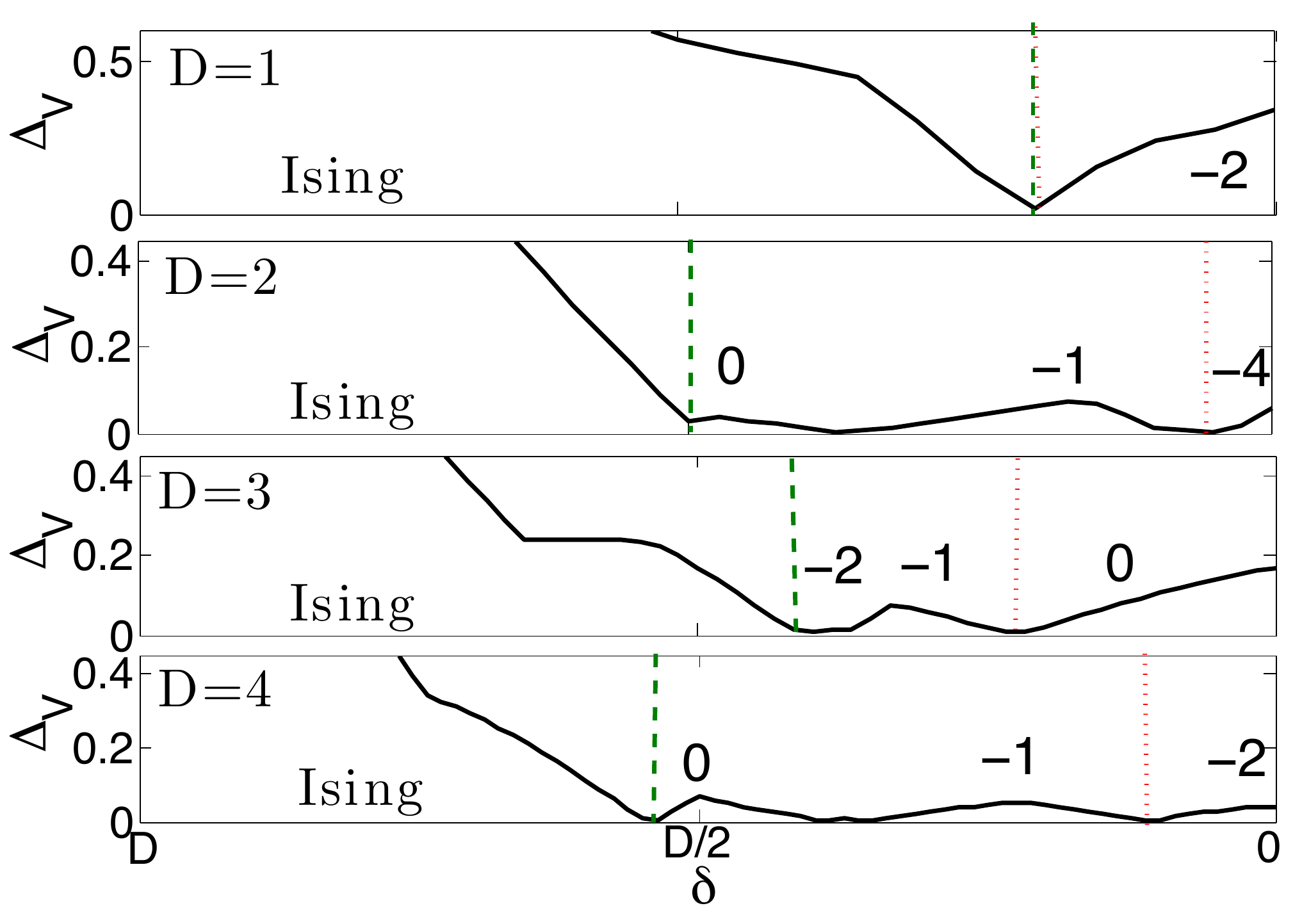}
\end{tabular}
\caption{\label{App_data_stagger} The nucleated gap $\Delta_V$ as a function of the dimerization $\delta$ for some different vortex lattice spacings $D$. The numbers in the plots are the Chern numbers of the intermediate phases. The dashed green line denotes the critical deformation $\delta_C^{NA}$ that recovers the non-Abelian phase in the absence of the vortex lattice, while the dotted red line is the critical deformation $\delta_c$ that drives the system out of the nucleated phase in the presence of the unperturbed vortex lattice. The data is for $K=0.1$.}
\end{figure}

Fig.~\ref{App_data_stagger} shows data for various dimerized vortex lattices. In each case the system is driven out of the nucleated phase for some $D$ dependent critical dimerization $\delta_c$. There will in general be some intermediate nucleated phases before the underlying non-Abelian phase is recovered around $\delta \gtrsim D/2$.

\subsection{Random local disorder}

\begin{figure}[h]
\begin{tabular}{cc}
\includegraphics[width=4cm]{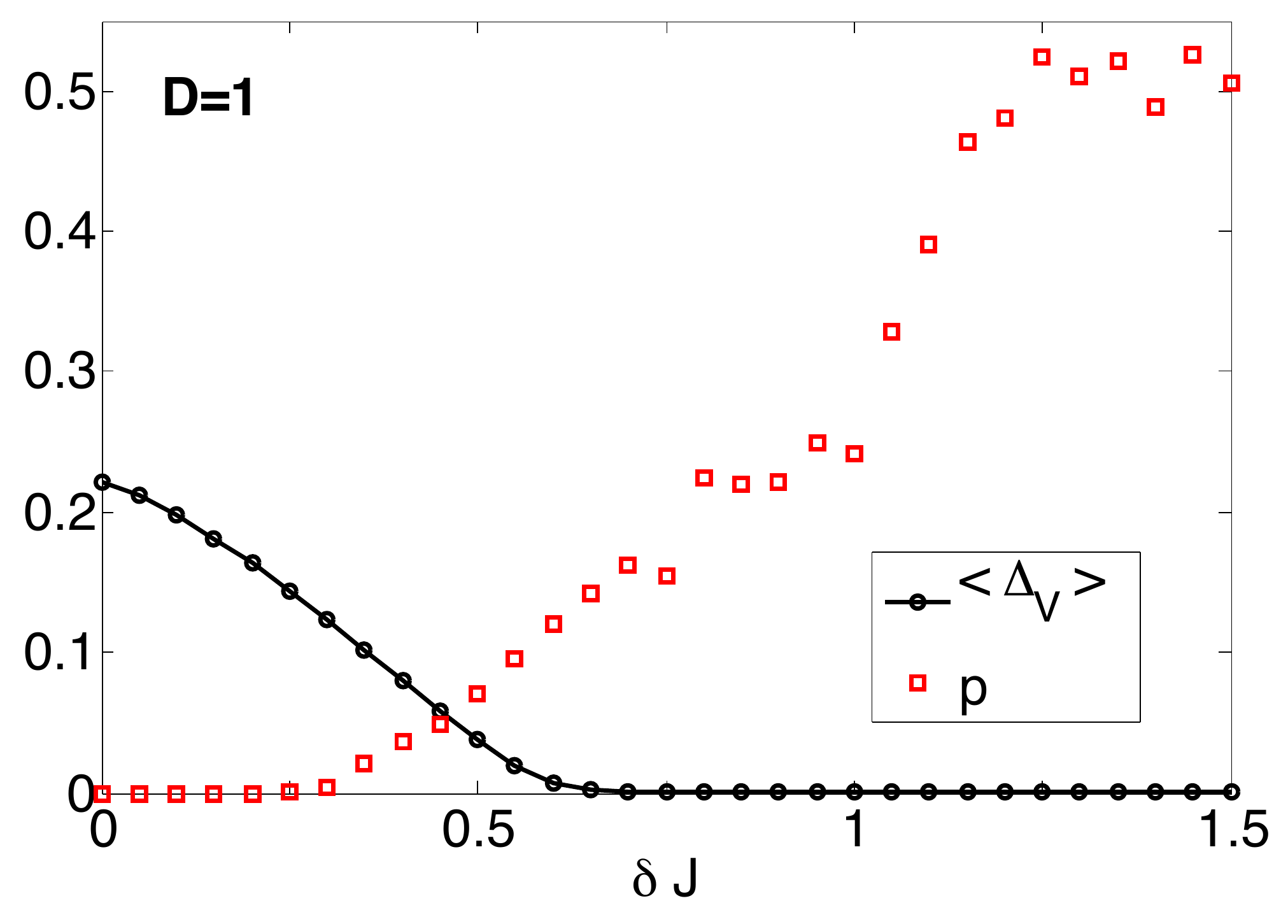} & \includegraphics[width=4cm]{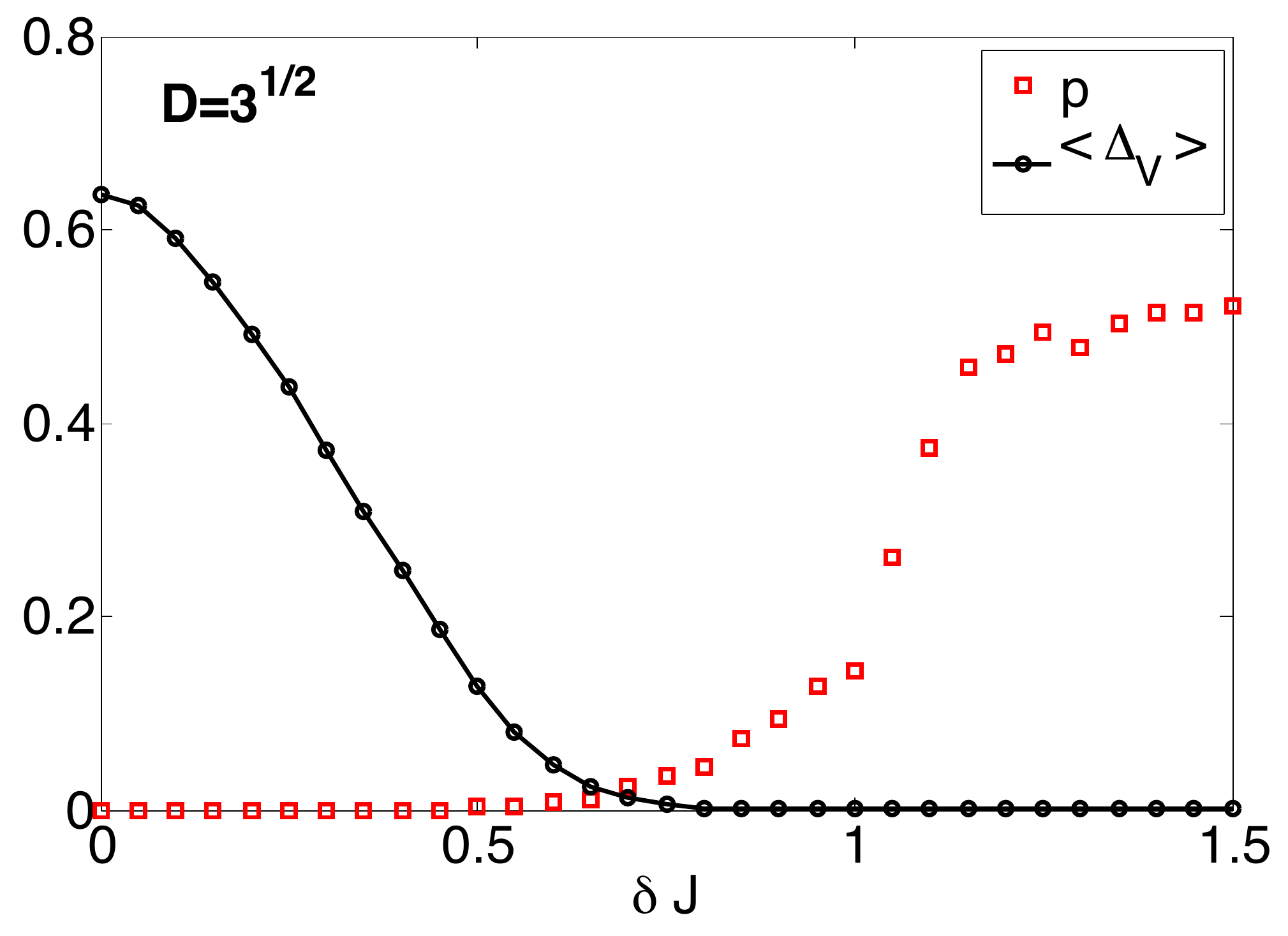}  \\
\includegraphics[width=4cm]{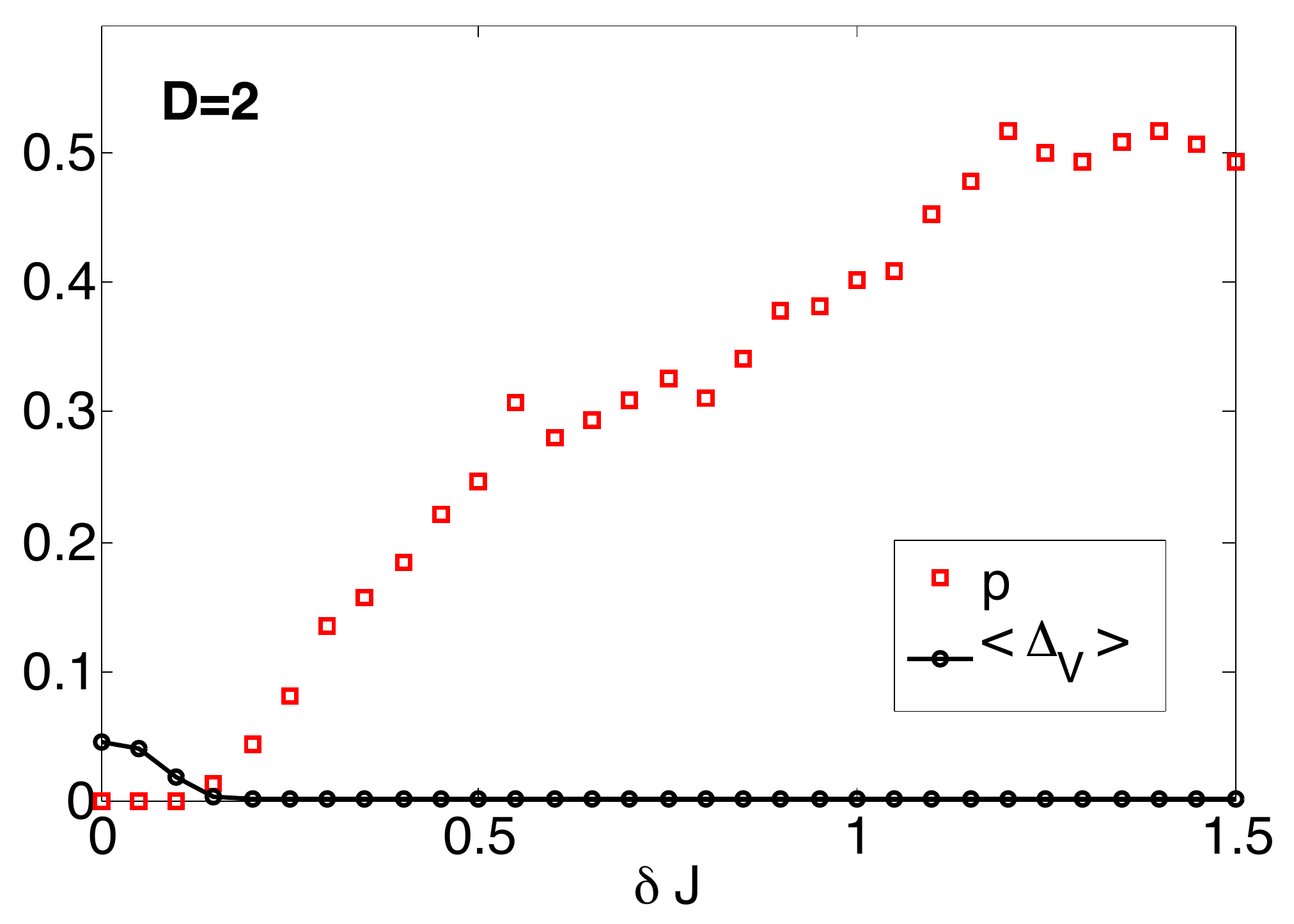} & \includegraphics[width=4cm]{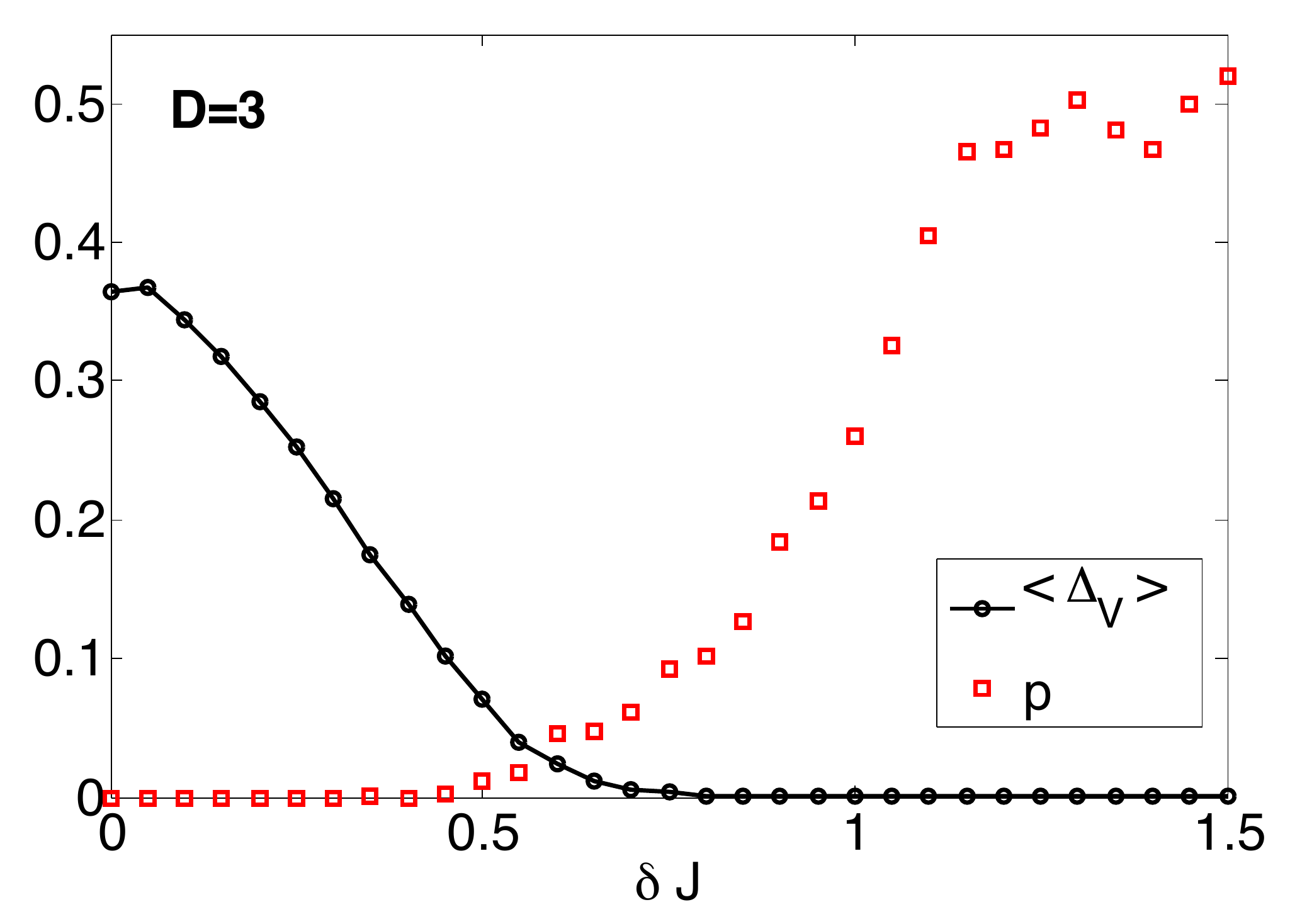} 
\end{tabular} 
\caption{\label{App_p} The correlation between the onset of finite tunneling sign flip probability $p$ and the closure of the disorder averaged gap $\langle \Delta_V \rangle$ in the presence of various vortex lattices of spacing $D$.  The data is for $K=0.05$, averaged over $200$ disorder samples and calculated using a finite system of $L \times L$ plaquettes ($L=40$, $3.2 \cdot 10^3$ sites, $L/\xi \approx 15$ in the clean limit).  }
\end{figure}

Fig.~\ref{App_p} shows data on the correlation between the onset of finite $t_{ij}$ sign flip propability $p$ and the collapse of the of the disorder averaged nucleated gap $\langle \Delta_V \rangle$. As the disorder strength $\delta J$ is increased, a finite probability for the tunneling amplitudes to flip signs always develops before the disorder averaged energy gap closes. 

\begin{thebibliography}{99}
\bibitem{Pachos12}
J. K. Pachos, {\em Introduction to Topological Quantum Computation}, Cambridge University Press (2012).

\bibitem{Alicea11}
J. Alicea, Y. Oreg, G. Refael, F. von Oppen and M. P. A. Fisher, Nat. Phys. {\bf 7}, 412-417 (2011).

\bibitem{Halperin12}
B. I. Halperin, Y. Oreg, A. Stern, G. Refael, J. Alicea and F. von Oppen, Phys. Rev. B {\bf 85}, 144501 (2012).

\bibitem{Hyart13}
T. Hyart, B. van Heck, I. C. Fulga, M. Burrello, A. R. Akhmerov and C. W. J. Beenakker, Phys. Rev. B {\bf 88}, 035121 (2013).

\bibitem{Bonderson09}
P. Bonderson, Phys. Rev. Lett. {\bf 103}, 110403 (2009).

\bibitem{Baraban09}
M. Baraban, G. Zikos, N. Bonesteel and S.H. Simon, Phys. Rev. Lett. {\bf 103}, 076801 (2009).

\bibitem{Cheng09}
M. Cheng, R.M. Lutchyn, V. Galitski and S. Das Sarma, Phys. Rev. Lett. {\bf 103}, 107001 (2009).

\bibitem{Lahtinen11}
V. Lahtinen, New. J. Phys. {\bf 13}, 075009 (2011).

\bibitem{Sau10}
A.Y.~Kitaev, Phys. Usp. {\bf 44}, 131 (2001).

\bibitem{Rosenow12}
B. Rosenow and S.H. Simon, Phys. Rev. B {\bf 85}, 201302, (2012).

\bibitem{Ben-Shach13}
G. Ben-Shach, C. R. Laumann, I. Neder, A. Yacoby and B. I. Halperin, Phys. Rev. Lett. {\bf 110}, 106805 (2013).

\bibitem{Gils09}
C. Gils, E. Ardonne, S. Trebst, A.W.W. Ludwig, M. Troyer and Z. Wang, Phys. Rev. Lett. {\bf 103}, 070401 (2009).

\bibitem{Ludwig11}
\newblock A.W.W. Ludwig, D. Poilblanc, S. Trebst and M. Troyer, New J. Phys. {\bf 13}, 045014 (2011).

\bibitem{Grosfeld06}
E. Grosfeld and A. Stern, Phys. Rev. B {\bf 73}, 201303 (2006).

\bibitem{Silaev13}
M. A. Silaev, Phys. Rev. B {\bf 88}, 064514 (2013).

\bibitem{Biswas13}
R. Biswas, Phys. Rev. Lett. {\bf 111}, 136401 (2013).

\bibitem{Zhu10}
H. Zhu {\it et al.}, Phys. Rev. Lett. {\bf 105}, 126803 (2010).

\bibitem{Kells12}
G. Kells, D. Meidan and P.W. Brouwer, Phys. Rev. B {\bf 85}, 060507(R) (2012).

\bibitem{Kitaev06}
\newblock A.Y. Kitaev,  Ann. Phys. {\bf 321}, 2 (2006).

\bibitem{Lahtinen12}
\newblock V. Lahtinen, A.W.W. Ludwig, J.K. Pachos and S. Trebst, Phys. Rev. B {\bf 86}, 075115 (2012). 

\bibitem{Laumann12}
C.R. Laumann, A.W.W. Ludwig, D.A. Huse and S. Trebst, Phys. Rev. B {\bf 85}, 161301(R) (2012).

\bibitem{Yao07}
H. Yao and S.A. Kivelson, Phys. Rev. Lett. {\bf 99}, 247203 (2007).

\bibitem{Kells11}
G. Kells, J. Kailasvuori, J. Slingerland and J. Vala, New J. Phys. {\bf 13}, 095014 (2011).

\bibitem{Read00}
N. Read and D. Green, Phys. Rev. B {\bf 61}, 10267 (2000).

\bibitem{Moore91}
G. Moore and N. Read, Nucl. Phys. B {\bf 360}, 362 (1991).

\bibitem{Kells13}
G. Kells, V. Lahtinen and J. Vala,  Phys. Rev. B {\bf 89}, 075122 (2014).

\bibitem{Dalibard13}
N. R. Cooper and J. Dalibard, Phys. Rev. Lett. {\bf 110}, 185301 (2013).

\bibitem{Komineas12}
S. Komineas and N.R. Cooper, Phys. Rev. A {\bf 85}, 053623 (2012).

\bibitem{Yu08}
Y. Yu and Z. Wang, Europhys. Lett. {\bf 84}, 57002 (2008).

\bibitem{Kells09}
G. Kells, J.K. Slingerland and J. Vala, Phys. Rev. B {\bf 80}, 125415 (2009).

\bibitem{Lahtinen08}
V. Lahtinen {\it et al.}, Ann. Phys. {\bf 323}, 2286 (2008).

\bibitem{Pedrocchi11}
F. Pedrocchi, S. Chesi and D. Loss, Phys. Rev. B {\bf 84}, 165414 (2011).

\bibitem{Lahtinen09}
V. Lahtinen and J.K. Pachos, New J. Phys. {\bf 11}, 093027 (2009).

\bibitem{Bolukbasi12}
A.T. Bolukbasi and J. Vala, New J. Phys. {\bf 14}, 045007 (2012).

\bibitem{Lieb94}
\newblock E.H. Lieb,  Phys. Rev. Lett. {\bf 73}, 2158 (1994).

\bibitem{Kells10-2}
G. Kells and J. Vala, Phys. Rev. B {\bf 82}, 125122 (2010).

\bibitem{Lahtinen10}
V. Lahtinen and J.K. Pachos, Phys. Rev. B {\bf 81}, 245132 (2010).

\bibitem{Pachos07}
J.~K. Pachos, Ann. Phys. {\bf 322}, 1254 (2007).

\bibitem{Kamfor09}
M. Kamfor, S. Dusuel, J. Vidal and K.P. Schmidt, J. Stat. Mech. P08010 (2010).

\bibitem{Nash09}
C. Nash and D. O'Connor, Phys. Rev. Lett. {\bf 102}, 147203 (2009).

\bibitem{Kamfor11}
M. Kamfor, S. Dusuel, K.P. Schmidt and J. Vidal, Phys. Rev. B {\bf 84}, 153404 (2011).

\bibitem{Bais09}
F.A. Bais and J.K. Slingerland, Phys. Rev. B {\bf 79}, 045316 (2009).

\bibitem{Gurarie07}
V. Gurarie and L. Radzihovsky, Phys. Rev. B {\bf 75}, 212509 (2007).

\bibitem{Varney12}
C. N. Varney, K. A. H. Sellin, Q. Wang, H. Fangohr, E. Babaev, J. Phys.: Cond. Matt. {\bf 25}, 415702 (2013).

\bibitem{Kraus11}
Y.E. Kraus and A. Stern, New J. Phys. {\bf 13}, 105006 (2011).

\bibitem{Willans10}
A.J. Willans, T.J. Chalker and R. Moessner, Phys. Rev. Lett. {\bf 104}, 237203 (2010).

\bibitem{Chua11}
V. Chua and G.A. Fiete, Phys. Rev. B {\bf 84}, 195129 (2011).

\bibitem{Altland97}
A. Altland and M. R. Zirnbauer, Phys. Rev. B {\bf 55}, 1142 (1997).

\bibitem{Bauer12}
B. Bauer, R.M. Lutchyn, M.B. Hastings and M. Troyer, Phys. Rev. B {\bf 87}, 014503 (2013).

\bibitem{Micheli05}
\newblock A. Micheli, G.~K. Brennen, and P. Zoller, Nature Physics {\bf 2}, 341 (2005).

\bibitem{Duan03}
\newblock L.M. Duan, E.~Demler, and M.D. Lukin, Phys. Rev. Lett. {\bf 91}, 090402 (2003).

\bibitem{Alba13}
J. K. Pachos, E. Alba, V. Lahtinen, J. J. Garcia-Ripoll, Phys. Rev. A {\bf 88}, 013622 (2013).

\bibitem{Mourik12}
V.~Mourik, K.~Zuo, S.~M.~Frolov, S.~R.~Plissard, E.~P.~A.~M.~Bakkers, and L.~P.~Kouwenhoven,
\newblock Science {\bf 336}, 1003 (2012).

\bibitem{Das12}
A.~Das, Y.~Ronen, Y.~Most, Y.~Oreg, M.~Heiblum and H.~Shtrikman,
Nature Physics {\bf 8}, 887-895 (2012).

\bibitem{Churchill13}
H. O. H. Churchill, V. Fatemi, K. Grove-Rasmussen, M. T. Deng, P. Caroff, H. Q. Xu, C. M. Marcus,
Phys. Rev. B {\bf 87}, 241401(R) (2013).

\bibitem{Chaloupka10}
J. Chaloupka, G. Jackeli and G. Khaliullin, Phys. Rev. Lett. {\bf 105}, 027204 (2010).

\bibitem{Jiang11}
H.-C. Jiang, Z.-C. Gu, X.-L. Qi, S. Trebst, Phys. Rev. B {\bf 83}, 245104 (2011).

\end{thebibliography}
\end{document}